\documentclass[twocolumn,caps,showpacs,superscriptaddress]{revtex4}

\usepackage{graphicx,color,soul}
\usepackage{latexsym}
\usepackage{psfrag}
\usepackage{amsmath,amssymb}        
\usepackage[draft=false]{hyperref}

\begin{document}


\title{Exact solution of the 1D Riemann problem in Newtonian and relativistic hydrodynamics}


\author{F. D. Lora-Clavijo, J. P. Cruz-P\'erez, F. S. Guzm\'an, J. A. Gonz\'alez}
\affiliation{Instituto de F\'{\i}sica y Matem\'{a}ticas, Universidad
              Michoacana de San Nicol\'as de Hidalgo. Edificio C-3, Cd.
              Universitaria, 58040 Morelia, Michoac\'{a}n,
              M\'{e}xico.}


\date{\today}


\begin{abstract} 
Some of the most interesting scenarios that can be studied in astrophysics, contain fluids and plasma moving under the influence of strong gravitational fields. To study these problems it is required to implement numerical algorithms robust enough to deal with the equations describing such scenarios, which usually involve hydrodynamical shocks. It is traditional that the first problem a student willing to develop research in this area is to numerically solve the one dimensional Riemann problem, both Newtonian and relativistic. Even a more basic requirement is the construction of the exact solution to this problem in order to verify that the numerical implementations are correct. We describe in this paper the construction of the exact solution and a detailed procedure of its implementation.
\end{abstract}




\maketitle

\section{Introduction}
\label{sec:introduction}

High energy astrophysics has become one of the most important subjects in astrophysics because it involves phenomena associated to high energy radiation, modeled with sources traveling at high speeds or sources under the influence of strong gravitational fields like those due to black holes or compact stars. Current models involve a hydrodynamical description of the luminous source, and therefore hydrodynamical equations have to be solved.

In this scenario, due to the complexity of the system of equations it is required to apply numerical methods able to control the physical discontinuities arising during the evolution of initial configurations, for example the evolution of the front shock in a supernova explosion, the front shock of a jet propagating in space, the edges of an accretion disk, or any shock formed during a violent process. The study of these systems involve the implementation of advanced numerical methods, being two of the most efficient and robust ones the high resolution shock capturing methods and smooth particle hydrodynamics which are representative of Eulerian and Lagrangian descriptions of hydrodynamics, each one with pros and cons.

It is traditional that a first step to evaluate how appropriate the implementation of a numerical method is, requires the comparison of numerical results with an exact solution in a simple situation. The simplest problem in hydrodynamics is the 1D Riemann problem. This is an excellent test case because it has an exact solution in the Newtonian case (e.g. \cite{Toro}) and also in the relativistic regime \cite{MartiLR,Marti}, where codes dealing with high Lorentz factors are expected to work properly. 
From our experience we have found that the existent literature about the construction of the exact solution is not as explicit as it may be expected by students having their first contact with this subject. This is the reason why we present a paper that is very detailed in the construction and implementation of the solution. We focus on the solution of the problem and omit some of the mathematical background that is actually very well described in the literature.

The paper is organized as follows. In section \ref{sec:newtonian} we present the Newtonian Riemann problem and how to implement it; in section \ref{sec:relativistic} we present the exact solution to the relativistic case and how to implement it. Finally in section \ref{sec:final} we present some final comments.

\section{Riemann problem for the Newtonian Euler equations}
\label{sec:newtonian}

The Riemann problem is an initial value problem for a gas with discontinuous initial data, whose evolution is ruled by Euler's equations. The set of Euler's equations determine the evolution of the density of gas, its velocity field and either its pressure or total energy. A comfortable way of writing such equations involves a flux balance form as follows

\begin{equation}
\partial_t {\bf u} + \partial_x {\bf F}({\bf u}) = 0 \label{eq:Euler}
\end{equation}

\noindent where ${\bf u}=(u_1,u_2,u_3)^T=(\rho,\rho v,E)^T$ is a set of conservative variables and ${\bf F}$ is a flux vector,   where $\rho$ is the mass density of the gas, $v$ its velocity and $E=\rho(\frac{1}{2}v^2 + \varepsilon)$, with $\varepsilon$ the specific internal energy of the gas. The enthalpy of the system is given by the expression $H=\frac{1}{2}v^2 + h$, where $h$ is the specific internal enthalpy given by $h=\varepsilon + p/\rho$, where $p$ is the pressure of the gas. The fluxes are explicitly in terms of the primitive variables $\rho,v,p$ and the conservative variables \cite{Toro}

\begin{equation}
{\bf F}({\bf u}) = \left(
	\begin{array}{c}
	\rho v \\
	\rho v^2 + p \\
	v(E+p)
	\end{array}
		\right) = 
	 \left(
	\begin{array}{c}
	u_2 \\
	\frac{1}{2}(3-\Gamma)\frac{u_2^2}{u_1}+(\Gamma-1)u_3 \\
	\Gamma\frac{u_2}{u_1}u_3-\frac{1}{2}(\Gamma-1)\frac{u_2^3}{u_1^2}
	\end{array}
		\right).
\nonumber
\end{equation}

The initial data of the Riemann problem is defined as follows

\begin{equation}
{\bf u} = \left\{
        \begin{array}{ll}
                {\bf u}_L, & x<x_0 \\
                {\bf u}_R, & x >x_0,
        \end{array}
        \right.\nonumber
\end{equation}

\noindent where ${\bf u}_L$ and ${\bf u}_R$ represent the values of the gas properties on a chamber at the left and at the right from an interface between the two states at $x=x_0$ that exists only at initial time.

The evolution of the initial data is described by the characteristic information of the system of equations, and this is why the properties of the Jacobian matrix are important. The Jacobian matrix of the system of equations is $A({\bf u})=\frac{\partial{\bf F}}{\partial{\bf u}}$ and explicitly reads

\begin{equation}
{\bf A} = \left(
	\begin{array}{ccc}
	0 &1 & 0\\
	\frac{1}{2}(\Gamma-3)v^2 & (3-\Gamma)v & \Gamma -1\\
	(\Gamma -1)v^3 -  \frac{\Gamma v E}{\rho} & \frac{\Gamma E}{\rho} - \frac{3}{2}(\Gamma -1)v^2 &  \Gamma v\\
	\end{array}
	\right).
	\nonumber
\end{equation}

\noindent Its eigenvalues satisfy the condition $\lambda_1({\bf u}) < \lambda_2 ({\bf u}) < \lambda_3 ({\bf u})$ and are given by

\begin{eqnarray}
\lambda_1 &=& v-a\label{eq:lambda_minus}  \\
\lambda_2 &=& v\label{eq:lambda_0}  \\
\lambda_3 &=& v+a\label{eq:lambda_plus}
\end{eqnarray}

\noindent where $a=\sqrt{\frac{\partial p}{\partial \rho}}|_{s}$ is the speed of sound in the gas, which depends on the equation of state. For the ideal gas $p=\rho \varepsilon (\Gamma-1)$, where $\Gamma$ is the ratio between the specific heats at constant pressure and volume $\Gamma=c_p / c_v$, the speed of sound is $a=\sqrt{\frac{\Gamma p} {\rho}}$. On the other hand, the eigenvectors of the Jacobian matrix read

\begin{equation}
{\bf r}_1 = \left(
	\begin{array}{c}
	1\\ v-a \\H-av
	\end{array}
	\right), ~
{\bf r}_2 = \left(
	\begin{array}{c}
	1\\ v \\\frac{1}{2}v^2
	\end{array}
	\right), ~
{\bf r}_3 = \left(
	\begin{array}{c}
	1\\v+a\\H+av
	\end{array}
	\right).
	\nonumber
\end{equation}

\noindent The eigenvectors ${\bf r}_1,~{\bf r}_2,~{\bf r}_3$ are classified in the following way:

\begin{itemize}
\item they are called genuinely non-linear when satisfy the condition $\nabla_u \lambda_i \cdot {\bf r}_i({\bf u}) \ne 0$.

\item and linearly degenerate when $\nabla_u \lambda_i \cdot {\bf r}_{i}({\bf u}) = 0$.
\end{itemize}

It happens that ${\bf r}_2$ is linearly degenerate and represents a contact discontinuity, however the other two are genuinely non-linear.

Depending on the particular region of the solution we will use both the Riemann invariant conditions for rarefaction waves and the Rankine Hugoniot conditions for shocks and contact discontinuities. The Riemann invariants are based on the self-similarity property of the solution in some regions, in the sense that the solution depends on the spatial and time coordinates $(x,t)$ with the combination $(x-x_0)/t$; it can be seen that such behavior implies that the following conditions  hold \cite{LeVeque}

\begin{equation}
\frac{d u_1}{{\bf r}^{i}_{1}} = \frac{d u_2}{{\bf r}^{i}_{2}} = \frac{d u_3}{{\bf r}^{i}_{3}} 
\label{eq:riemann_invariants}
\end{equation}

\noindent where $i$ indicates the component of a given eigenvector. On the other hand,  the Rankine Hugoniot conditions relate states on both sides of a shock wave or a contact discontinuity

\begin{equation}
\Delta {\bf F} = V \Delta {\bf u},\label{eq:RHConditions}
\end{equation}

\noindent which are simply jump conditions, where $\Delta {\bf u}$ is the size of the discontinuity in the variables, $V$ is the velocity of  either the contact discontinuity or shock and $\Delta {\bf F}$ is the change of the flux across the discontinuity.

\subsection{Contact discontinuity waves}

The contact discontinuity is described by the second eigenvector and evolves with velocity $\lambda_2$. Let us then analyze the second eigenvector. In this case the Riemann invariant conditions read

\begin{equation}
\frac{d\rho}{1} = \frac{d(\rho v)}{v} = \frac{dE}{\frac{1}{2}v^2}.
\nonumber
\end{equation}

\noindent These relations implies that $d(\rho \varepsilon)=dv=0$, further implying that $p=constant$ and $v=constant$ across the contact wave. In order to relate the two sides from the contact discontinuity we use the Rankine-Hugoniot conditions, which are given by 

\begin{eqnarray}
 \rho_L v_L- \rho_R v_R &=& V_c (\rho_L-\rho_R), \label{eq:1st_contact_newtonian}\\
 \rho_L v^2_L +  p^2_L - \rho_R v^2_R +  p^2_R &=& V_c (\rho_L v_L - \rho_R v_R), \label{eq:2nd_contact_newtonian}\\
 v_L (E_L +p_L ) -v_R(E_R+p_R) &=& V_c (v_L(E_L+p_L)  \nonumber \\
 	&-&v_R(E_R+p_R)). \label{eq:3rd_contact_newtonian}
\end{eqnarray}

\noindent Here $V_c$ is the velocity of propagation of the contact discontinuity. 

The discontinuity travels at speed $\lambda^0=v$ therefore the $V_c=v$. For this reason from equation (\ref{eq:1st_contact_newtonian}) follows that 
$v_L=v_R=V_c$. As a consequence of this, equation (\ref{eq:2nd_contact_newtonian}) gives the condition $p_L=p_R$, which implies (\ref{eq:3rd_contact_newtonian}) is satisfied. Notice that no condition on the density arises, which allows the density to be discontinuous.

\subsection{Rarefaction waves}

At this point we do not know the nature of  waves 1 and 3, and we can assume they may be rarefaction waves. Once again we use the Riemann invariant equalities, which for vectors 1 and 3 read

\begin{eqnarray}
\frac{d \rho}{1} &=& \frac{d(\rho v)}{v-a} = \frac{dE}{H-av}, \nonumber\\
\frac{d \rho}{1} &=& \frac{d(\rho v)}{v+a} = \frac{dE}{H+av}. \nonumber
\end{eqnarray}

\noindent  Manipulation of these equalities results in the following equations

\begin{eqnarray}
\frac{d\rho}{dv} &=& -\frac{\rho}{a}~~~ {\rm for} ~\lambda_1,\label{eq:rar_ri_1}\\ 
\frac{d\rho}{dv} &=& \frac{\rho}{a}~~~ {\rm for} ~\lambda_3,\label{eq:rar_ri_3}\\
\frac{d\varepsilon}{d\rho} &=& \frac{p}{\rho^2}~~~ {\rm for ~ both}~ \lambda_1 ~{\rm and}~
\lambda_3.\label{eq:rar_ri_eps}
\end{eqnarray}

The next step is to integrate these equations assuming an equation of state, in our case the ideal gas. From (\ref{eq:rar_ri_eps}) we obtain 

\begin{equation}
p = K\rho^{\Gamma}\label{eq:isentropic}
\end{equation}

\noindent where $K$ is a constant. A rarefaction process is isentropic (unlike a shock), and therefore the states at the left and at the right from the wave obey (\ref{eq:isentropic}) with the same constant $K$.

Using this expression for $p$ in the speed of sound we have $a=\sqrt{K\Gamma \rho^{\Gamma-1}}=\sqrt{\Gamma p /\rho}$, which substituted into (\ref{eq:rar_ri_1},\ref{eq:rar_ri_3}) results in

\begin{equation}
v = \pm \int \sqrt{K\Gamma\rho^{\Gamma-3}} d\rho +k = \pm \frac{2a}{\Gamma-1}+k,\label{eq:vel_rarefaction}
\end{equation}

\noindent where $+$ stands for the wave moving to the right (the case of $\lambda_3$ and ${\bf r}_3$ corresponding to a rarefaction wave) and $-$ when moving to the left (the case of $\lambda_1$ and ${\bf r}_1$ corresponding to a rarefaction wave), where $k$ is an integration constant and therefore the velocity is constant as well. This property allows us to set relations between the velocity of the gas on the state at the left and at the right from the rarefaction wave, explicitly there are two possible cases:

\begin{itemize}
\item[i)] When the wave is moving to the left, condition (\ref{eq:vel_rarefaction}) implies that 

\begin{equation} 
v_L + \frac{2a_L}{\Gamma-1} = v_R + \frac{2a_R}{\Gamma-1}.\label{eq:rar_vel_left}
\end{equation}

\item[ii)] When the wave is moving to the right, condition (\ref{eq:vel_rarefaction}) implies

\begin{equation} 
v_L - \frac{2a_L}{\Gamma-1} = v_R - \frac{2a_R}{\Gamma-1}\label{eq:rar_vel_right}
\end{equation}

\end{itemize}

When the wave is moving to the left, we assume information from the left state is available and we look for expression of the variables on the state to the right from the wave. For the velocity of the fluid at the right state we then have from (\ref{eq:rar_vel_left})

\begin{equation}
v_R = v_L - \frac{2}{\Gamma-1}[a_R-a_L],
\end{equation}

\noindent now considering that the speed of sound on both sides obeys $a=\sqrt{K\Gamma \rho^{\Gamma-1}}=\sqrt{\Gamma p /\rho}$ (see (\ref{eq:isentropic}))

\begin{equation}
a_R = a_L \left( \frac{p_R}{p_L} \right)^\frac{\Gamma-1}{2\Gamma},\label{eq:sound_rar_left}
\end{equation}

\noindent a useful expression for $v_R$ arises

\begin{equation}
v_R = v_L - \frac{2a_L}{\Gamma-1} \left[ \left( \frac{p_R}{p_L}\right)^{\frac{\Gamma-1}{2\Gamma}}-1\right].
\label{eq:R_L}
\end{equation}

\noindent The only unknown quantity is $p_R$.

On the other hand, when the wave is moving to the right we assume we know the information at the state at the right from the wave, then we search for expressions of the variables on the state at the left. For the velocity we find according to (\ref{eq:rar_vel_right})

\begin{equation}
v_L = v_R - \frac{2}{\Gamma-1}[a_R-a_L],
\end{equation}

\noindent and the speed of sound on both sides obeys

\begin{equation} \label{eq:aaL}
a_L = a_R \left( \frac{p_L}{p_R} \right)^\frac{\Gamma-1}{2\Gamma},
\end{equation}
\noindent which finally implies

\begin{equation}
v_L = v_R - \frac{2a_R}{\Gamma-1} \left[ 1-\left( \frac{p_L}{p_R}\right)^{\frac{\Gamma-1}{2\Gamma}}\right].
\label{eq:v_L_rarefactionR}
\end{equation}

\noindent The only unknown quantity in this case is $p_L$.

The rarefaction zone has a finite size, bounded by two curves, the tail and the head. The head of the wave is the line of the front of the wave and the tail is the boundary left behind the wave. The region in the middle is called the fan of the rarefaction wave.

The velocity of all the particles between the head and the tail obeys the following expression 

\begin{equation} \label{eq:vfan}
\frac{x-x_0}{t} = v \pm a,
\end{equation}

\noindent where + is used when the wave is propagating to the right and the $-$ when it is moving to the  left. Then, when the wave is moving to the left, using this expression we have $a_R = v_R - (x-x_0)/t$, which substituted into (\ref{eq:R_L}) provides the following expression for the velocity of the gas on the state at the right from the wave is

\begin{equation}
v_R = \frac{2}{\Gamma+1}\left[ a_L + \frac{1}{2} (\Gamma-1) v_L + \frac{x-x_0}{t}\right].\label{eq:vR_rarL}
\end{equation}

Then it is possible to calculate the pressure and density as well. Substituting (\ref{eq:vR_rarL}) into (\ref{eq:rar_vel_left}) and (\ref{eq:sound_rar_left}) we obtain an expression for the pressure also at the state to the right

\begin{equation}
p_R = p_L \left[ \frac{2}{\Gamma + 1} + \frac{\Gamma-1}{a_L (\Gamma+1)} \left( v_L - \frac{x-x_0}{t} \right)\right]^{\frac{2\Gamma}{\Gamma-1}}.\label{eq:pR_rarL}
\end{equation}

\noindent Now, using this into (\ref{eq:isentropic}) implies the expression for the density

\begin{equation}
\rho_R = \rho_L \left[ \frac{2}{\Gamma + 1} + \frac{\Gamma-1}{a_L (\Gamma+1)} \left( v_L - \frac{x-x_0}{t} \right)\right]^{\frac{2}{\Gamma-1}}.\label{eq:rhoR_rarL}
\end{equation}

\noindent Then finally we have expressions for the velocity, pressure and density on the state at the right when the wave is moving to the left.

\noindent Similarly when the wave is moving to the right we have from (\ref{eq:vfan}) that $a_L=v_L + (x-x_0)/t$, which substituted into (\ref{eq:v_L_rarefactionR}) implies the following for the velocity on the state at the left from the wave

\begin{equation}
v_L = \frac{2}{\Gamma+1}\left[ -a_R + \frac{1}{2} (\Gamma-1) v_R + \frac{x-x_0}{t}\right].\label{eq:vL_rarR}
\end{equation}

\noindent In order to obtain the expressions for the pressure and the density, we substitute this last expressions into (\ref{eq:rar_vel_right}) in order to relate the speeds of sound, and then using (\ref{eq:aaL}) we finally obtain the expression for the pressure at the left 

\begin{equation}
p_L = p_R \left[ \frac{2}{\Gamma + 1} - \frac{\Gamma-1}{a_R (\Gamma+1)} \left( v_R - \frac{x-x_0}{t} \right)\right]^{\frac{2\Gamma}{\Gamma-1}}.\label{eq:pL_rarR}
\end{equation}

\noindent Finally using the equation (\ref{eq:isentropic}) we obtain the density

\begin{equation}
\rho_L = \rho_R \left[ \frac{2}{\Gamma + 1} - \frac{\Gamma-1}{a_R (\Gamma+1)} \left( v_R - \frac{x-x_0}{t} \right)\right]^{\frac{2}{\Gamma-1}}.\label{eq:rhoL_rarR}
\end{equation}

\noindent In this way we have relations between the variables on to the state at the left and at the right from a rarefaction wave. These relations will be useful when solving the Riemann problem.






\subsection{Shock waves}

Similar to the previous case, the shock can move either to the right (if $\lambda_3$ and ${\bf r}_3$ correspond to a shock wave) or to the left (if $\lambda_1$ and ${\bf r}_1$ correspond to a shock wave), and for each of the two cases there is known and unknown information. When a shock is moving to the right one is expected to have information of the state at the right from the shock and conversely, when the shock is moving to the left one accounts with information of the state at the left.

Shocks require the use of Rankine Hugoniot conditions (\ref{eq:RHConditions}). We express these conditions in terms of the primitive variables as follows

\begin{eqnarray}
\rho_L v_L - \rho_R v_R &=& S (\rho_L - \rho_R),\nonumber\\
\rho_L v_{L}^{2} + p_L - \rho_R v_{R}^{2} - p_R &=& S (\rho_L v_L - \rho_R v_R),\nonumber\\
v_L (E_L +p_L) - v_R(E_R +p_R) &=& S (E_L - E_R),\nonumber
\end{eqnarray}

\noindent where $S$ is the speed of the wave, which may take the values $v-a$ or $v+a$ depending on whether the wave moves to the left or to the right respectively. Manipulating these equations one gets

\begin{eqnarray}
\rho_L \hat{v}_L &=& \rho_R \hat{v}_R,\label{eq:rh_1}\\
\rho_L \hat{v}_L^2 + p_L &=& \rho_R \hat{v}_R^2 + p_R,\label{eq:rh_2}\\
\hat{v}_L(\hat{E}_L + p_L) &=& \hat{v}_R(\hat{E}_R + p_R), \label{eq:rh_3}
\end{eqnarray}

\noindent where $\hat{v}_L=v_L - S$,  $\hat{v}_R=v_R - S$ are velocities in the rest frame of the shock and $\hat{E}_L=\rho_L\left(\frac{1}{2}\hat{v}_L^2 + \varepsilon_L \right)$ and $\hat{E}_R=\rho_R\left(\frac{1}{2}\hat{v}_R^2 + \varepsilon_R \right)$.  These expressions correspond to the Rankine Hugoniot jump conditions  measured  by an observer located in the rest frame of the shock wave.

From equation (\ref{eq:rh_1}), we introduce the mass flux definition 

\begin{equation}
j = \rho_L \hat{v}_L  = \rho_R\hat{v}_R. \label{eq:MassFlux}
\end{equation}

\noindent Then, from equation (\ref{eq:rh_2}) and the mass flux definition before mentioned , we can get an expression for $j$, which is given by 

\begin{equation}
j=-\frac{p_R-p_L}{\hat{v}_R-\hat{v}_L}=-\frac{p_R-p_L}{v_R-v_L}, \label{eq:MassFlux2}
\end{equation}

\noindent which is a consequence of $j$ being invariant under Galilean transformations. Considering the shock is moving to the left, we would be interested in constructing the variables on the state at the right from the shock and we can start with the velocity, which can be written as  

\begin{equation}\label{eq:thefantastic}
v_R = v_L-\frac{p_R-p_L}{j} .
\end{equation}

\noindent Now, in order  to express the velocity in terms of the pressure and the variables of the state at the left from the shock, we can rewrite (\ref{eq:MassFlux})  as follows 

\begin{equation}\label{eq:pareja}
v_R-S=\frac{j}{\rho_R}, ~~~ v_L-S=\frac{j}{\rho_L}.
\end{equation}

\noindent Thus, substituting this into (\ref{eq:MassFlux2}) we obtain 

\begin{equation}\label{eq:j2}
	j^2=-\frac{p_R-p_L}{\frac{1}{\rho_R}-\frac{1}{\rho_L}}.
\end{equation}

\noindent On the other hand, using equation (\ref{eq:rh_3}) and the expression for the specific internal enthalpy $h$ we can easily get the following expression for the difference of internal specific enthalpies 

\begin{equation}\label{eq:hrmhl}
	h_R-h_L=\frac{1}{2}\left[ \hat{v}_L^2-\hat{v}_R^2 \right],
\end{equation} 

\noindent where $h_L=\varepsilon_L+p_L/\rho_L$  and  $h_R=\varepsilon_R+p_R/\rho_R$.  Now, from equations (\ref{eq:rh_1}) and (\ref{eq:rh_2}) we give expressions for the velocitites measured by the observer located in the rest frame of the shock wave 

\begin{eqnarray}
	\hat{v}_R^2 &=& \frac{\rho_L}{\rho_R} \frac{p_L-p_R}{\rho_L-\rho_R}, \nonumber \\
	\hat{v}_L^2 &=& \frac{\rho_R}{\rho_L} \frac{p_L-p_R}{\rho_L-\rho_R}. \nonumber
\end{eqnarray} 

\noindent  With the substitution of these last equations into (\ref{eq:hrmhl}) and considering the definitions for the specific internal enthalpy mentioned above, we obtain

\begin{equation}
	\varepsilon_R-\varepsilon_L= \frac{1}{2} \frac{(p_L+p_R)(\rho_R-\rho_L)}{\rho_L \rho_R}.
	\nonumber
\end{equation}

\noindent Assuming the gas obeys an ideal equation of state we get an expression for the density as follows

\begin{equation}
	\frac{\rho_R}{\rho_L}=\frac{p_L(\Gamma-1)+p_R(\Gamma+1)}{p_R(\Gamma-1)+p_L(\Gamma+1)}.
	\label{eq:rhoR_shockL}
\end{equation}

\noindent Notice that this expression relates the density among the two sides from the shock. Now, substituting this expression into (\ref{eq:j2}) we obtain

\begin{equation}\label{eq:pareja2}
j^2=\frac{p_R+B_L}{A_L}, ~~~A_L=\frac{2}{(\Gamma+1)\rho_L}, ~~~B_L=\frac{\Gamma-1}{\Gamma+1} p_L.
\end{equation} 

\noindent Thus, the expression for the velocity (\ref{eq:thefantastic}) can be written as follows

\begin{equation}
v_R=v_L-(p_R-p_L) \sqrt{\frac{A_L}{p_R+B_L}}. \label{eq:vR_shockL}
\end{equation}

\noindent From expression (\ref{eq:pareja}) and using (\ref{eq:pareja2}) we express the shock velocity as follows 

\begin{equation}
S = v_L - \sqrt{\frac{p_R(\Gamma +1) + p_L (\Gamma -1)}{2\rho_L}}. \nonumber
\end{equation} 

\noindent Finally, using the sound speed expression $a_L=\sqrt{\frac{p_L \Gamma}{\rho_L}}$ we obtain the final expression for the shock velocity

\begin{equation}
S = v_L - a_L\sqrt{\frac{(\Gamma +1)p_R}{2p_L\Gamma}+\frac{\Gamma-1}{2\Gamma}} \label{eq:SL}.
\end{equation}  

Analogously, when the shock moves to the right, it is possible to construct the expressions for the variables for the state at the left from the shock

\begin{eqnarray}
v_L&=&v_R+(p_L-p_R) \sqrt{\frac{A_R}{p_L+B_R}}\label{eq:L_R},\\
\rho_L&=&\rho_R \frac{p_R(\Gamma-1)+p_L(\Gamma+1)}{p_L(\Gamma-1)+p_R(\Gamma+1)},\label{eq:rho_L_R}\\
S &=& v_R + a_R\sqrt{\frac{(\Gamma +1)p_L}{2p_R\Gamma}+\frac{\Gamma-1}{2\Gamma}}. \label{eq:SR}
\end{eqnarray}

\noindent and we let this as an exercise to the reader.

\subsection{Classical Riemann Problem}

The Riemann problem is physically a tube filled with gas which is divided into two chambers separated by a removable membrane at $x=x_0$. At the initial time the membrane is removed and the gas begins to flow. Once the membrane is removed, the discontinuity decays into two elementary, 
non-linear waves that move in opposite directions.

Depending on the values of the thermodynamical variables in each chamber, four cases can occur. Considering the fluid is described on a one-dimensional spatial domain, rarefaction and shock waves can evolve toward the left or right from the location of the membrane.

In general the solution in all the cases can be studied in six following regions:

\begin{itemize}
\item[] Region 1: initial left state that has not been yet influenced by rarefaction or shock waves
\item[] Region 2: wave traveling to the left (may be rarefaction or shock)
\item[] Region 3: region between the wave moving to the left and the contact discontinuity, called region star-left
\item[] Contact discontinuity
\item[] Region 4: region between the contact discontinuity and the wave moving to the right, called region star-right
\item[] Region 5: wave traveling to the right (may be rarefaction or shock)
\item[] Region 6: initial right state that has not been yet influenced by rarefaction or shock waves
\end{itemize}

\noindent Regions 2 and 5 are special. If the wave propagating in such regions is a rarefaction wave the region involves a head-fan-tail structure, whereas if it is a shock the region becomes only a discontinuity. 
Counting from left to right on the spatial domain, the results can be reduced to the following four possible combinations of waves:

\begin{itemize}
\item[1)] rarefaction-shock
\item[2)] shock-rarefaction
\item[3)] rarefaction-rarefaction
\item[4)] shock-shock
\end{itemize}

\noindent with a contact discontinuity between the two waves in all cases. It is worth noticing that these combinations can occur under a wide variety of possible combinations of the initial values of the thermodynamical variables. In this paper we illustrate each of these scenarios using particular sets of initial conditions.

\subsubsection{Case 1: Rarefaction-Shock}

This case corresponds to the typical case used to test numerical codes, a test called the Sod's shock tube problem \cite{SodShockTube}. A traditional set of initial values that produces this scenario corresponds to a gas with higher density and pressure in the left chamber than in the right chamber, and the velocity is set initially to zero in both.  

A rarefaction wave travels into the high density region (moves to the left), whereas a shock moves into the low density region (moves to the right).

Summarizing, the problem then involves five regions only. Regions 1 correspond to the initial state to the left that has not been influenced by the evolution of the system. Region 2 corresponds to a rarefaction wave containing the head-fan-tail structure, region 3 and 4 are the left and right states separated by the contact discontinuity. Region 5 reduces to the shock. Finally region 6 is the initial state at the right chamber that has not been influenced by the evolution of the system.

The goal is to determine the state in all the regions using the relations between the thermodynamical quantities constructed before.

The starting point to construct the solution happens at the contact discontinuity, where the velocity and pressure obey the conditions $p_3=p_4=p^{*}$ and $v_3=v_4=v^{*}$.

Region 3 plays the role of the state at the right from the rarefaction wave and region 1 the state at the left. Then we can use (\ref{eq:R_L}) to obtain an expression for $v_3$

\begin{equation}
v_3 = v_1 - \frac{2a_1}{\Gamma-1}\left[ \left( \frac{p_3}{p_1} \right)^{\frac{\Gamma-1}{2\Gamma}} - 1\right].\label{eq:v_star_Raf_shock1}
\end{equation}

\noindent On the other hand, region 4 plays the role of a state at the left from the shock wave and region 6 the role of the state at the right. Then we use (\ref{eq:L_R}) to calculate $v_4$:

\begin{equation}
v_4 = v_6 + (p_4 -p_6) \sqrt{\frac{A_6}{p_4+B_6}}.\label{eq:v_star_Raf_shock2}
\end{equation}

\noindent where $A_6=2/\rho_6/(\Gamma+1)$ and $B_6=p_6(\Gamma-1)/(\Gamma+1)$. Given that $v_3=v_4=v^{*}$, equating both expressions one obtains a trascendental equation for $p^{*}$:

\begin{equation}
(p^{*} - p_6)\sqrt{\frac{A_6}{p^{*}+B_6}} + \frac{2a_1}{\Gamma-1}\left[ \left( \frac{p^{*}}{p_1} \right)^{\frac{\Gamma-1}{2\Gamma}} - 1\right]+ v_6-v_1 = 0. \label{eq:p_star_RarShock}
\end{equation}

\noindent Unfortunately as far as we can tell, no exact solution is known for $p^{*}$, and then we proced to construct its solution numerically. Once this equation is solved, $p_3$ and $p_4$ are automatically known, and $v_3$ and $v_4$ can be calculated using (\ref{eq:v_star_Raf_shock1}) and (\ref{eq:v_star_Raf_shock2}) respectively.

\begin{figure}
\psfrag{x}{$x$}
\psfrag{t}{$t$}
\psfrag{I}{$1$}
\psfrag{II}{$2$}
\psfrag{III}{$3$}
\psfrag{IV}{$4$}
\psfrag{V}{$5$}
\psfrag{VI}{$6$}
\begin{center}
\includegraphics[width=0.5 \textwidth]{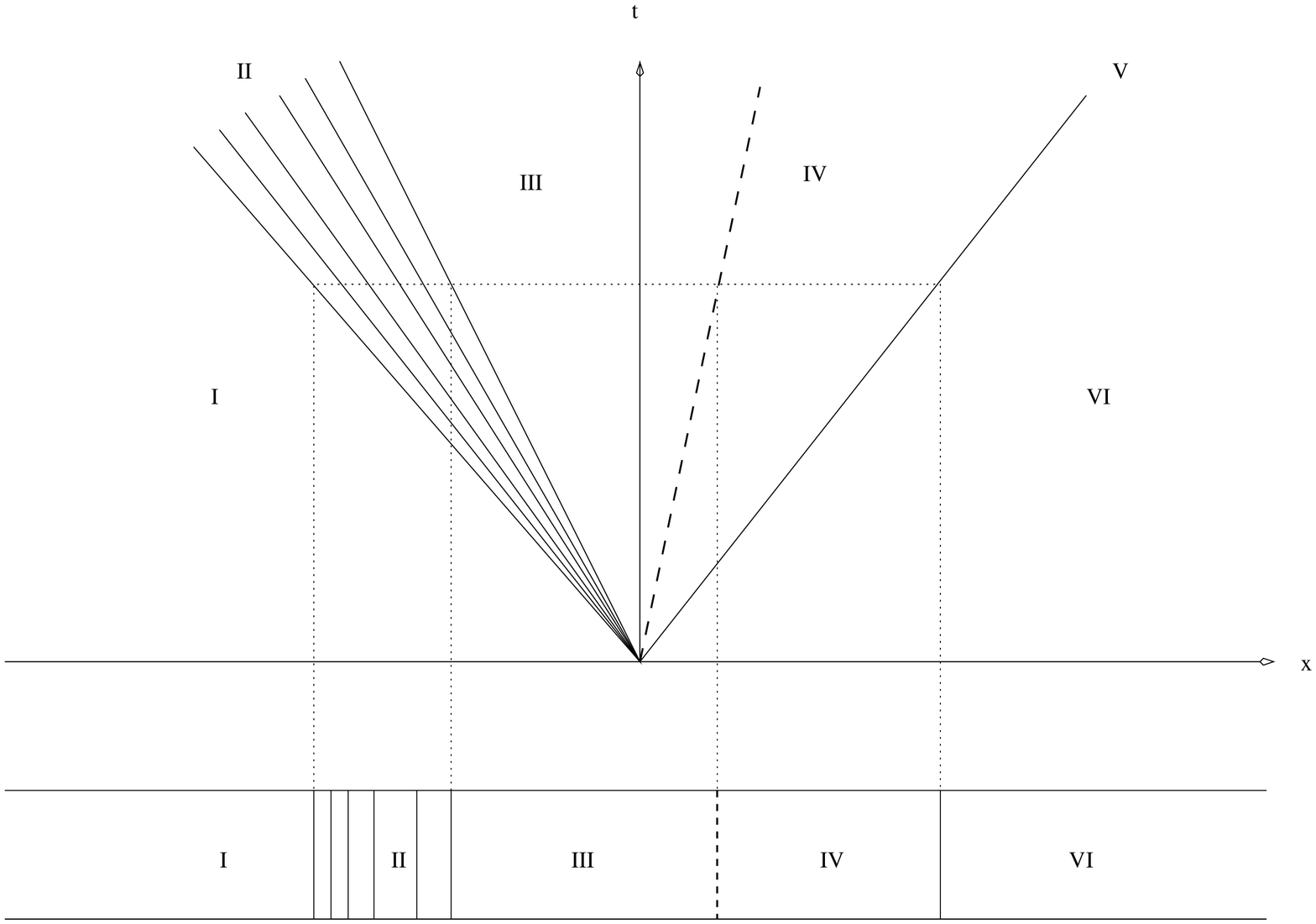}
\caption{\label{fig:regions_RS} Description of the relevant regions for the Rarefaction-Shock case.}
\end{center}
\end{figure}

Then, it is possible to calculate $\rho_3$ using (\ref{eq:isentropic}) at both sides of the rarefaction zone, given $C$ is the same on both sides because it is an isentropic process:

\begin{equation}
\rho_3=\rho_1 \left( \frac{p_3}{p_1}\right)^{1/\Gamma}\label{eq:rho_3_Rar_Shock}
\end{equation}

\noindent where now $p_1$, $\rho_1$ and $p_3$ are known. On the other hand one can also calculate 
$\rho_4$ using (\ref{eq:rho_L_R}) 

\begin{equation}
\rho_4 = \rho_6 \left(\frac{p_6 (\Gamma-1)+p_4 (\Gamma+1)}{p_4(\Gamma-1) + p_6(\Gamma+1)}\right)\label{eq:rho_4_Rar_Shock}
\end{equation}

\noindent also in terms of known information. With this information it is already possible to construct the solution in the whole domain. We explain how to do it region by region. A scheme of how the regions are distributed is shown in Fig. \ref{fig:regions_RS}. 

\begin{enumerate}
\item Region 1 is defined by the condition $x-x_0 < t V_{head}$, where $V_{head}$ is the velocity of the head of the rarefaction wave given by the characteristic value of the Jacobian matrix evaluated at the location next to the head from the left side, that is, considering (\ref{eq:lambda_minus}) $V_h = v_1-a_1$. The solution there is simply

\begin{eqnarray}
p_{exact} &=& p_1, \noindent \nonumber\\
v_{exact} &=& v_1, \noindent \nonumber\\
\rho_{exact} &=& \rho_1. \noindent \nonumber
\end{eqnarray}

\item Region 2 is defined by the condition $t V_{head} < x-x_0 < t V_{tail}$, where $V_{tail}$ is the same characteristic value again, but this time evaluated at the tail curve, that is $V_{tail}=v_3-a_3$. This is the fan region for a rarefaction wave moving to the left, for which we simply use expressions (\ref{eq:vR_rarL},\ref{eq:pR_rarL},\ref{eq:rhoR_rarL})  that need only information from region 1 and obtain

\begin{eqnarray}
\rho_{exact} &=& \rho_1 \left[ \frac{2}{\Gamma + 1} + \frac{\Gamma-1}{a_1 (\Gamma+1)} 	\left( v_1 -\frac{x-x_0}{t} \right)	\right]^{\frac{2}{\Gamma-1}},\nonumber\\
p_{exact} &=& p_1 \left[ \frac{2}{\Gamma + 1} + \frac{\Gamma-1}{a_1 (\Gamma+1)} \left( v_1 - 	\frac{x-x_0}{t} \right)\right]^{\frac{2\Gamma}{\Gamma-1}},\nonumber\\
v_{exact} &=& \frac{2}{\Gamma+1}\left[ a_1 + \frac{1}{2} (\Gamma-1) v_1 + \frac{x-x_0}{t}\right].
\end{eqnarray}

\item Region 3 is defined by the condition $t V_{tail} < x-x_0 < t V_{contact}$, where $V_{contact}$ is the velocity of the contact discontinuity, which is the second eigenvalue (\ref{eq:lambda_0}) of the Jacobian matrix evaluated at this region, that is $V_{contact}= v_3=v_4$. The solution there finally reads

\begin{eqnarray}
p_{exact} &=& p_3, \noindent \nonumber\\
v_{exact} &=& v_3, \noindent \nonumber\\
\rho_{exact} &=& \rho_3. \noindent \nonumber
\end{eqnarray}

\item Region 4 is defined by the condition $t V_{contact}< x-x_0 < t V_{shock}$, where according to (\ref{eq:SR}), the velocity of a shock moving to the right separating regions 4 and 6 is $V_{shock}=v_6+a_6\sqrt{\frac{(\Gamma+1)p_4}{2\Gamma p_6} + 	\frac{\Gamma-1}{2\Gamma}}$, where $a_6=\sqrt{p_6 \Gamma / \rho_6}$. Then the solution in this region is

\begin{eqnarray}
p_{exact} &=& p_4, \noindent \nonumber\\
v_{exact} &=& v_4, \noindent \nonumber\\
\rho_{exact} &=& \rho_4. \noindent \nonumber
\end{eqnarray}

\noindent as calculated 

\item There is no region 5.

\item Region 6 is defined by $t V_{shock} < x-x_0$. In this region the solution is simply

\begin{eqnarray}
p_{exact} &=& p_6, \noindent \nonumber\\
v_{exact} &=& v_6, \noindent \nonumber\\
\rho_{exact} &=& \rho_6. \noindent \nonumber
\end{eqnarray}

\end{enumerate}

\begin{figure}[htp]
\includegraphics[width=4cm]{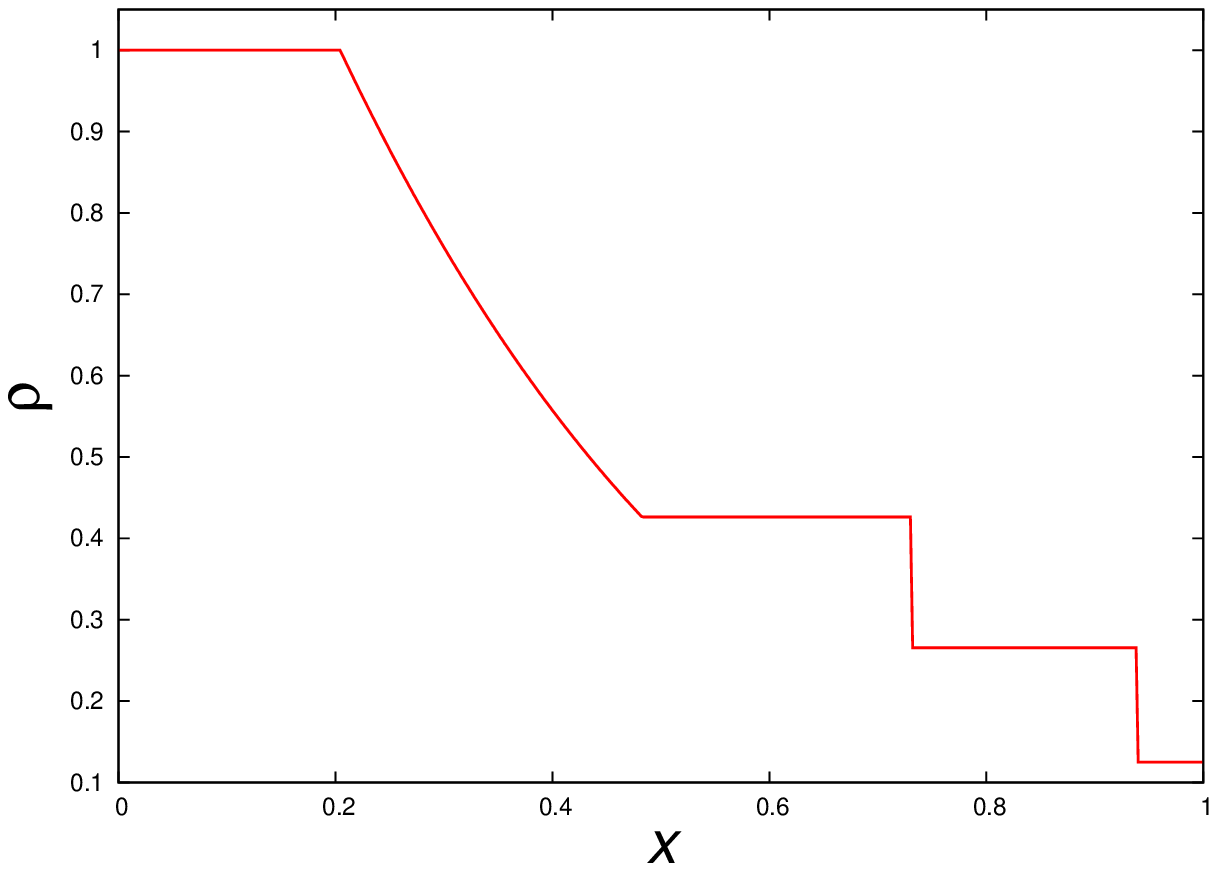}
\includegraphics[width=4cm]{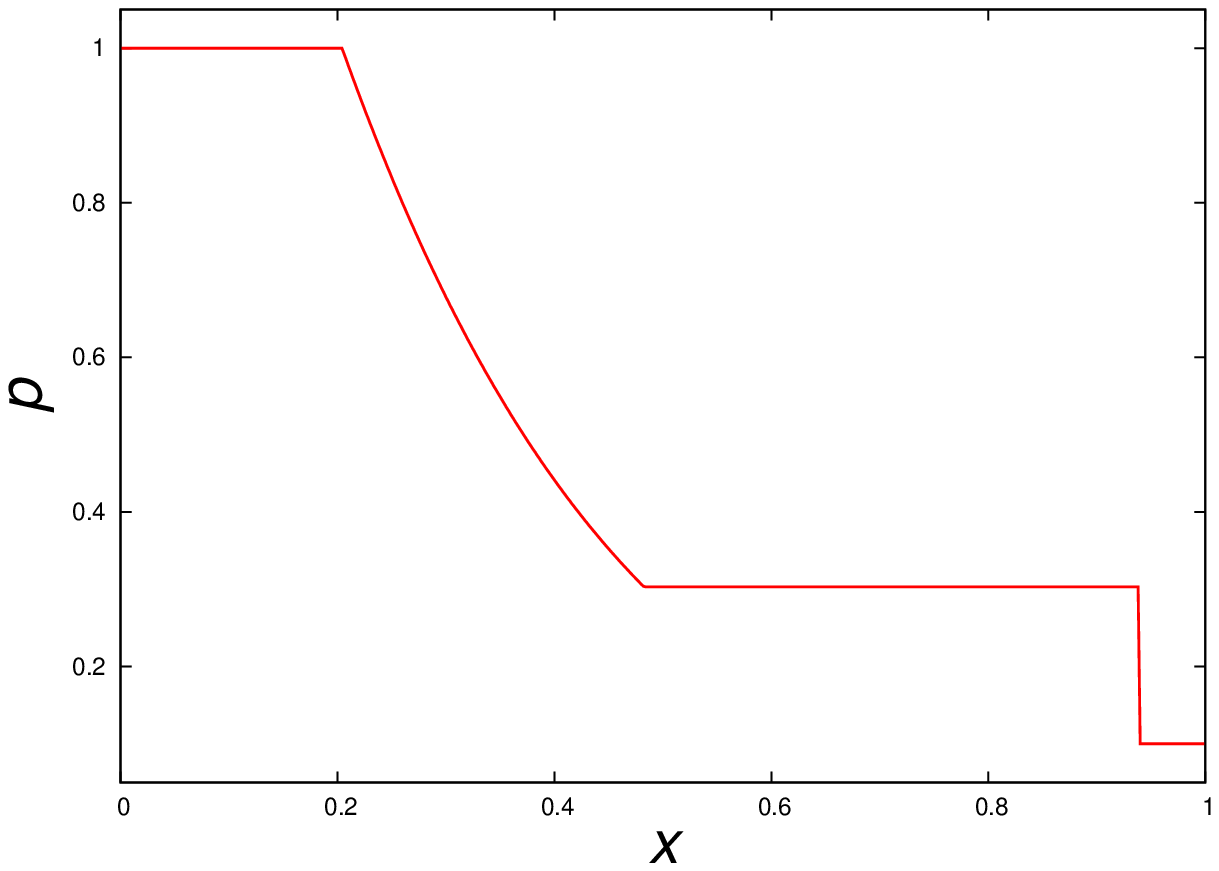}
\includegraphics[width=4cm]{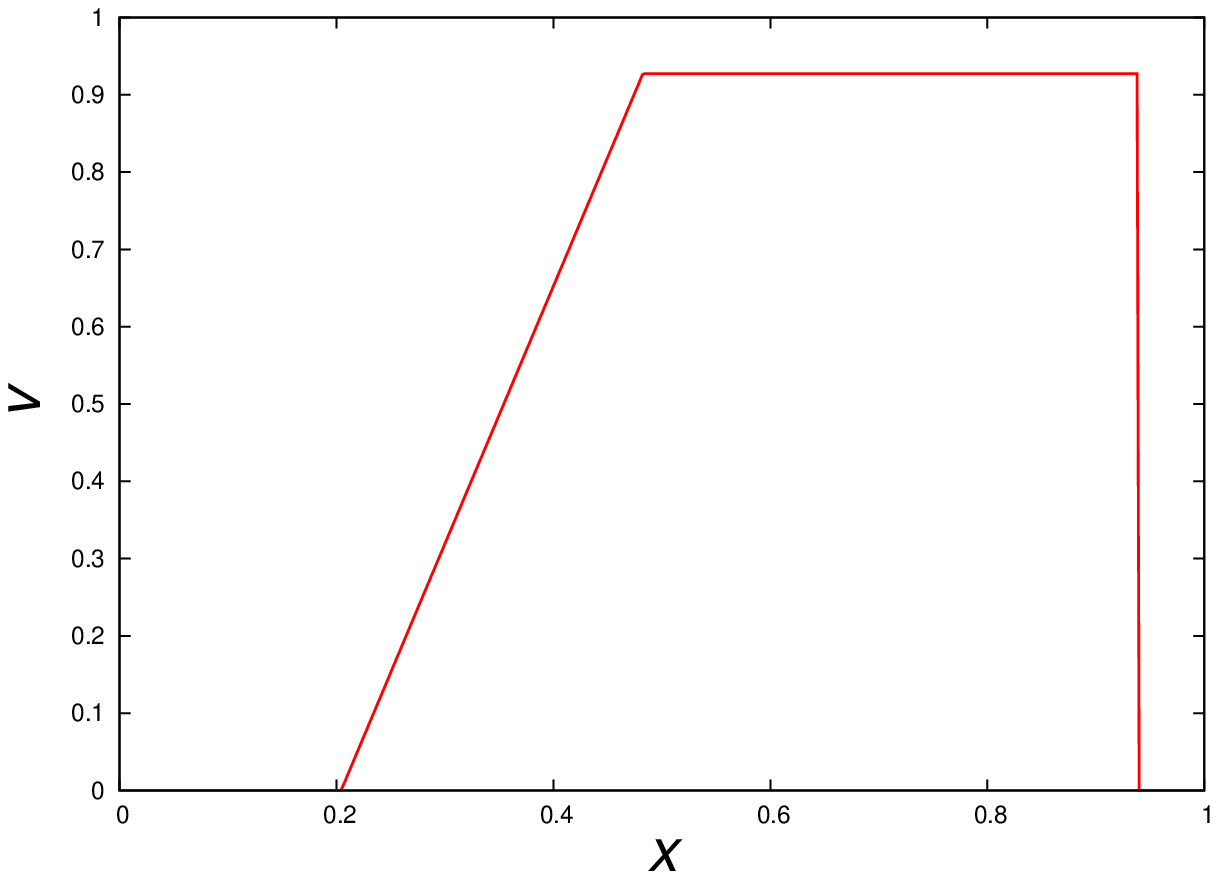}
\includegraphics[width=4cm]{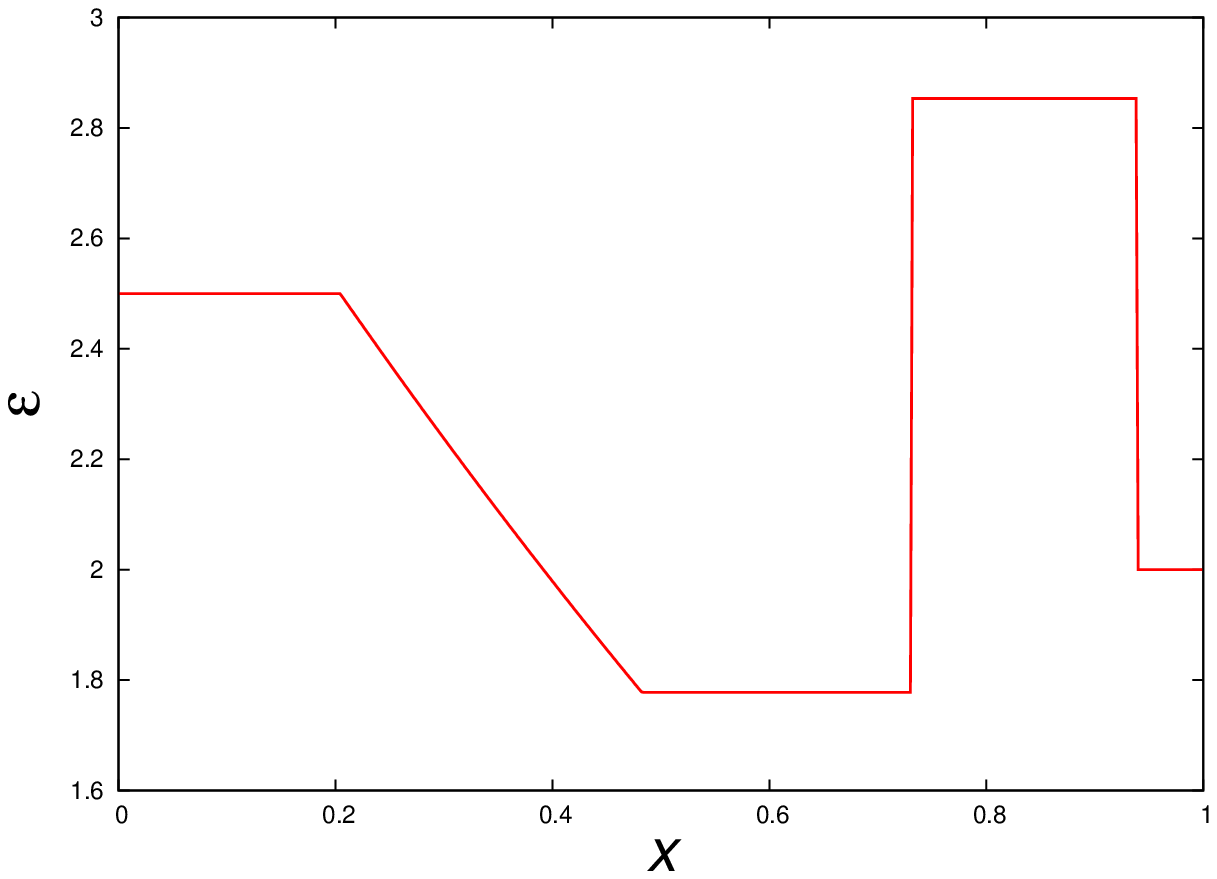}
\caption{\label{fig:Newtonian_RS} Exact solution for the Rarefaction-Shock case at time $t=0.25$ for the parameters in Table \ref{tab:newtonian}.}
\end{figure}

An example of how the solution looks like is shown in Fig. \ref{fig:Newtonian_RS} for initial data in Table \ref{tab:newtonian}. 

\begin{table}
\begin{tabular}{|c|c|c|c|c|c|c|}\hline
Case	& $p_L$	& $p_R$	& $v_L$	& $v_R$	& $\rho_L$	&	$\rho_R$\\\hline
Rarefaction-Shock & 1.0	& 0.1	& 0.0		& 0.0	 &	1.0	&	0.125 \\
Shock-Rarefaction	& 0.1		& 1.0		& 0.0	&0.0	&0.125	& 1.0	\\
Rarefaction-Rarefaction	& 0.4	&0.4	&-1.0	& 1.0	&1.0	&1.0\\
Shock-Shock	& 0.4	&0.4	&1.0	&-1.0	&1.0	& 1.0\\\hline
\end{tabular}
\caption{Table with the initial data for the four different cases. We choose the spatial domain to be $x\in [0,1]$ and the location of the membrane at $x_0=0.5$. In all cases we use $\Gamma=1.4$.}
\label{tab:newtonian}
\end{table}

\subsubsection{Case 2: Shock-Rarefaction}

This case is identical to the previous one, except that we choose the initial pressure and density are higher on the right chamber. After initial time, the wave traveling to the left is a shock, while the one moving to the right is a rarefaction wave. This implies that region 2 plays the role of region 5 in the previous case and region 5 has the tail-fan-head structure of a rarefaction wave.

Starting from the contact discontinuity, the conditions $v_3=v_4=v^{*}$ and $p_3=p_4=p^{*}$ hold. The conditions on a shock wave moving to the left imply according to (\ref{eq:vR_shockL}) that the velocity of the state at the right is

\begin{equation}
v_3 = v_1 - (p_3-p_1)\sqrt{\frac{A_1}{p3+B_1}},\label{eq:v_3_SR}
\end{equation}

\noindent and information from the rarefaction wave interface can be obtained from (\ref{eq:vL_rarR}) for $v_4$ as the velocity on the state at the left from a rarefaction wave moving to the right

\begin{equation}
v_4 = v_6 -\frac{2a_6}{\Gamma-1}\left[ 1 - \left( \frac{p_4}{p_6}\right)^{\frac{\Gamma-1}{2\Gamma}} \right]. \label{eq:v_4_SR}
\end{equation}

\begin{figure}
\psfrag{x}{$x$}
\psfrag{t}{$t$}
\psfrag{I}{$1$}
\psfrag{II}{$2$}
\psfrag{III}{$3$}
\psfrag{IV}{$4$}
\psfrag{V}{$5$}
\psfrag{VI}{$6$}
\begin{center}
\includegraphics[width=0.5 \textwidth]{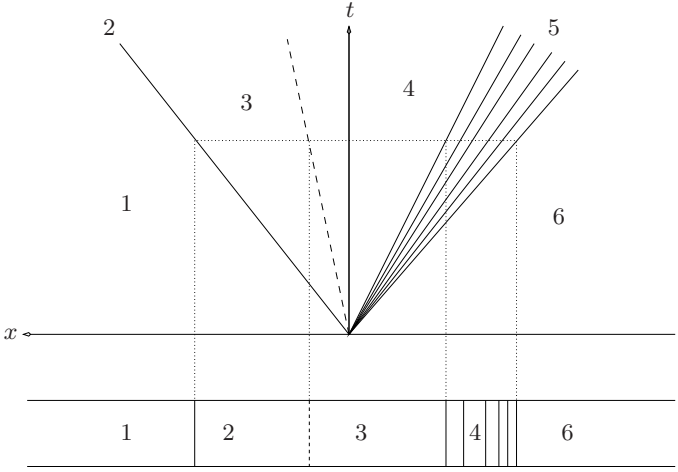}
\caption{\label{fig:regions_SR} Description of the relevant regions for the Shock-Rarefaction case.}
\end{center}
\end{figure}

\noindent Equating these two expression one obtains a trascendental equation for $p^{*}$:

\begin{equation}
-(p^{*} - p_1)\sqrt{\frac{A_1}{p^{*}+B_1}} + \frac{2a_6}{\Gamma-1}\left[ 1 - \left(\frac{p^{*}}{p_6} \right)^{\frac{\Gamma-1}{2\Gamma}} \right] + v_1 - v_6 =0 \label{eq:trascencental_SR}
\end{equation}

\noindent that one solves numerically for $p^{*}$. This information provides the necessary information to construct the solution in the whole domain as described below. The different regions are illustrated in Fig. \ref{fig:regions_SR} and the exact solution region by region is as follows.

\begin{enumerate}
\item Region 1 is defined by $x-x_0 < t V_{shock}$, where the velocity of the shock is given by (\ref{eq:SL}) because the shock is traveling to the left:

\begin{equation}
V_{s} = v_1 - a_1 \sqrt{\frac{(\Gamma+1)p_3}{2p_1 \Gamma} + \frac{\Gamma-1}{2\Gamma}}, \nonumber
\end{equation}

\noindent and the exact solution here reads

\begin{eqnarray}
p_{exact} &=& p_1, \noindent \nonumber\\
v_{exact} &=& v_1, \noindent \nonumber\\
\rho_{exact} &=& \rho_1. \noindent \nonumber
\end{eqnarray}

\item There is no region 2.

\item Region 3 is defined by the condition $t V_{s} < x - x_0 < t V_{contact}$. 
$V_{contact}$ is the characteristic value (\ref{eq:lambda_0}) evaluated at this region: $V_{contact} = v_3 = v_4 = v^{*}$. Using  (\ref{eq:rhoR_shockL}) explicitly for the density and (\ref{eq:v_3_SR}) for the velocity, the solution in this region reads

\begin{eqnarray}
p_{exact} &=& p_3,\nonumber\\
v_{exact} &=& v_3,\nonumber\\
\rho_{exact} &=& \rho_1 \frac{p_1 (\Gamma-1) + p_3 (\Gamma+1)}{p_3 (\Gamma-1) + p_1 (\Gamma+1)}.\nonumber
\end{eqnarray}

\item Region 4  is defined by the condition $t V_{contact} < x - x_0 < t V_{t}$, where the velocity of the tail of the rarefaction wave $V_{t}$ is the third eigenvalue (\ref{eq:lambda_plus}) evaluated at the region behind the tail $V_{t} = v_4 + a_4$.

One uses (\ref{eq:v_4_SR}) to calculate $v_4$ and (\ref{eq:isentropic}) implies $p_4/p_6 = (\rho_4/\rho_6)^{\Gamma}$ for a constant value of $K$, which implies an expression for $\rho_4$. The resulting exact solution is

\begin{eqnarray}
p_{exact} &=& p_4,\nonumber\\
v_{exact} &=& v_4\nonumber\\
\rho_{exact} &=& \rho_6 \left( \frac{p_4}{p_6} \right)^{1/\Gamma}.\nonumber
\end{eqnarray}

\item Region 5 is  a fan region defined by the condition $t V_{t} < x - x_0 < t V_{h}$ where the velocity of the head of the wave is again the third eigenvalue, but this time evaluated at the head $V_{h}=v_6 +a_6$. One uses the expressions for a fan region of a rarefaction wave moving to the right (\ref{eq:vL_rarR},\ref{eq:pL_rarR},\ref{eq:rhoL_rarR}) to calculate the exact solution

\begin{eqnarray}
p_{exact} &=& p_6 \left[ \frac{2}{\Gamma+1} - \frac{\Gamma-1}{a_6(\Gamma+1)} \left(v_6 - \frac{x-x_0}{t}\right)\right]^{\frac{2\Gamma}{\Gamma-1}},\nonumber\\
v_{exact} &=& \frac{2}{\Gamma+1} \left[ -a_6 + \frac{1}{2}(\Gamma-1) v_6 + \frac{x-x_0}{t}\right],\nonumber\\
\rho_{exact} &=& \rho_6 \left[ \frac{2}{\Gamma+1}-  \frac{\Gamma-1}{a_6(\Gamma+1)}\left( v_6 - \frac{x-x_0}{t}\right) \right]^{\frac{2}{\Gamma-1}}.\nonumber 
\end{eqnarray}

\item Region 6 is defined by the condition $t V_{h} < x-x_0$. The exact solution is given by the initial states at the right chamber.

\begin{eqnarray}
p_{exact} &=& p_6, \noindent \nonumber\\
v_{exact} &=& v_6, \noindent \nonumber\\
\rho_{exact} &=& \rho_6. \noindent \nonumber
\end{eqnarray}

\end{enumerate}

\begin{figure}[htp]
\includegraphics[width=4cm]{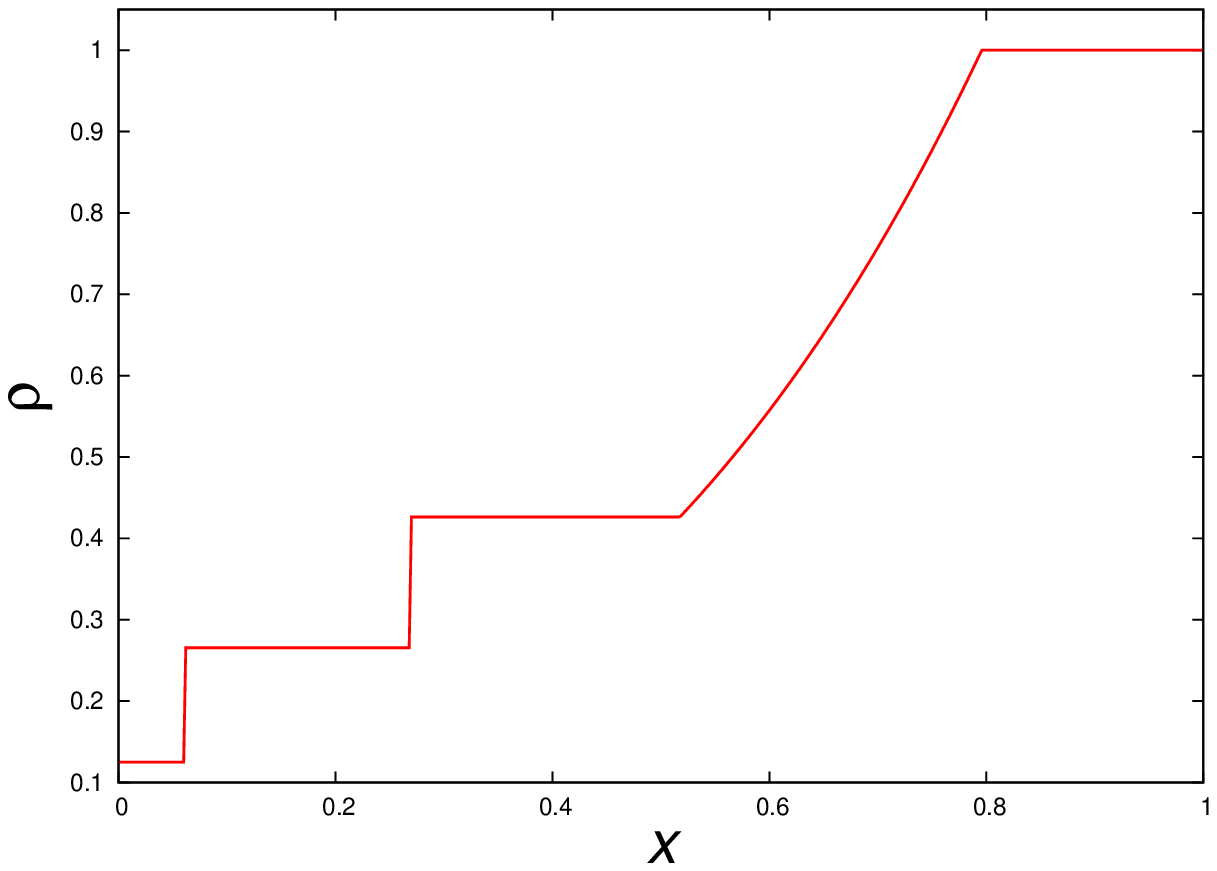}
\includegraphics[width=4cm]{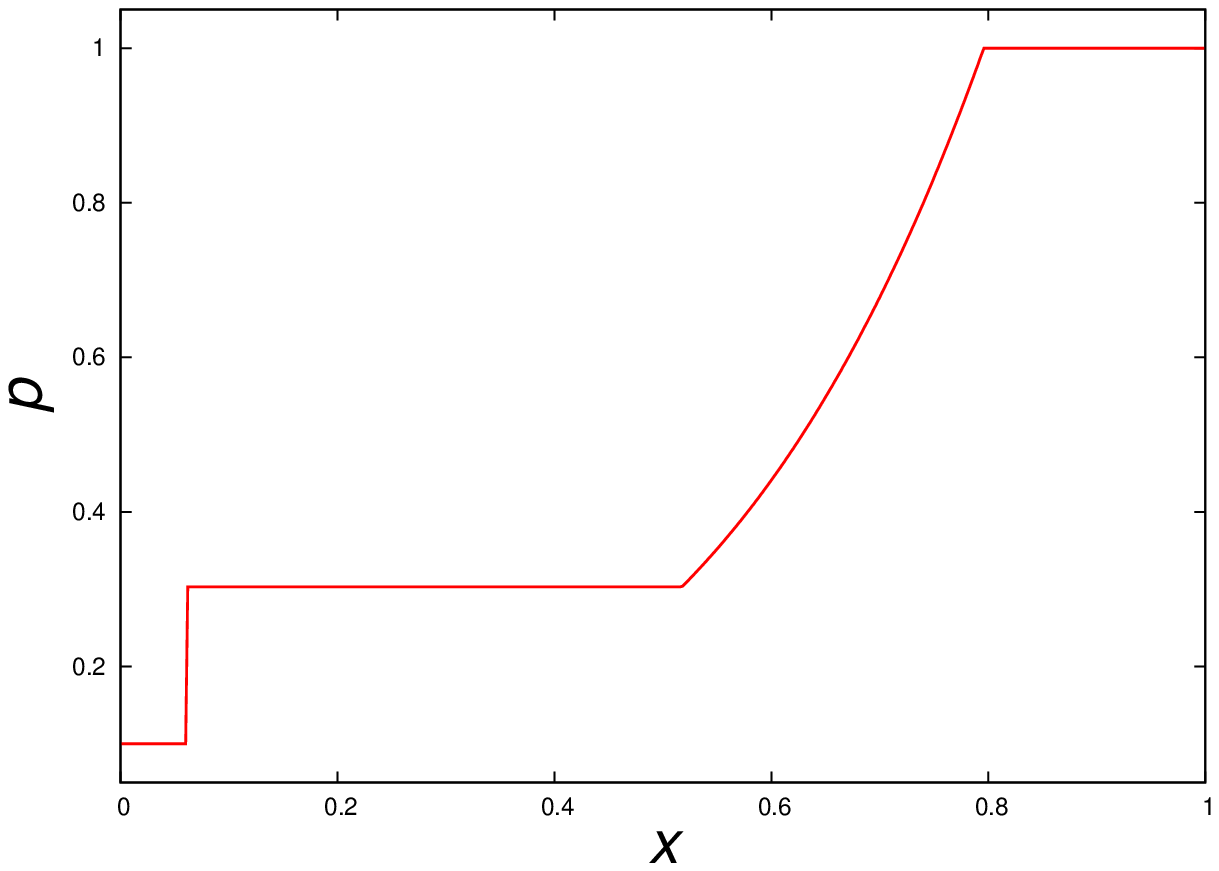}
\includegraphics[width=4cm]{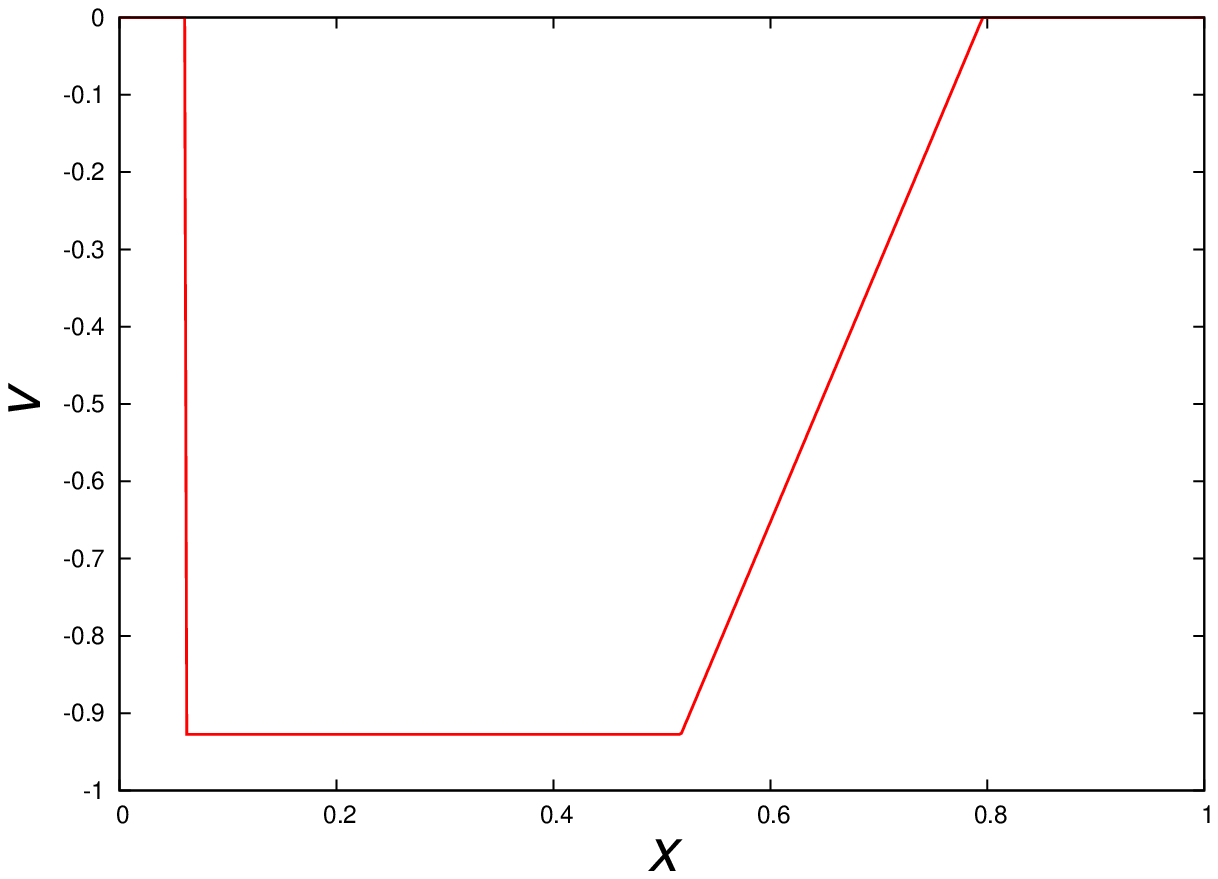}
\includegraphics[width=4cm]{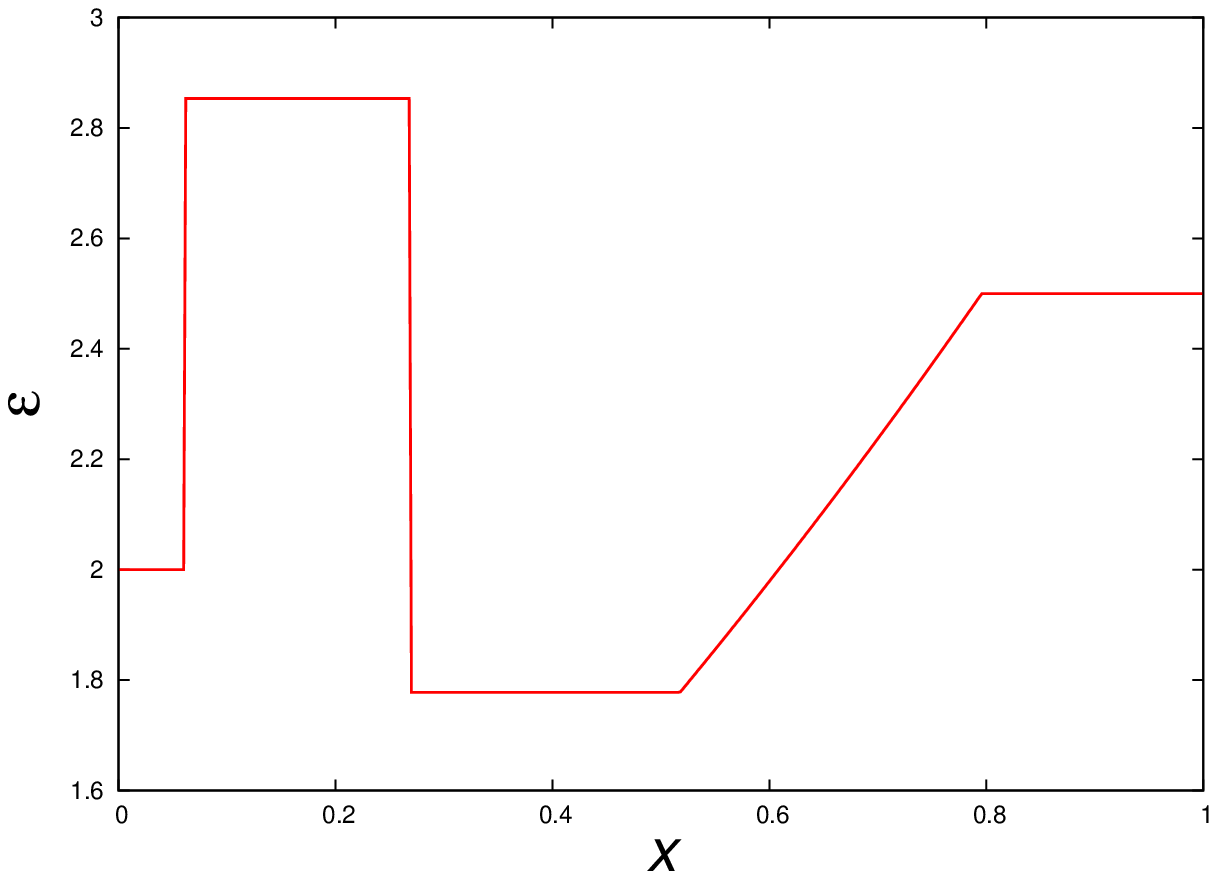}
\caption{\label{fig:Newtonian_SR} Exact solution for the Shock-Rarefaction case at time $t=0.25$ for the parameters in Table \ref{tab:newtonian}.}
\end{figure}

An example is shown in Fig. \ref{fig:Newtonian_SR} for initial data in Table \ref{tab:newtonian}.

\subsubsection{Case 3: Rarefaction-Rarefaction}

A physical situation that provides this scenario is $p_L=p_R$, $\rho_L = \rho_R$ and $-v_L=+v_R >0$. In this case both, regions 2 and 5 correspond to rarefaction waves. In this particular case since one of the rarefaction waves moves to the left and the other one to the right, we distinguish them using the labels for each of their parts.

Again the contact discontinuity defines a relationship between velocity and pressure. In the present case, there is an expression for $v_3$ in terms of $v_1$ for a rarefaction wave moving to the left given by (\ref{eq:R_L}) and another one for $v_4$ in terms of $v_6$ for a rarefaction wave moving to the right (\ref{eq:v_L_rarefactionR}):

\begin{eqnarray}
v_3 &=& v_1 -\frac{2a_1}{\Gamma-1}\left[ \left( \frac{p_3}{p_1}\right)^{\frac{\Gamma-1}{2\Gamma}} - 1\right],\label{eq:v_3_RR}\\
v_4 &=& v_6 -\frac{2a_6}{\Gamma-1}\left[ 1 - \left( \frac{p_4}{p_6}\right)^{\frac{\Gamma-1}{2\Gamma}} \right]. \label{eq:v_4_RR}
\end{eqnarray}

\noindent The condition $v_3=v_4=v^{*}$ at the contact discontinuity implies a trascendental equation for $p^{*}=p_3=p_4$:

\begin{equation}
\frac{2a_6}{\Gamma-1}\left[ 1 - \left(\frac{p^{*}}{p_6}\right)^{\frac{\Gamma-1}{2\Gamma}} \right]  -\frac{2a_1}{\Gamma-1} \left[ \left( \frac{p^{*}}{p_1}\right)^{\frac{\Gamma-1}{2\Gamma}} -1\right]+ v_1 - v_6 = 0
\label{eq:trascencental_RR}
\end{equation}

\begin{figure}
\psfrag{x}{$x$}
\psfrag{t}{$t$}
\psfrag{I}{$1$}
\psfrag{II}{$2$}
\psfrag{III}{$3$}
\psfrag{IV}{$4$}
\psfrag{V}{$5$}
\psfrag{VI}{$6$}
\begin{center}
\includegraphics[width=0.5 \textwidth]{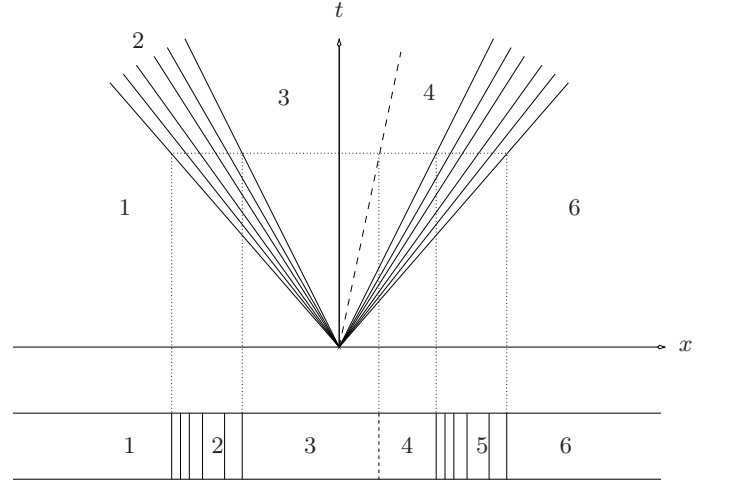}
\caption{\label{fig:regions_RR} Description of the relevant regions for the Rarefaction-Rarefaction case.}
\end{center}
\end{figure}

Again, once $p^{*}$ is calculated numerically, the solution in all the regions of the domain can be calculated as follows. The first implication is that $p_3=p_4=p^{*}$, and thus $v_3$ and $v_4$ can be calculated using (\ref{eq:v_3_RR}) and (\ref{eq:v_4_RR}). The different regions are illustrated in Fig. \ref{fig:regions_RR}.

\begin{enumerate}
\item Region 1 is defined by the condition $x-x_0 < t V_{h,2}$, where $V_{h,2}$ is the velocity of the head of the wave moving to the left, and is obtained from the characteristic value of such rarefaction wave evaluated at the left interface, that is $V_{h,2} = v_1 - a_1$. In this region the gas has not affected the initial state on the left, then the solution is

\begin{eqnarray}
p_{exact} &=& p_1, \nonumber\\
v_{exact} &=& v_1, \nonumber\\
\rho_{exact} &=& \rho_1. \nonumber
\end{eqnarray}

\item Region 2 is a fan region defined by the condition $t V_{h,2} < x-x_0 < t V_{t,2}$, where the velocity of the tail $V_{t,2}$ is that of the state left behind by the wave, that is $V_{t,2} = v_3 - a_3$.

\noindent The exact solution is that of a fan region of a rarefaction wave moving to the left (\ref{eq:vR_rarL},\ref{eq:pR_rarL},\ref{eq:rhoR_rarL})

\begin{eqnarray}
p_{exact} &=& p_1 \left[ \frac{2}{\Gamma + 1} + \frac{\Gamma-1}{a_1 (\Gamma+1)} \left( v_1 - 	\frac{x-x_0}{t} \right)\right]^{\frac{2\Gamma}{\Gamma-1}},\nonumber\\
v_{exact} &=& \frac{2}{\Gamma+1}\left[ a_1 + \frac{1}{2} (\Gamma-1) v_1 + \frac{x-x_0}{t}\right],\nonumber\\
\rho_{exact} &=& \rho_1 \left[ \frac{2}{\Gamma + 1} + \frac{\Gamma-1}{a_1 (\Gamma+1)} 	\left( v_1 - \frac{x-x_0}{t} \right)	\right]^{\frac{2}{\Gamma-1}}.\nonumber
\end{eqnarray}

\item Region 3 is defined by the condition $tV_{t,2} < x-x_0 < t V_{contact}$. The velocity of the contact discontinuity is $V_{contact}=v_3=v_4=v^{*}$ according to the eigenvalue (\ref{eq:lambda_0}). In this region $p_3=p^{*}$ and $v_3=v^{*}$ are already known from $p^{*}$. Finally, the density is obtained from (\ref{eq:isentropic}) for an isentropic process like the rarefaction wave for a constant $C$ on both sides of such wave as found in the previous two cases. Thus the solution is

\begin{eqnarray}
p_{exact} &=& p_3,\nonumber\\
v_{exact} &=& v_3.\nonumber\\
\rho_{exact} &=& \rho_1 \left( \frac{p_3}{p_1}\right)^{1/\Gamma},\nonumber
\end{eqnarray}

\item Region 4 is defined by the condition $t V_{contact} < x-x_0 < t V_{t,5}$, where the velocity of the tail of the wave moving to the right $V_{t,5}$ is given by the eigenvalue (\ref{eq:lambda_plus}) evaluated at the state left behind the rarefaction wave moving to the right, that is $V_{t,5}=v_4+a_4$, where again we point out that $v_4=v^{*}$ and $p_4=p^{*}$ are known once $p^{*}$ is calculated. The solution is obtained in the same way as for the previous region, but now the wave relates states in regions 4 and 6:

\begin{eqnarray}
p_{exact} &=& p_4,\nonumber\\
v_{exact} &=& v_4.\nonumber\\
\rho_{exact} &=& \rho_6 \left( \frac{p_4}{p_6}\right)^{1/\Gamma},\nonumber
\end{eqnarray}

\item Region 5 is defined by the condition $t V_{t,5} < x-x_0 < V_{h,5}$, where the velocity of the head of the wave moving to the right is $V_{h,5}=v_6+a_6$, and the solution is obtained using the values of the state variables for the fan of a rarefaction wave moving to the right (\ref{eq:vL_rarR},\ref{eq:pL_rarR},\ref{eq:rhoL_rarR}):

\begin{eqnarray}
p_{exact} &=& p_6 \left[ \frac{2}{\Gamma+1} - \frac{\Gamma-1}{a_6(\Gamma+1)} \left(v_6 - \frac{x-x_0}{t}\right)\right]^{\frac{2\Gamma}{\Gamma-1}},\nonumber\\
v_{exact} &=& \frac{2}{\Gamma+1} \left[ -a_6 + \frac{1}{2}(\Gamma-1) v_6 + \frac{x-x_0}{t}\right],\nonumber\\
\rho_{exact} &=& \rho_6 \left[ \frac{2}{\Gamma+1}-  \frac{\Gamma-1}{a_6(\Gamma+1)}\left( v_6 - \frac{x-x_0}{t}\right) \right]^{\frac{2}{\Gamma-1}}.\nonumber 
\end{eqnarray}

\item Finally, region 6 is defined by the condition $V_{h,5} < x-x_0$. The exact solution is given by the initial values at the chamber on the right because in this region the gas has not been affected yet by the dynamics of the gas:

\begin{eqnarray}
p_{exact} &=& p_6, \nonumber\\
v_{exact} &=& v_6, \nonumber\\
\rho_{exact} &=& \rho_6. \nonumber
\end{eqnarray}

\end{enumerate}

\begin{figure}[htp]
\includegraphics[width=4cm]{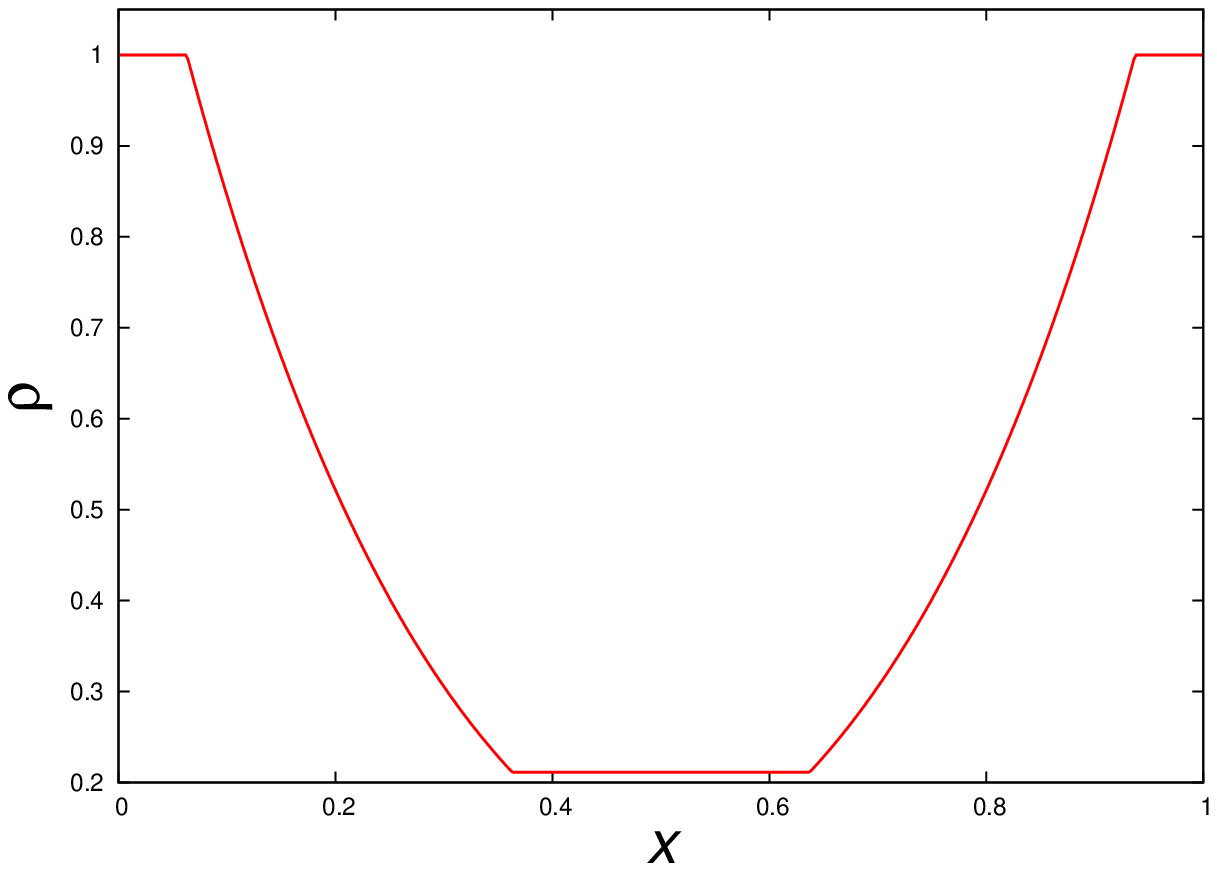}
\includegraphics[width=4cm]{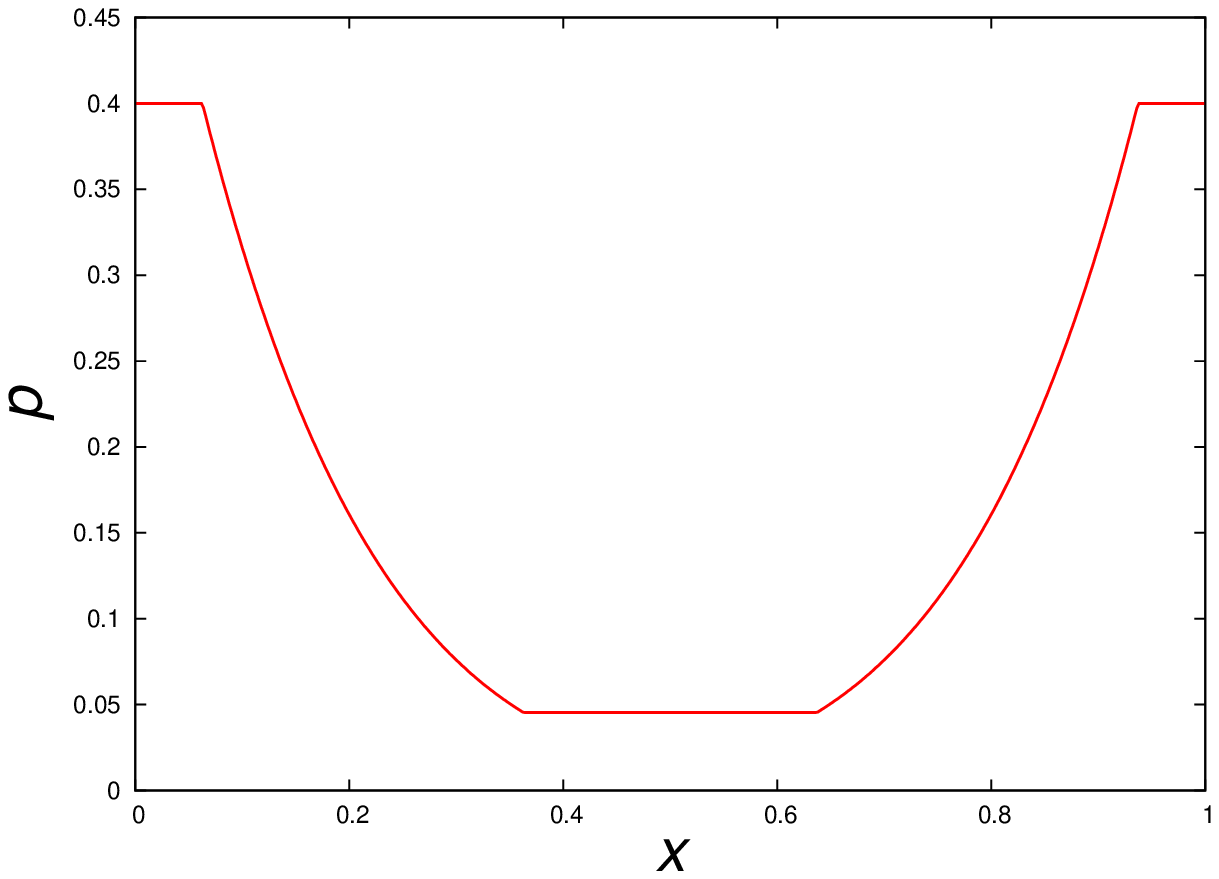}
\includegraphics[width=4cm]{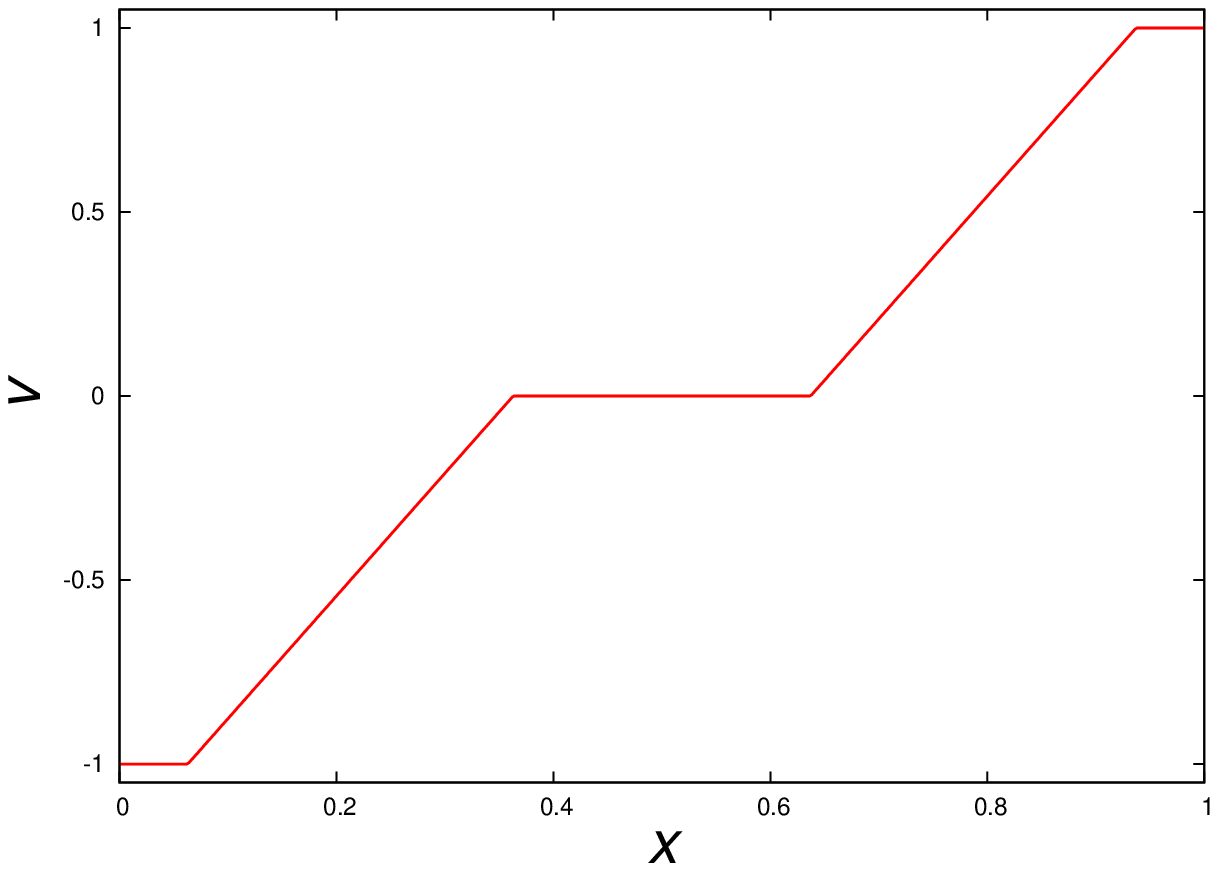}
\includegraphics[width=4cm]{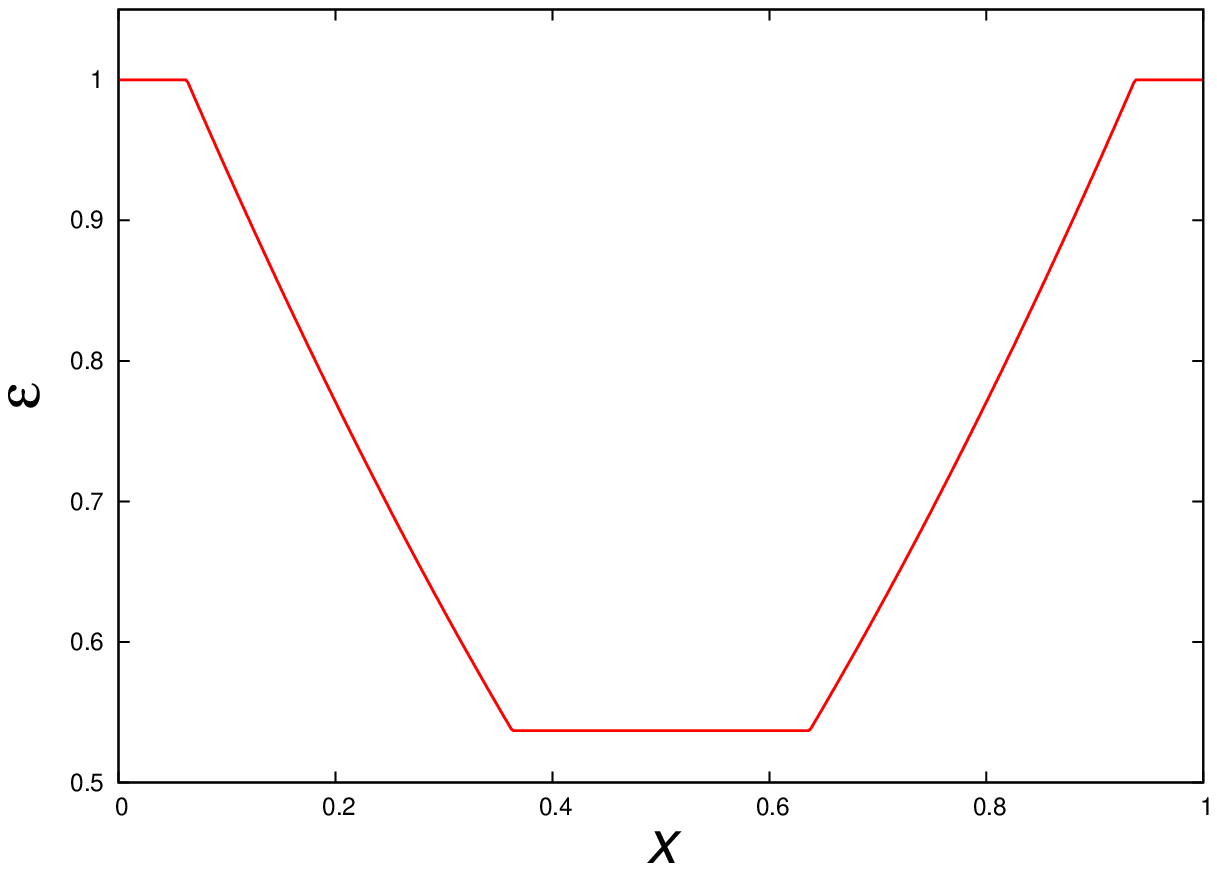}
\caption{\label{fig:Newtonian_RR} Exact solution for the Rarefaction-Rarefaction case at time $t=0.25$ for the parameters in Table \ref{tab:newtonian}.}
\end{figure}

An example is shown in Fig. \ref{fig:Newtonian_RR} for initial data in Table \ref{tab:newtonian}.

\subsubsection{Case 4: Shock-Shock}

A physical situation that provides this scenario corresponds to two streams colliding with opposite directions. We choose in this case $p_L=p_R$, $\rho_L = \rho_R$ and $-v_L=+v_R <0$. In this case regions 2 and 5 are shock waves.

Again the contact discontinuity defines a relationship between velocity and pressure. In the present case there is an expression for $v_3$ in terms of $v_1$ for a shock-wave moving to the left given by (\ref{eq:vR_shockL}) and another one for $v_4$ in terms of $v_6$ for a shock-wave moving to the right (\ref{eq:L_R}):

\begin{eqnarray}
v_3 &=& v_1 - (p_3-p_1)\sqrt{\frac{A_1}{p_3+B_1}},\label{eq:v_3_SS}\\
v_4 &=& v_6 + (p_4-p_6)\sqrt{\frac{A_6}{p_4+B_6}}. \label{eq:v_4_SS}
\end{eqnarray}

\noindent The condition $v_3=v_4=v^{*}$ at the contact discontinuity implies a trascendental equation for $p^{*}=p_3=p_4$:

\begin{equation}
-(p^{*} - p_1)\sqrt{\frac{A_1}{p^{*}+B_1}} - (p^{*}-p_6)\sqrt{\frac{A_6}{p^{*}+B_6}}+v_1-v_6=0.
\label{eq:trascencental_SS}
\end{equation}

Again, once $p^{*}$ is calculated numerically, the solution in all the regions of the domain can be calculated as follows. Immediately one has that $p_3=p_4=p^{*}$ and $v_3$ and $v_4$ can be calculated using (\ref{eq:v_3_SS}) and (\ref{eq:v_4_SS}). 

\begin{figure}
\psfrag{x}{$x$}
\psfrag{t}{$t$}
\psfrag{I}{$1$}
\psfrag{II}{$2$}
\psfrag{III}{$3$}
\psfrag{IV}{$4$}
\psfrag{V}{$5$}
\psfrag{VI}{$6$}
\begin{center}
\includegraphics[width=0.5 \textwidth]{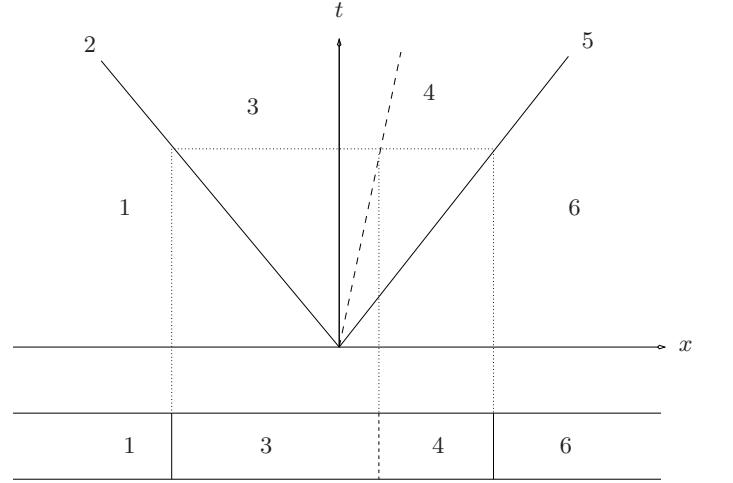}
\caption{\label{fig:regions_SS} Description of the relevant regions for the Shock-Shock case.}
\end{center}
\end{figure}

In this particular case regions 2 and 5 reduce to lines. The solution in each region reads as follows and the regions are illustrated in Fig. \ref{fig:regions_SS}.

\begin{enumerate}
\item Region 1 is defined by the condition $x-x_0 < t V_{s,2}$, where the velocity of the shock moving to the left $V_{s,2}$ is given by (\ref{eq:SL}) and reads $V_{s,2} = v_1 - a_1 \sqrt{\frac{(\Gamma+1)p_3}{2p_1 \Gamma} + \frac{\Gamma-1}{2\Gamma}}$. The solution there is that of the initial values of the variables on the left chamber:

\begin{eqnarray}
p_{exact} &=& p_1, \nonumber\\
v_{exact} &=& v_1, \nonumber\\
\rho_{exact} &=& \rho_1. \nonumber
\end{eqnarray}

\item There is no region 2.

\item Region 3 is defined by the condition $t V_{s,2} < x - x_0 < t V_{contact}$, where $V_{contact} = v_3 = v_4 = v^{*}$. Once (\ref{eq:trascencental_SR}) is solved one can calculate all the required information. 
Using (\ref{eq:v_3_SS}) for $v_3$ and (\ref{eq:rhoR_shockL}) for $\rho_3$  the solution in this region reads

\begin{eqnarray}
p_{exact} &=& p_3,\nonumber\\
v_{exact} &=& v_3\nonumber\\
\rho_{exact} &=& \rho_1 \frac{p_1 (\Gamma-1) + p_3 (\Gamma+1)}{p_3 (\Gamma-1) + p_1 (\Gamma+1)}.\nonumber
\end{eqnarray}
\item Region 4 is defined by $t V_{contact} < x-x_0 < tV_{s,5}$, where the velocity of the shock moving to the right is given by (\ref{eq:SR}) and reads $V_{s,5} =v_6 + a_6 \sqrt{\frac{(\Gamma+1)p_4}{2p_6 \Gamma} + \frac{\Gamma-1}{2\Gamma}}$. Finally, using (\ref{eq:v_4_SS}) for $v_4$ and (\ref{eq:rho_L_R}) for $\rho_4$  the solution in this region reads

\begin{eqnarray}
p_{exact} &=& p_4, \nonumber\\
v_{exact} &=& v_4, \nonumber\\
\rho_{exact} &=& \rho_6 \frac{p_6 (\Gamma-1) + p_4 (\Gamma+1)}{p_4 (\Gamma-1) + p_6 (\Gamma+1)}. \nonumber
\end{eqnarray}

\item There is no region 5.

\item Finally region 6 is defined by the condition $V_{s,5} < x-x_0$. The exact solution is given by the initial values at the chamber on the right:

\begin{eqnarray}
p_{exact} &=& p_6, \nonumber\\
v_{exact} &=& v_6, \nonumber\\
\rho_{exact} &=& \rho_6. \nonumber
\end{eqnarray}

\end{enumerate}

\begin{figure}[htp]
\includegraphics[width=4cm]{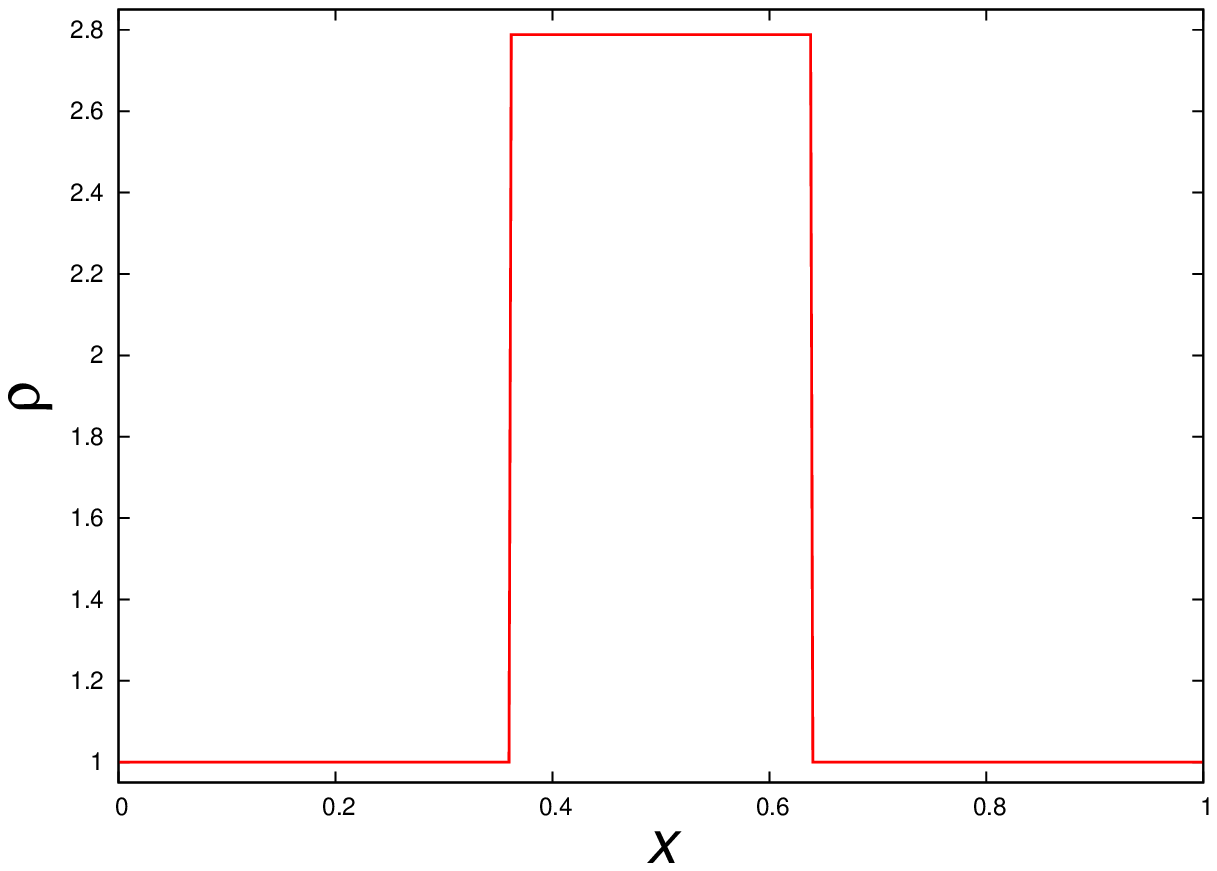}
\includegraphics[width=4cm]{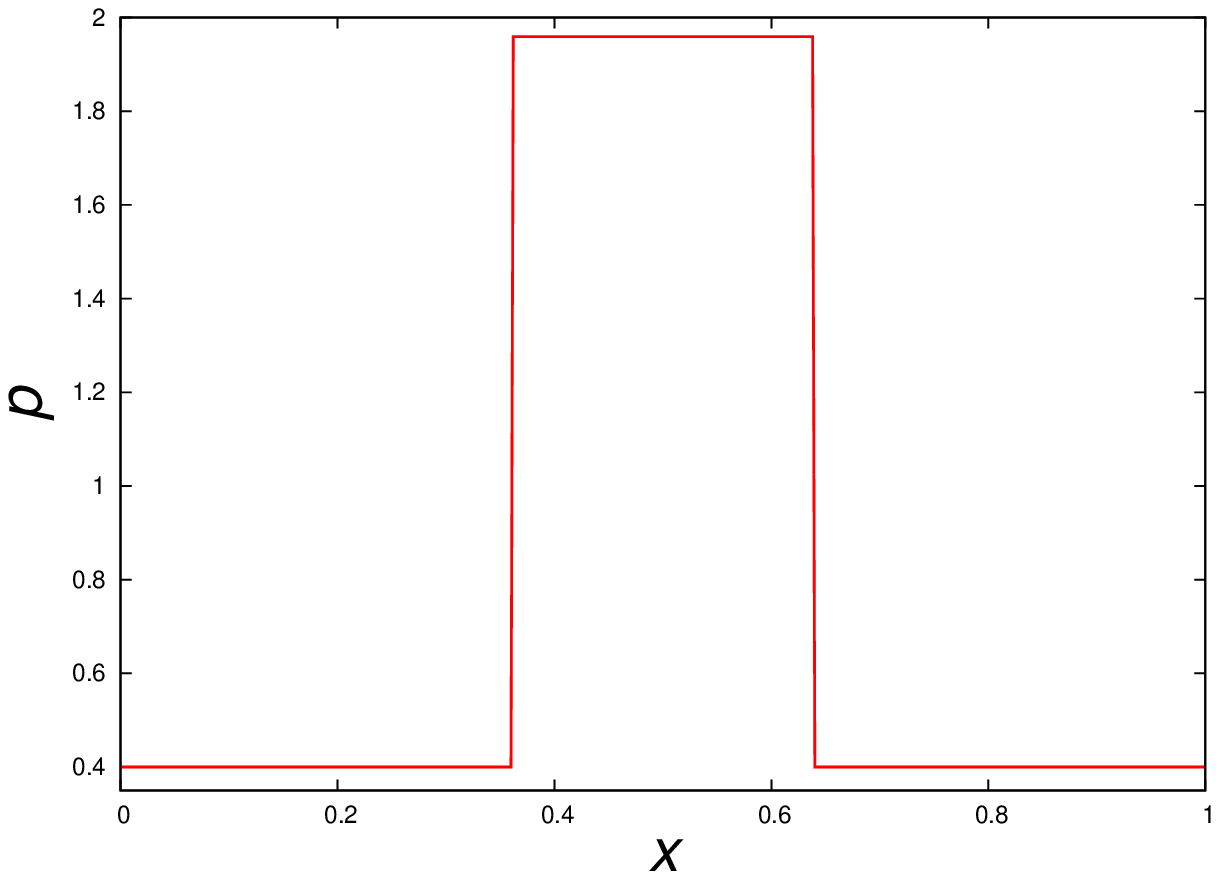}
\includegraphics[width=4cm]{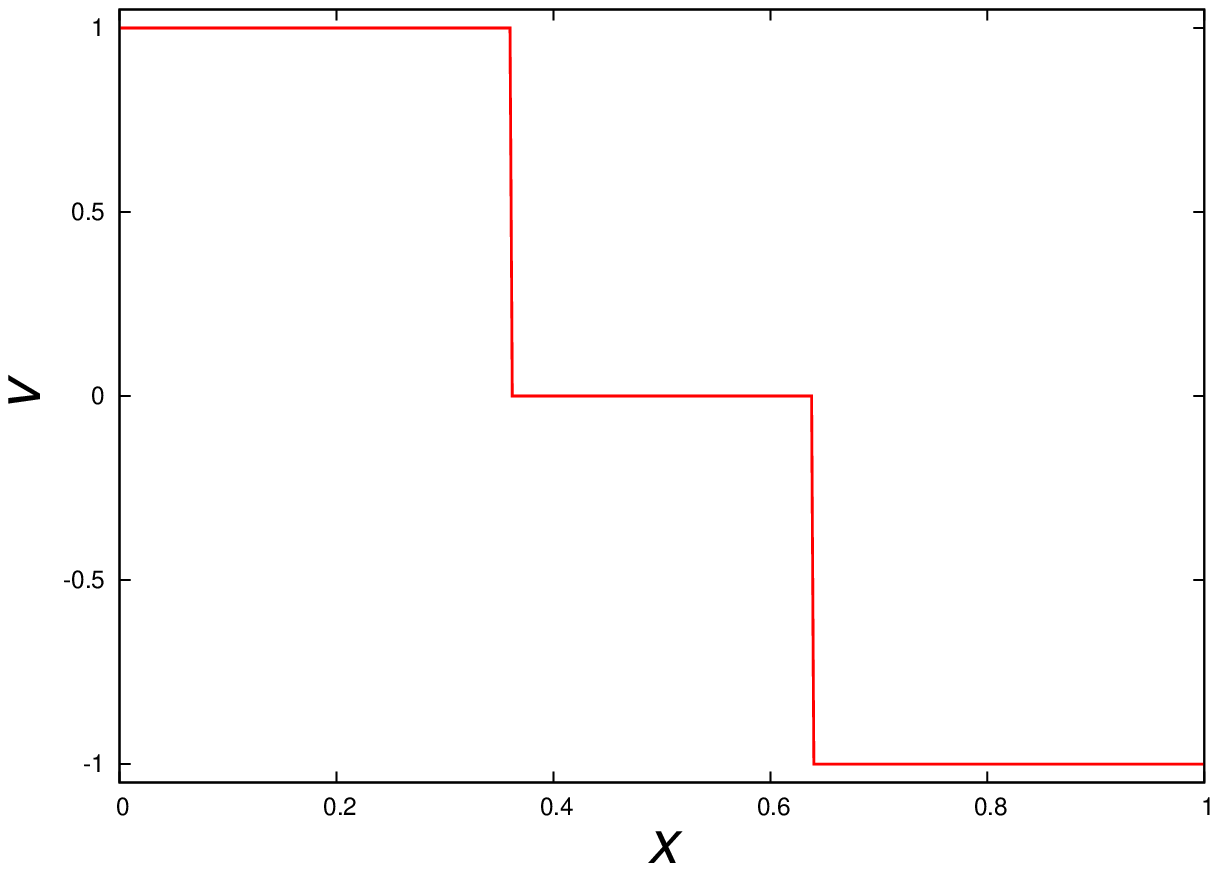}
\includegraphics[width=4cm]{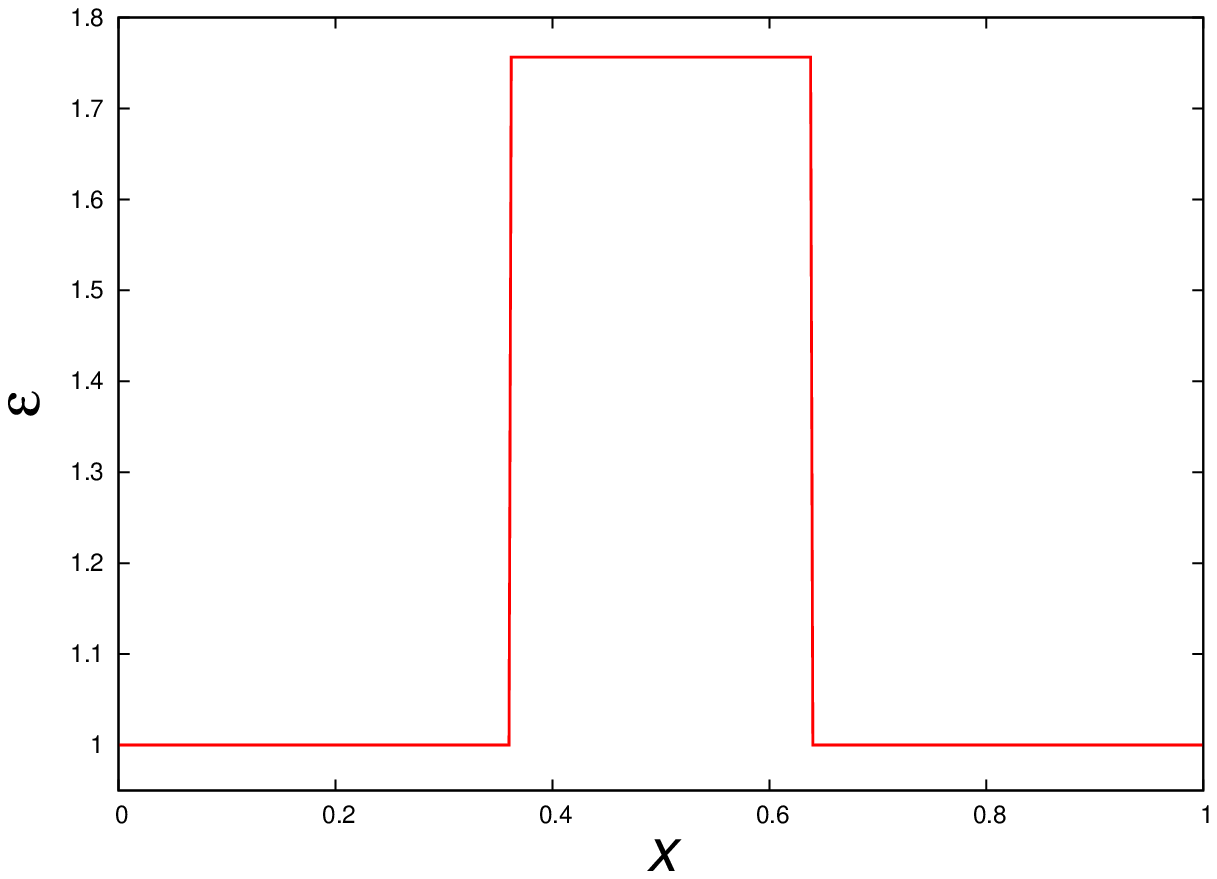}
\caption{\label{fig:Newtonian_SS} Exact solution for the Shock-Shock case at time $t=0.25$ for the parameters in Table \ref{tab:newtonian}.}
\end{figure}

An example is shown in Fig. \ref{fig:Newtonian_SS}.

\section{Relativistic shock tube}
\label{sec:relativistic}

First of all one needs to define a model for the gas. In our case we use the perfect fluid defined because it has no viscosity nor heat transfer, is shear free and is non-compressible. Such system is described by the stress energy tensor

\begin{equation}
T^{\mu\nu} = \rho_0 h u^{\mu}u^{\nu} + p \eta^{\mu\nu},
\end{equation}

\noindent where $\rho_0$ is the rest mass density of a fluid element, $u^{\mu}$ its four velocity, $p$ the pressure, $h=1+\varepsilon+p/\rho_0$ is the specific enthalpy and $\eta^{\mu\nu}$ are the components of the metric describing Minkowski space-time.

The set of relativistic Euler equations is obtained from the local conservation of the rest mass and the local conservation of the stress energy tensor of the fluid, which are respectively

\begin{eqnarray}
(\rho_0 u^{\mu})_{,\mu} &=&0,\nonumber\\
(T^{\mu\nu})_{,\nu} &=& 0,\nonumber
\end{eqnarray}

\noindent where $u^{\mu} = W(1,v^{x},0,0)$ and $W=\frac{1}{\sqrt{1-v^i v_i}}$ is the Lorentz factor and $v^x$ is the Eulerian velocity of the fluid elements. It is possible to arrange these equations as a flux balance set of equations as in the Newtonian case

\begin{equation}
\partial_t {\bf u} + \partial_x {\bf F}({\bf u}) = 0, \label{eq:Euler}
\end{equation}

\noindent where conservative variables are defined by ${\bf u} = (D,S^x,\tau)^T$ and the resulting fluxes are ${\bf F} = (Dv,S v+p, S)$, where we assume that specifically $v=v^x$ and $S=S^x$, since we are only considering one spatial dimension. The conservative variables are defined in terms of the primitive ones as follows

\begin{eqnarray} \label{eq:conservative}
D &=& \rho_0 W, \nonumber\\
S &=& \rho_0 h W^2 v, \nonumber\\
\tau &=& = \rho_0 h W - p.
\end{eqnarray}

\noindent The flux balance equations are explicitly:

\begin{eqnarray}
\partial_t D +  \partial_x (Dv)=0, \label{eq:rel_baryon_cons} \\
\partial_t S + \partial_x (Sv+p) = 0, \label{eq:rel_momentum_cons} \\
\partial_t \tau + \partial_x S =0. \label{eq:rel_energy_cons}
\end{eqnarray}

\noindent The eigenvalues of the Jacobian matrix of this system of equations are

\begin{equation}\label{eq:eigen_rel}
\lambda^o=v, \, \, \, \lambda^{\pm}= \frac{v \pm c_s}{1 \pm v c_s}.
\end{equation}

\noindent Each of the characteristic values (\ref{eq:eigen_rel}) may correspond to eigenvectors with different properties exactly as in the Newtonian case, that is, $\lambda^{0}$ corresponds to a contact discontinuity, whereas the eigenvalues $\lambda^{\pm}$ may correspond to rarefaction or shock waves. The shock tube problem in this case is defined as in the Newtonian case:

\begin{equation}
{\bf u} = \left\{
	\begin{array}{ll}
		{\bf u}_L, & x<x_0 \\
		{\bf u}_R, & x >x_0.
	\end{array}
	\right.
\end{equation}

Next we describe the treatment of each of the wave or discontinuities that develop during the evolution.





\subsection{Rarefaction Waves}
\label{subsec:Rarefaction_Waves}

Rarefaction waves are self-similar solutions of the flow equations \cite{LeVeque}. They are self-similar solutions in the sense that all quantities describing the fluid depend on the variable $\xi=(x-x_0)/t$. In order to explore the change of all physical quantities along the straight line $\xi$, we define the useful
change on the derivative operators 

\begin{equation}\label{eq:change_2_xi}
\partial_t =-\frac{1}{t} \xi \partial_{\xi}, \, \, \, \partial_x = \frac{1}{t} \partial_{\xi}. 
\end{equation}

Using the advective derivative $d_a=\partial_t+v\partial_x$, we obtain the expressions

\begin{eqnarray}
\partial_x p= -D d_a (hWv), \label{eq:1}\\
\partial_t p= D d_a(hW), \label{eq:2}
\end{eqnarray}

\noindent where we have used the rest mass conservation law to simplify the expressions.
From (\ref{eq:change_2_xi}) we obtain for the advective derivative $d_a=\frac{1}{t}(\xi-v) d/d\xi$, for which we will use $d:=d/d\xi$ from now on. With this in mind we obtain from (\ref{eq:1},\ref{eq:2}) the differential equation

\begin{equation}\label{eq:main_rarefaction}
(v-\xi) \rho h W^2 dv + (1-\xi v) dp=0.
\end{equation}

On the other hand, the change of variable in (\ref{eq:rel_baryon_cons}) from $t,x$ to $\xi$ implies

\begin{equation}\label{eq:baryon_cons_xi}
(v-\xi) d \rho + \rho W^2 (1-v \xi) dv = 0.
\end{equation}

\noindent and from equations (\ref{eq:main_rarefaction}) and (\ref{eq:baryon_cons_xi}) we obtain a relation between the density and pressure

\begin{equation}
dp= h \left[ \frac{v-\xi}{1-v \xi} \right]^2 d \rho.
\end{equation}

\noindent Since the process along $\xi$ is isentropic \cite{Taub} the sound speed is $ c^2_s= \frac{1}{h} \frac{\partial p}{\partial \rho} |_s$, which combined with the previous expression implies the speed of sound

\begin{equation}\label{eq:cs_v_xi}
c_s(v,\xi) = \left| \frac{v-\xi}{1-v\xi} \right|.
\end{equation}

\noindent Besides, we can find a useful expression for an isentropic process using $p=K \rho^\Gamma$ (we are using a politropic equation of state).

\begin{equation}\label{eq:cs_rel_classic}
c_s= \sqrt{\frac{\Gamma p}{\rho h}}.
\end{equation}

\noindent From system (\ref{eq:eigen_rel}) we obtain the speed of sound in terms of the eigenvalues of the Jacobian matrix

\begin{eqnarray}\label{eq:cs_dir}
c_s=\left \{  \begin{array}{l l}
    -(v-\lambda^+)/(1-v \lambda^+) & \quad \text{if $\xi=\lambda^+$}, \\
     (v-\lambda^-)/(1- v \lambda^-)& \quad \text{if $\xi=\lambda^-$}.\\
  \end{array} \right.
\end{eqnarray}

Comparing with (\ref{eq:cs_v_xi}) we find that $c_s(v,\lambda^+)$ is the speed of sound for a rarefaction wave traveling to the right and $c_s(v,\lambda^-)$ for a wave traveling to the left. 

According to this equation we get from (\ref{eq:baryon_cons_xi}) that

\begin{equation}\label{eq:diff_R_inv}
W^2 dv \pm \frac{c_s}{\rho} d \rho = 0.
\end{equation}

Here the $+$ sign refers to the wave traveling to the left and the $-$ sign when it travels to the right. From this equation we obtain the Riemann invariant because this differential equation is valid along a straight line along the $x-t$ plane, as long as it is not a shock. Integrating the first term of (\ref{eq:diff_R_inv}) we obatain

\begin{equation}\label{eq:middle_step}
\frac{1}{2} \ln \frac{1+v}{1-v} \pm \int \frac{c_s}{\rho} d \rho= constant.
\end{equation}

\noindent In order to calculate the integral we use the definition of the sound speed and the polytropic equation of state $p= K \rho^\Gamma$, from which we obtain

\begin{equation}\label{eq:cs_square_rho}
c_s^2(\rho)=\frac{K \Gamma (\Gamma - 1 ) \rho^{\Gamma-1}}{\Gamma-1 +  K \Gamma \rho^{\Gamma - 1}},
\end{equation}

\noindent or in terms of the pressure instead of the density the speed of sound  reads

\begin{equation}\label{eq:cs_square_p}
c_s^2(p)= \frac{\Gamma-1}{\frac{1-\Gamma}{K \Gamma} \left( \frac{p}{K}\right)^{\frac{\Gamma-1}{\Gamma}}+1}.
\end{equation}

Conversely, if the speed of sound is known one can calculate the density using (\ref{eq:cs_square_rho}):

\begin{equation}
\rho=\frac{1}{\left[ K \Gamma  \left( \frac{1}{c^2_s}- \frac{1}{\Gamma-1} \right) \right]^{\frac{1}{\Gamma-1}}}.\label{eq:rho_rel_rar}
\end{equation}

Then the integral can be written as

\begin{equation}
\int \frac{c_s}{\rho} d \rho = \int c_s {\left[ K \Gamma  \left( \frac{1}{c^2_s}- \frac{1}{\Gamma-1} \right) \right]^{\frac{1}{\Gamma-1}}} \frac{d \rho}{ d c_s} d c_s.
\end{equation}

\noindent Integrating by parts and using (\ref{eq:middle_step}) we find the useful constraint

\begin{equation}
\frac{1}{2} \ln{\frac{1+v}{1-v}} \pm \frac{1}{(\Gamma -1)^{1/2}} \ln{\left[ \frac{\sqrt{\Gamma-1}+c_s}{\sqrt{\Gamma-1}-c_s}\right]}= constant,
\end{equation}

\noindent which in turn simplifies as follows

\begin{equation}\label{eq:RI_rar}
\frac{1+v}{1-v} A^{\pm}= constant,
\end{equation}

\noindent where $A^{\pm}$ is

\begin{equation}
A^{\pm}=\left[ \frac{\sqrt{\Gamma-1}+c_s}{\sqrt{\Gamma-1}-c_s} \right]^{\pm 2 (\Gamma - 1)^{-1/2}}.\label{eq:Apm_gral}
\end{equation}

Equation (\ref{eq:RI_rar}) is valid only across straight lines arising from the origin $(x_0,t=0)$ and evolving along $\xi=(x-x_0)/t$ inside the rarefaction zone. For this family of straight lines the Riemann invariant is the same. This allows us to relate any two different states in the rarefaction zone, particularly we are going to take the states $L$ and $R$ as the states just next to the left and to the right from the rarefaction wave. 

\begin{equation}\label{eq:RL_rar}
\frac{1+v_L}{1-v_L} A_L^{\pm}= \frac{1+v_R}{1-v_R} A_R^{\pm}.
\end{equation}

Assuming that when the wave is propagating to the left we account with information from the left state, we can calculate the velocity of the fluid on the region at the right from the wave in terms of the state variables on the state at the left and $A^{+}$:

\begin{equation}
v_R= \frac{(1+v_L)A^{+}_L-(1-v_L)A^{+}_R}{(1+v_L)A^{+}_L+(1-v_L)A^{+}_R}.\label{eq:velR_rarL}
\end{equation}

Analogously when the wave is moving to the right we expect to account with information on the state to the right. Then we can express the velocity on the left in terms of the variables on the state at the right and $A^{-}$

\begin{equation}
v_L= \frac{(1+v_R)A^{-}_R-(1-v_R)A^{-}_L}{(1+v_R)A^{-}_R+(1-v_R)A^{-}_L}.\label{eq:rar_vel_L}
\end{equation}

\subsubsection{The fan}

The fan is the region where the rarefaction takes place, propagating with velocity either $\lambda^{+}$ if the wave is moving to the right or $\lambda^{-}$ when moving to the left. The fan will be bounded by two values of $\xi$ corresponding to the head and the tail of the wave:

\begin{eqnarray}
\xi_h = \frac{v_{L,R} \pm c^{(L,R)}_s}{1 \pm v c^{(L,R)}_s}, \label{eq:headxi}\\
\xi_t = \frac{v_{R,L} \pm c^{(R,L)}_s}{1 \pm v c^{(R,L)}_s},\label{eq:tailxi}
\end{eqnarray}

\noindent where the $-$ sign applies to waves traveling to the left and $+$ when the wave moves to the right. In order to construct the solution inside the fan, we use the constraint (\ref{eq:RL_rar}). We have two cases according to the direction of the rarefaction wave. If the rarefaction wave travels to left we use

\begin{equation}\label{eq:fan1_r}
\frac{1+v_L}{1-v_L} A_L^{+}- \frac{1+v_R}{1-v_R} A_R^{+}=0	
\end{equation}

\noindent and solve the equation for $v_R$. When the rarefaction wave travels to right we use

\begin{equation}\label{eq:fan2_r}
\frac{1+v_L}{1-v_L} A_L^{-}- \frac{1+v_R}{1-v_R} A_R^{-}	=0,
\end{equation}

\noindent and solve the equation for $v_L$. We calculate in each case $A^{\pm}$ using (\ref{eq:Apm_gral}) in the appropriate region

\begin{equation}\label{eq:fan1_r_A}
A^{\pm}_{(L,R)}=\left[ \frac{\sqrt{\Gamma-1}+c_{s,(L,R)}^{\pm}}{\sqrt{\Gamma-1}-c_{s,(L,R)}^{\pm}} \right]^{\pm 2 (\Gamma - 1)^{-1/2}},
\end{equation}

\noindent where the sound speed is given by (\ref{eq:cs_v_xi}) and (\ref{eq:cs_dir})

\begin{equation}\label{eq:fan_2}
c_{s,(L,R)}^{\pm}= \pm \frac{v_{(L,R)}-\xi}{1-v_{(L,R)} \xi},
\end{equation}

\noindent where the $+$ sign is used when the wave moves to the left and $-$ when moving to the right. 
Finally since we are in the rarefaction zone we can express a point $(x,t)$ with $\xi=(x-x_0)/t$ in (\ref{eq:fan_2}) and using this expression in (\ref{eq:fan1_r_A}) and substituting into  (\ref{eq:fan1_r}) or (\ref{eq:fan2_r}) depending on the direction of propagation we finally obtain a trascendental equation for the velocity $v_{(L,R)}$. We assume that if the wave moves to the left we know the variables on the state to the left $L$ and ignore those of the state to the right $R$ and viceversa. Then we look for a solution of $v_L$ when the wave moves to the left and of $v_R$ when moving to the right.
Instead of looking for a closed solution to this equation we solve it numerically to obtain $v_{(L,R)}$ assuming we know $v_{(R,L)}$. Once $v_{(L,R)}$ is calculated we can substitute back, and using equation (\ref{eq:fan_2}) obtain the sound speed; next, using (\ref{eq:rho_rel_rar}) obtain the density $\rho$; finally with the help of the EOS we can calculate the pressure $p=K\rho^{\Gamma}$. This completes the solution in the fan region.

The particular cases described later illustrate how to implement this procedure.

\subsection{Shock Waves}
\label{subsec:Shock_Waves}

Shocks require the use of the relativistic Rankine-Hugoniot jump conditions $[\rho_0 u^{\mu}] n_{\mu}=0$ and $[T^{\mu\nu}]n_{\nu}=0$ across the shock \cite{Taub}, where $n^{\mu}=(-V_s W_s,Ws,0,0)$ is a normal vector to the shock's front, $W_s$ is the shock's Lorentz factor and $V_s$ is the speed of the shock. Here we have used the notation $[F]=F_L -F_R$, where $F_L$ and $F_R$ are the values of the function $F$ at both sides of the shock's surface. These conditions reduce to the following system of equations, in terms of primitive and conservative variables, as 

\begin{eqnarray} 
D_L v_L - D_R v_R &=& V_s (D_L - D_R), \label{eq:RH_1} \\ 
S_L v_L + p_L - (S_R v_R + p_R) &=& V_s (S_L-S_R), \label{eq:RH_2}\\ 
S_L - S_R &=& V_s (\tau_L - \tau_R). \label{eq:RH_3}
\end{eqnarray}

\noindent The subindices $(L,R)$ represent  two arbitrary states at left and at the right from the shock. These equations can be written in the reference rest frame of the shock by considering a Lorentz transformation, that is

\begin{eqnarray} 
\hat{D}_L \hat{v}_L  &=& \hat{D}_R \hat{v}_R, \label{eq:S1} \\ 
\hat{S}_L \hat{v}_L  + p_L  &=&  \hat{S}_R \hat{v}_R + p_R , \label{eq:S2} \\ 
\hat{S}_L &=& \hat{S}_R , \label{eq:S3}
\end{eqnarray} 

\noindent where the hatted quantities are evaluated at the rest frame of the shock. Here $\hat{v}_{(L,R)}=\frac{V_s - v_{(L,R)}}{1-V_s v_{(L,R)}}$, $\hat{D}_{(L,R)}=\rho_{(L,R)} \hat{W}_{L,R}$, $\hat{S}_{(L,R)}=\rho_{(L,R)} h_{(L,R)} 	\hat{W}^2_{(L,R)} \hat{v}_{(L,R)}$ and $\hat{W}_{(L,R)} = \frac{1}{\sqrt{1 - \hat{v}^2_{(L,R)}}}$.

From (\ref{eq:S1}), we can introduce the invariant relativistic mass flux across the shock as

\begin{equation} 
j = W_s D_L (V_s - v_L) = W_s D_R (V_s - v_R), \label{eq:IRMFS}
\end{equation}

\noindent where $W_s = \frac{1}{\sqrt{1-V^2_s}}$. It is important to point out that when the shock moves to the right the mass flux is positive $j>0$, whereas when the shock moves to the left it has to be negative $j<0$.

Now, using the expression for the mass flux (\ref{eq:IRMFS}) into the Rankine-Hugoniot conditions (\ref{eq:RH_1}, \ref{eq:RH_2}, \ref{eq:RH_3}) we can obtain the following system of equations in terms of a combination of primitive and conservative variables  

\begin{eqnarray}
v_L - v_R &=& - \frac{j}{W_s} \left( \frac{1}{D_L} - \frac{1}{D_R} \right), \label{eq:R_a} \\
p_L - p_R &=&  \frac{j}{W_s} \left( \frac{S_L}{D_L} - \frac{S_R}{D_R} \right), \label{eq:R_b} \\
v_L p_L - v_R p_R &=&  \frac{j}{W_s} \left( \frac{\tau_L}{D_L} - \frac{\tau_R}{D_R} \right). \label{eq:R_c} 
\end{eqnarray}

Considering the shock is moving to the right and thus that the state $R$ is known, we will write an expression for the velocity $v_L$ in terms of the state variables $R$ and also in terms of $j$, $V_s$ and $p_L$. In order to do this, we rewrite expressions (\ref{eq:R_b}) and (\ref{eq:R_c}) using the definitions for the conservative variables in terms of the primitive variables (\ref{eq:conservative}) as follows

\begin{eqnarray}
\frac{W_s}{j v_L}(p_L - p_R) &=& h_L W_L - h_R W_R \frac{v_R}{v_L}, \label{eq:S_b} \\
\frac{W_s}{j }(v_L p_L - v_R p_R) &=&  h_L W_L - \frac{p_L}{\rho_L W_L} \\ \nonumber  
&-& h_R W_R + \frac{p_R}{\rho_R W_R} . \label{eq:S_c} 
\end{eqnarray}
 
 \noindent Subtracting these expressions and dividing by $p_L$ we get
 
\begin{eqnarray}
&& \frac{W_s}{j}\left( v_L - \frac{v_R p_R}{p_L} - \frac{1}{v_L} + \frac{p_R}{v_Lp_L} \right) = \\ \nonumber 
&& \frac{h_R W_R}{p_L}\left(\frac{v_R}{v_L} -1 \right) + \frac{p_R}{p_L \rho_R W_R} - \frac{1}{\rho_L W_L}.
\end{eqnarray} 
 
\noindent Inserting this into (\ref{eq:R_a}) we finally obtain an expression for the velocity $v_L$ 

\begin{equation}
v_L = \frac{h_R W_R v_R  + \frac{W_s}{j} (p_L - p_R)}{h_R W_R + (p_L - p_R) \left( \frac{W_s v_R}{j} + \frac{1}{\rho_R W_R} \right)}. \label{eq:vleft}
\end{equation} 
 
 \noindent When the shock moves to the left and the state $L$ is known, the velocity on the state to the right is
 
\begin{equation}
v_R = \frac{h_L W_L v_L  + \frac{W_s}{j} (p_R - p_L)}{h_L W_L + (p_R - p_L) \left( \frac{W_s v_L}{j} + \frac{1}{\rho_L W_L} \right)}, \label{eq:vRight}
\end{equation} 
 
\noindent where the condition $j<0$ has to be satisfied. 
 
In order to obtain the shock velocity $V_s$, we start form the mass flux conservation across the shock (\ref{eq:IRMFS}), which relates the shock velocity with the mass flux. Substituting $W_s=1/\sqrt{1-V_{s}^{2}}$, it is possible to solve the resulting quadratic equation and obtain the two roots for the shock velocity

\begin{eqnarray}
V_s &=& \frac{\rho^2_R W^2_R v_R + \sqrt{j^4 + j^2 \rho^2_R}}{\rho^2_R W^2_R + j^2}, \label{eq:ShockVelR} \\
V_s &=& \frac{\rho^2_L W^2_L v_L - \sqrt{j^4 + j^2 \rho^2_L}}{\rho^2_L W^2_L + j^2}, \label{eq:ShockVelL}
\end{eqnarray} 

\noindent which correspond respectively to a shock moving to the right and to the left. The signs of the quadratic formula are chosen such that they are physically possible, that is, for the case of a shock moving to the right $j>0$ we use (\ref{eq:ShockVelR}) and for a shock moving to the left $j<0$ we use (\ref{eq:ShockVelL}) \cite{Marti}.
 
In order to solve completely the problem across the shock, we first express equation (\ref{eq:S2}) as 

\begin{eqnarray}
\nonumber  \frac{\rho_L h_L (V_s-v_L)^2}{1-V^2_s-v^2_L + V^2_s v^2_L} &-&  \frac{\rho_R h_R (V_s-v_R)^2}{1-V^2_s-v^2_R + V^2_s v^2_R} =\\
 &-& (p_L - p_R).
\end{eqnarray}

\noindent Considering that $W_s W_{(L,R)} = \frac{1}{\sqrt{1-V^2_s}\sqrt{1-v^2_{(L,R)}}} = \frac{1}{\sqrt{1-V^2_s-v^2_{(L,R)} + V^2_s v^2_{(L,R)}}}$ the last equation takes the following form

\begin{equation}
\rho_L h_L W^2_s W^2_L (V_s - v_L)^2 - \rho_R h_R W^2_s W^2_R (V_s - v_R)^2 = -(p_L - p_R).
\end{equation}

\noindent As we can see from this equation, the definition of the conserved mass flux is present, then using equation (\ref{eq:IRMFS}) in this last equation, we obtain a useful expression for the square of the flux

\begin{equation}
j^2 = \frac{-(p_L - p_R)}{\left(\frac{h_L}{\rho_L} - \frac{h_R}{\rho_R} \right)}, \label{eq:Taubs1}
\end{equation}

\noindent where the positive root corresponds to a shock moving to the right whereas the negative root to a shock moving to the left. 

Another useful expression comes from equation (\ref{eq:S3}), which can be rewritten directly in the form 

\begin{equation}
h_L \hat{W}_L = h_R \hat{W}_R, \label{eq:Taubs2}
\end{equation} 

\noindent which combined with equation  (\ref{eq:Taubs1}) implies 

\begin{equation}
h^2_L - h^2_R = (p_L - p_R)\left( \frac{h_L}{\rho_L} + \frac{h_R}{\rho_R} \right) . \label{eq:Taubs3}
\end{equation}

\noindent This last equation is commonly called the {\it Taub's adiabat}. Moreover  equations (\ref{eq:Taubs1}), (\ref{eq:Taubs2}) and (\ref{eq:Taubs3}) are known as relativistic {\it Taub's} junction conditions for shock waves \cite{Taub,KThorne}.

Finally, in order to obtain the density $\rho_L$ and pressure $p_L$ for a shock moving to the right in terms of the variables in the region to the right, we consider the definition of the specific internal enthalpy and that the fluid obeys and ideal gas equation of state. With these assumptions equation (\ref{eq:Taubs3}) can be rewritten in the form 

\begin{eqnarray}
\nonumber && \frac{1}{\rho_L}[p_L (2\sigma -1) + p_R] + \frac{\sigma}{\rho^2_L} [p^2_L (\sigma-1)+p_L p_R] = \\
&& \frac{1}{\rho_R}[p_R (2\sigma -1) + p_L] + \frac{\sigma}{\rho^2_R} [p^2_R (\sigma-1)+p_L p_R] , 
\end{eqnarray}

\noindent where $\sigma=\frac{\Gamma}{\Gamma-1}$. The solution for the quadratic equation reads

\begin{widetext}
\begin{eqnarray}
\frac{1}{\rho_L} &=& \frac{-[p_L (2\sigma -1) + p_R] \pm \sqrt{[p_L (2\sigma -1) + p_R]^2 + 4\zeta_{L}\sigma [p^2_L (\sigma-1)+p_L p_R]}}{2\sigma[p^2_L (\sigma-1)+p_L p_R]}, \label{eq:densityL}  \\
\frac{1}{\rho_R} &=& \frac{-[p_R (2\sigma -1) + p_L] \pm \sqrt{[p_R (2\sigma -1) + p_L]^2 + 4\zeta_{R}\sigma [p^2_R(\sigma-1)+p_R p_L]}}{2\sigma[p^2_R (\sigma-1)+p_R p_L]}, \label{eq:densityR}
\end{eqnarray} 
\end{widetext}

\noindent where $\zeta_{L}=\frac{1}{\rho_R}[p_R (2\sigma -1) + p_L] + \frac{\sigma}{\rho^2_R} [p^2_R (\sigma-1)+p_L p_R]$, and $\zeta_{R}=\frac{1}{\rho_L}[p_L (2\sigma -1) + p_R] + \frac{\sigma}{\rho^2_L} [p^2_L (\sigma-1)+p_R p_L]$.  A physically acceptable solution requires $\rho>0$, which restricts the sign to be positive one in both cases.

\subsection{Contact Wave}

The equations describing the jump conditions (\ref{eq:RH_1},\ref{eq:RH_2},\ref{eq:RH_3}) admit the solution using $V_s=v_R=v_L=\lambda^o=V_{contact}$ where $v_R$ and $v_L$ are the values of the velocity of the fluid at the right and at the left from the contact discontinuity. This represents the contact wave traveling along the line $x-x_0=\lambda^0 t$.

Then (\ref{eq:RH_1}) is trivial and (\ref{eq:RH_2}) reads

\begin{equation}
(S_L  -S_R) V_s+ p_L - p_R =  (S_L-S_R) V_s,
\end{equation}

\noindent which implies $p_R=p_L$ and equation (\ref{eq:RH_3}) is satisfied.

We are now in the position of analyzing each of the possible combinations of shock and rarefaction waves in a Riemann problem. We then proceed in the same way as in the Newtonian case studying each combination.
\subsection{The four different cases}

In what follows, as we did for the Newtonian case, we present the four combinations of rarefaction and shock waves associated to the relativistic Riemann problem. We illustrate each case with a particular set of parameters contained in Table \ref{tab:relativistic}.

\begin{table*}
\begin{tabular}{|c|c|c|c|c|c|c|}\hline
Case	& $p_L$	& $p_R$	& $v_L$	& $v_R$	& $\rho_L$	&	$\rho_R$\\\hline
Rarefaction-Shock & 13.33	& 0	& 0		& 0 &	10	&	1 \\
Shock-Rarefaction	& 0		& 13.33		& 0.0	&0.0	& 1	& 10	\\
Rarefaction-Rarefaction	& 0.05	&-0.05	&-0.2	& 0.2	& 0.1	&0.1\\
Shock-Shock	& 3.333e-9	&-3.333e-9	&0.999999	&0.999999	&0.001	& 0.001\\\hline
\end{tabular}
\caption{Initial data for the four different cases. We choose the spatial domain to be $x\in [0,1]$ and the location of the membrane at $x_0=0.5$. In all cases we use $\Gamma=4/3$.}
\label{tab:relativistic}
\end{table*}

\subsubsection{Case 1: Rarefaction-Shock}
\label{subsec:RS}

The contact wave conditions are $v_3=v_4=v^*$ and $p_3=p_4=p^*$. The velocity in region 3 is given by equation (\ref{eq:velR_rarL}) that provides the velocity on the state at the right from a rarefaction wave moving to the left:

\begin{equation}\label{eq:rsr-v3}
v_3= \frac{(1+v_1)A^{+}_1-(1-v_1)A^{+}_3}{(1+v_1)A^{+}_1+(1-v_1)A^{+}_3}.
\end{equation}

\noindent where according to (\ref{eq:fan1_r_A})

\begin{equation}
A^{+}_{(1,3)}=\left[ \frac{\sqrt{\Gamma-1}+c_{s,(1,3)}^{+}}{\sqrt{\Gamma-1}-c_{s,(1,3)}^{+}} \right]^{+ 2 (\Gamma - 1)^{-1/2}}.
\end{equation}

\noindent Here $c^{+}_{s,1}:=c_s(p_1)=\sqrt{\Gamma p_1/(\rho_1 h_1)}$, $h_1=1+\frac{p_1 \Gamma}{\rho_1(\Gamma-1)}$
and $c^{+}_{s,3}:=c_s(p_3)$ is given by equation (\ref{eq:cs_square_p})

\begin{equation}
c^+_{s,3}(p_3)= \sqrt{\frac{\Gamma-1}{\frac{\Gamma-1}{K \Gamma} \left( \frac{p_3}{K}\right)^{\frac{1-\Gamma}{\Gamma}}+1}},\,\, ~~K=\frac{p_1}{\rho^\Gamma_1},
\end{equation}

\noindent where we remind the reader that in the rarefaction region the polytopic constant remains the same during the process, that is, it is the same in regions 1, 2 and 3. On the other hand the velocity of the gas in region 4 corresponds to the velocity on the state at the left of a shock moving to the right (\ref{eq:vleft})

\begin{equation}\label{eq:rsr-v4}
v_4= \frac{h_6 W_6 v_6 + \frac{W_{s,5}}{j} (p_4-p_6)}{h_6W_6+(p_4-p_6)\left( \frac{W_{s,5} v_6}{j}+ \frac{1}{\rho_6 W_6}\right)},
\end{equation}

\noindent where $W_{s,5}=1/\sqrt{1-V_{s,5}^2}$ is the Lorentz factor of the shock, where we use the subindex 5 in order to denote the shock occurring in region 5. In order to obtain $v_4$ in terms of $p_4$ we need to perform the following steps:

\begin{itemize}

\item The rest mass density $\rho_4$ is given in terms of $p_4$ and other known information can be expressed using (\ref{eq:densityL}) as

\begin{widetext}
\begin{eqnarray}
\frac{1}{\rho_4} &=& \frac{-[p_4 (2\sigma -1) + p_6] + \sqrt{[p_4 (2\sigma -1) + p_6]^2 + 4\zeta_{4}\sigma [p^2_4 (\sigma-1)+p_4 p_6]}}{2\sigma[p^2_4 (\sigma-1)+p_4 p_6]}, \label{eq:rho4_rs}\\
\zeta_{4}&=&\frac{1}{\rho_6}[p_6 (2\sigma -1) + p_4] + \frac{\sigma}{\rho^2_6} [p^2_6 (\sigma-1)+p_4 p_6],\qquad \text{where $\sigma=\frac{\Gamma}{\Gamma-1}$.}
\end{eqnarray}
\end{widetext}

\item Once $\rho_4$ is given in terms of $p_4$ it is possible to compute the enthalpy in region 4 as $h_4=1+\sigma \frac{p_4}{\rho_4}$. 

\item Then equation (\ref{eq:Taubs1}) reads

\begin{equation}\label{eq:j_rs}
j^2=-\frac{(p_4-p_6)}{\frac{h_4}{\rho_4}-\frac{h_6}{\rho_6}},
\end{equation}

\noindent where $h_6=1+\sigma \frac{p_6}{\rho_6}$. Something to remember here is the fact that as the shock moves to the right, we consider $j$ to be the positive square root.

\item Once $j$ is obtained, the shock velocity can be found from expression (\ref{eq:ShockVelR}) as

\begin{equation}\label{eq:vs_r_sRS}
V_{s,5}=\frac{\rho^2_6 W^2_6 v_6+ |j| \sqrt{j^2+\rho^2_6}}{j^2+ \rho^2_6 W^2_6}.
\end{equation}

\item Finally one calculates $W_{s,5}=\frac{1}{\sqrt{1-V^2_{s,5}}}$ and in this way $v_4$ in terms of $p_4$ and the known state in region 6 using (\ref{eq:rsr-v4}).
\end{itemize}

According to the contact discontinuity condition $v_3=v_4=v^*$, we equate (\ref{eq:rsr-v3}) and (\ref{eq:rsr-v4}) and obtain a transcendental equation for $p^*$:

\begin{eqnarray}\label{eq:newrap_rs}
\frac{(1+v_1)A^{+}_1-(1-v_1)A^{+}_3(p^*)}{(1+v_1)A^{+}_1+(1-v_1)A^{+}_3(p^*)} &-& \nonumber \\
\frac{h_6 W_6 v_6 + \frac{W_s}{j} (p^*-p_6)}{h_6W_6+(p^*-p_6)\left( \frac{W_s v_6}{j}+ \frac{1}{\rho_6 W_6}\right)}&=&0,
\end{eqnarray}

\noindent which has to be solved using a root finder.

Once this equation is solved, $p_3$ and $p_4$ are automatically known and $v_3$ and $v_4$ can be calculated using (\ref{eq:rsr-v3}) and (\ref{eq:rsr-v4}), respectively. It is possible to calculate $\rho_3$ using the fact that in the rarefaction zone the process is adiabatic and then $\rho_3=\rho_1(p_3/p_1)^{1/\Gamma}$. On the other hand we can also  calculate $\rho_4$ using (\ref{eq:rho4_rs}). With this information it is already possible to construct the solution in the whole domain. 

Up to this point we account with the known initial states $(p_1,v_1,\rho_1)$ and $(p_6,v_6,\rho_6)$, the solution in regions 3 and 4 given by $(p_3,v_3,\rho_3)$ and $(p_4,v_4,\rho_4)$, and $V_{s,5}$ which represents the velocity of propagation of the shock 5. The exact solution region by region is described next.

\begin{enumerate}

\item Region 1 is defined by the condition $x-x_0 < t \xi_{h}$, where according to (\ref{eq:headxi}) $\xi_{h}$ is the velocity of the head of the rarefaction wave traveling to the left $\xi_h=\frac{v_1 - c_{s,1}}{1 - v_1 c_{s,1}}$. The values of the physical variables are known from the initial conditions:

\begin{eqnarray}
p_{exact} &=& p_1, \noindent\\
v_{exact} &=& v_1, \noindent\\
\rho_{exact} &=& \rho_1. \noindent
\end{eqnarray}

\item Region 2 is defined by the condition $t \xi_h < x-x_0 < t \xi_t$, where according to (\ref{eq:tailxi}) $\xi_t$ is the characteristic value again, but this time evaluated at the tail of the rarefaction wave, that is $\xi_{t}=\frac{v_3-c_{s,3}}{1 - v_3 c_{s,3}}$.  In order to compute $v_2$ we use (\ref{eq:fan1_r})

\begin{equation}\label{eq:forv2RS}
\frac{1+v_1}{1-v_1} A_1^{+}- \frac{1+v_2}{1-v_2} A_2^{+}(v_2)=0	
\end{equation}

\noindent considering equations (\ref{eq:cs_rel_classic}), (\ref{eq:fan1_r_A}) and (\ref{eq:fan_2}) as follows

\begin{eqnarray}
A^{+}_{(1,2)} &=& \left[ \frac{\sqrt{\Gamma-1}+c_{s,(1,2)}^{+}}{\sqrt{\Gamma-1}-c_{s,(1,2)}^{+}} \right]^{+ 2 (\Gamma - 1)^{-1/2}}, \\
c^+_{s,1} &=& \sqrt{\frac{\Gamma p_1}{ \rho_1 h_1}}, \, \, h_1=1+\frac{p_1}{\rho_1}\left( \frac{\Gamma}{\Gamma -1}\right)\\
c^+_{s,2} &=& \frac{v_{2}-\xi}{1-v_{2} \xi} ~~\Rightarrow~~ v_2 = \frac{\xi + c^+_{s,2}}{1+ c^+_{s,2} \xi}. \label{eq:cs_square_2RS}
\end{eqnarray}

\noindent where $\xi=(x-x_0)/t$.  In this way, equation (\ref{eq:forv2RS}) is transcendental and has to be solved 
equivalently for $v_2$ or for $c^+_{s,2}$ using a root finder for each point of region 2. We recommend solving for $c^+_{s,2}$ and then construct $v_2$ using (\ref{eq:cs_square_2RS}). Finally we calculate $\rho_2$ using equation (\ref{eq:rho_rel_rar}):

\begin{equation}
\rho_2=\frac{1}{\left[ K \Gamma  \left( \frac{1}{(c^+_{s,2})^2}- \frac{1}{\Gamma-1} \right) \right]^{\frac{1}{\Gamma-1}}},\, \, ~~K= \frac{p_1}{\rho_1^\Gamma}.
\end{equation}

\noindent Finally we obtain $p_2$ using the fact that in the process $K$ is constant 

\begin{equation}
p_2=p_1\left( \frac{\rho_2}{\rho_1} \right)^\Gamma.
\end{equation}

\item Region 3 is defined by the condition $t \xi_{t} < x-x_0 < t V_{contact}$, where $V_{contact}=\lambda_o=v_3=v_4$. The solution there reads

\begin{eqnarray}
p_{exact} &=& p_3, \noindent\\
v_{exact} &=& v_3, \noindent\\
\rho_{exact} &=& \rho_3. \noindent
\end{eqnarray}

\item Region 4 is defined by the condition $t V_{contact}< x-x_0 < t V_{s,5}$, where $V_{s,5}$ is given by (\ref{eq:vs_r_sRS}) and explicitly

\begin{eqnarray}
p_{exact} &=& p_4, \noindent\\
v_{exact} &=& v_4, \noindent\\
\rho_{exact} &=& \rho_4. \noindent
\end{eqnarray}

\item There is no region 5. Only the shock traveling with speed $V_{s,5}$.

\item Region 6 is defined by $t V_{s,5} < x-x_0$. In this region the solution is simply

\begin{eqnarray}
p_{exact} &=& p_6, \noindent\\
v_{exact} &=& v_6, \noindent\\
\rho_{exact} &=& \rho_6. \noindent
\end{eqnarray}
\end{enumerate}

\begin{figure}[htp]
\includegraphics[width=4cm]{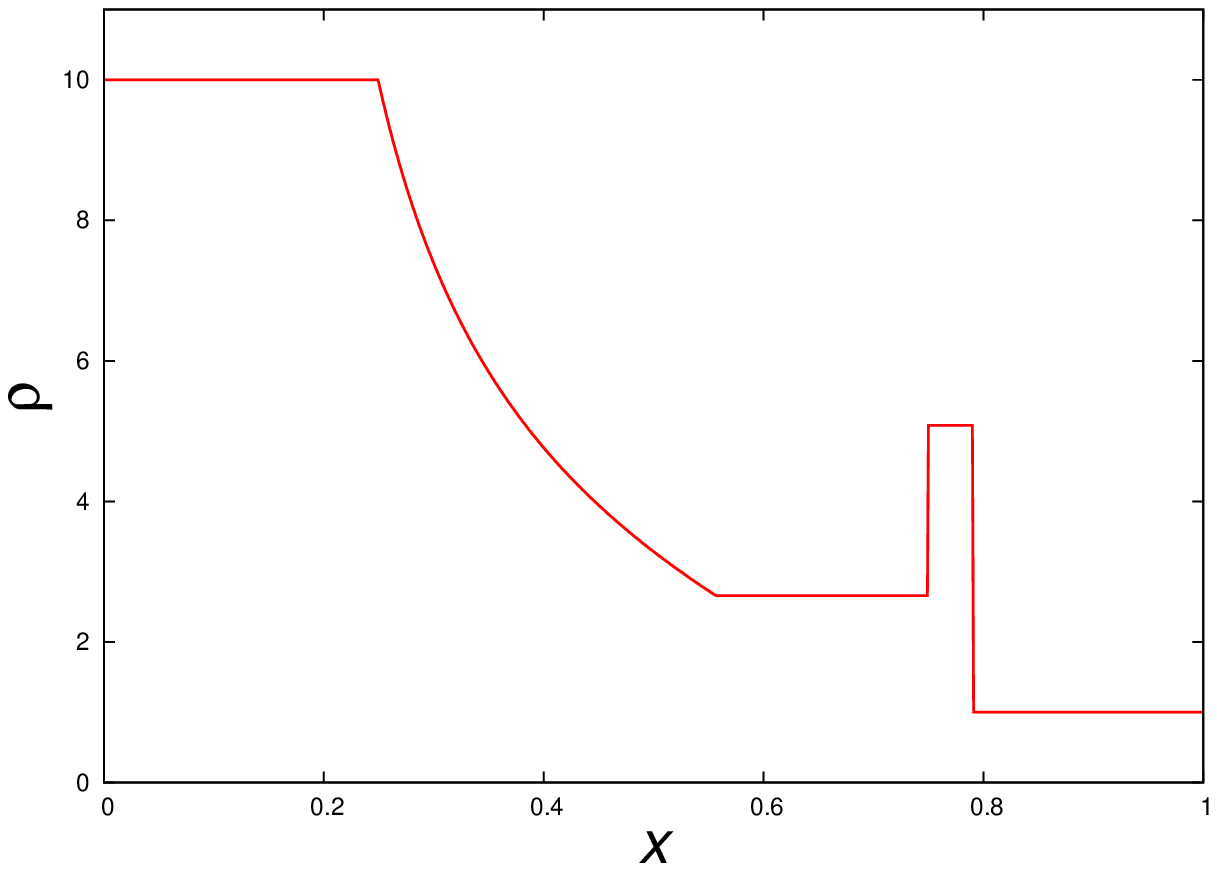}
\includegraphics[width=4cm]{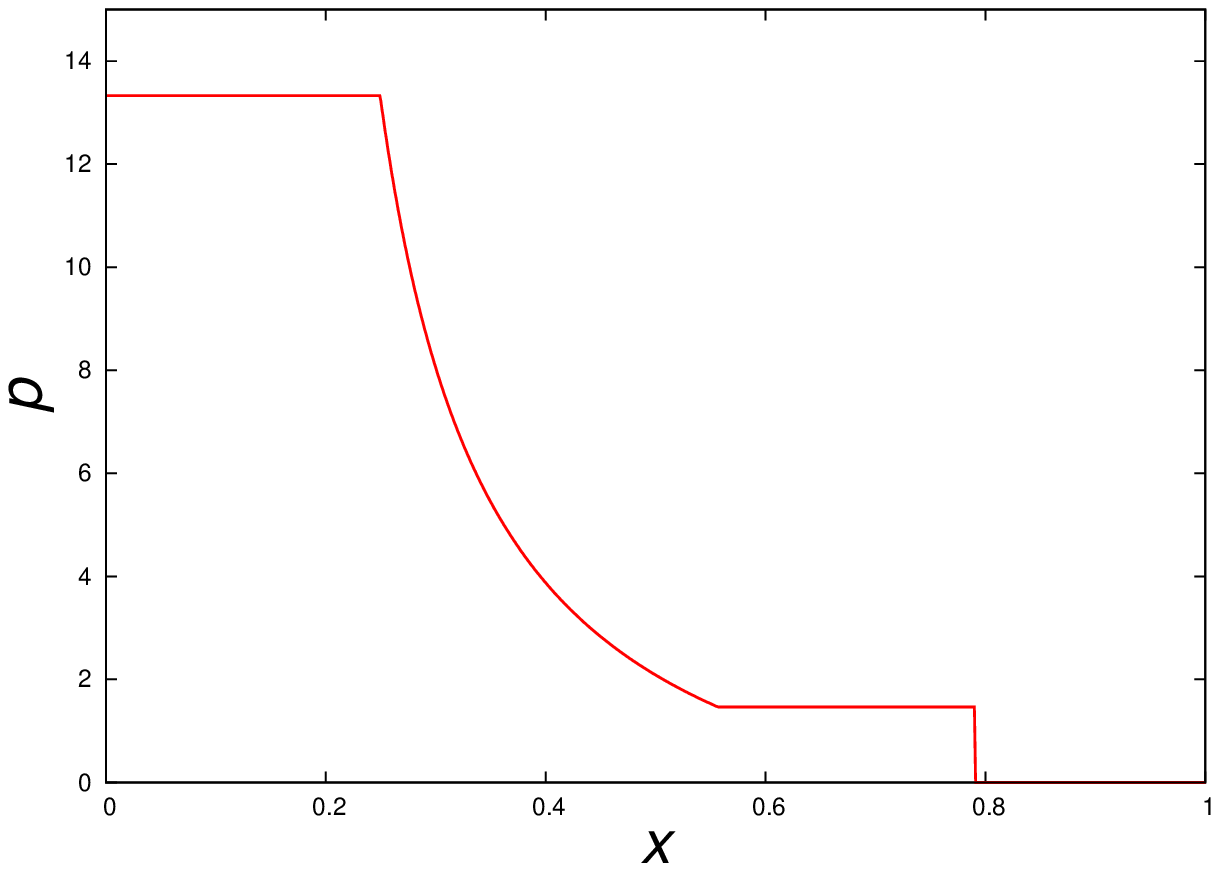}
\includegraphics[width=4cm]{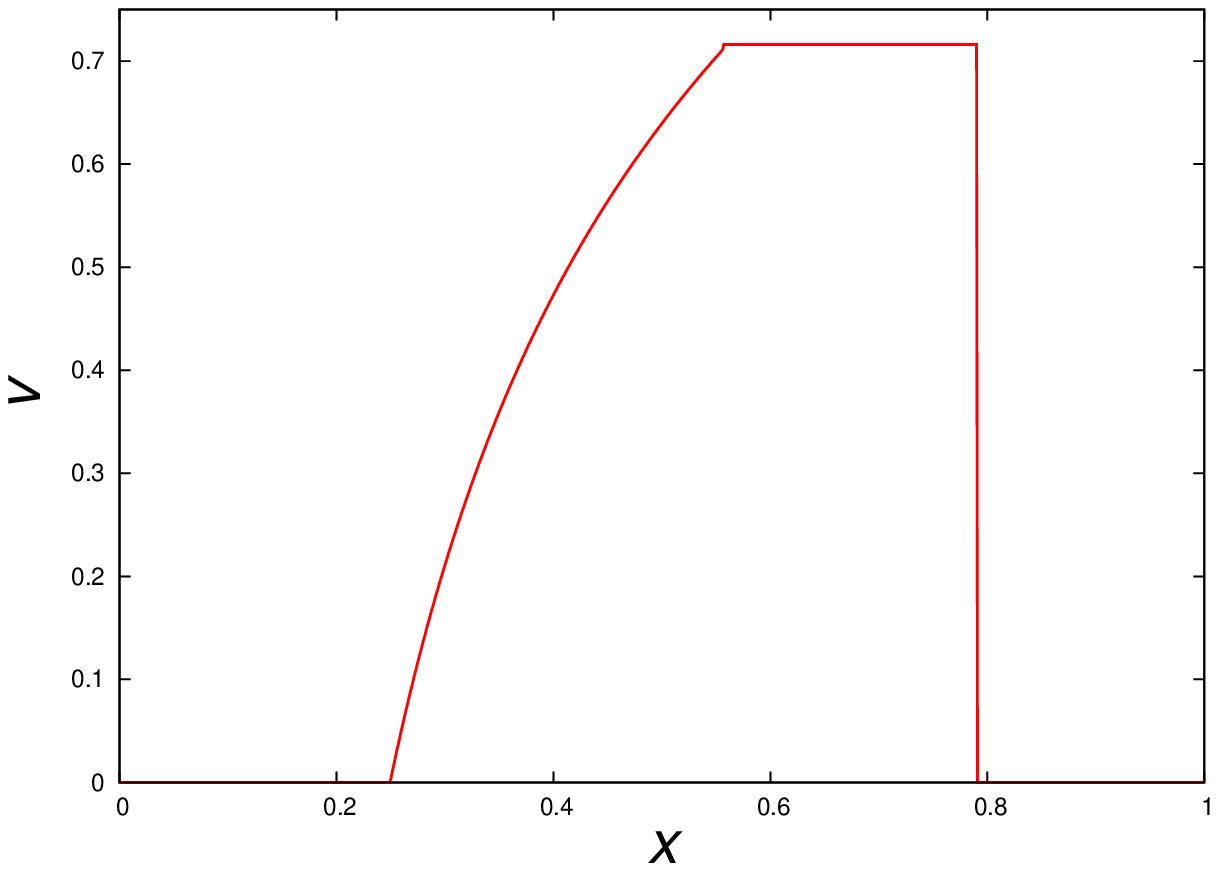}
\includegraphics[width=4cm]{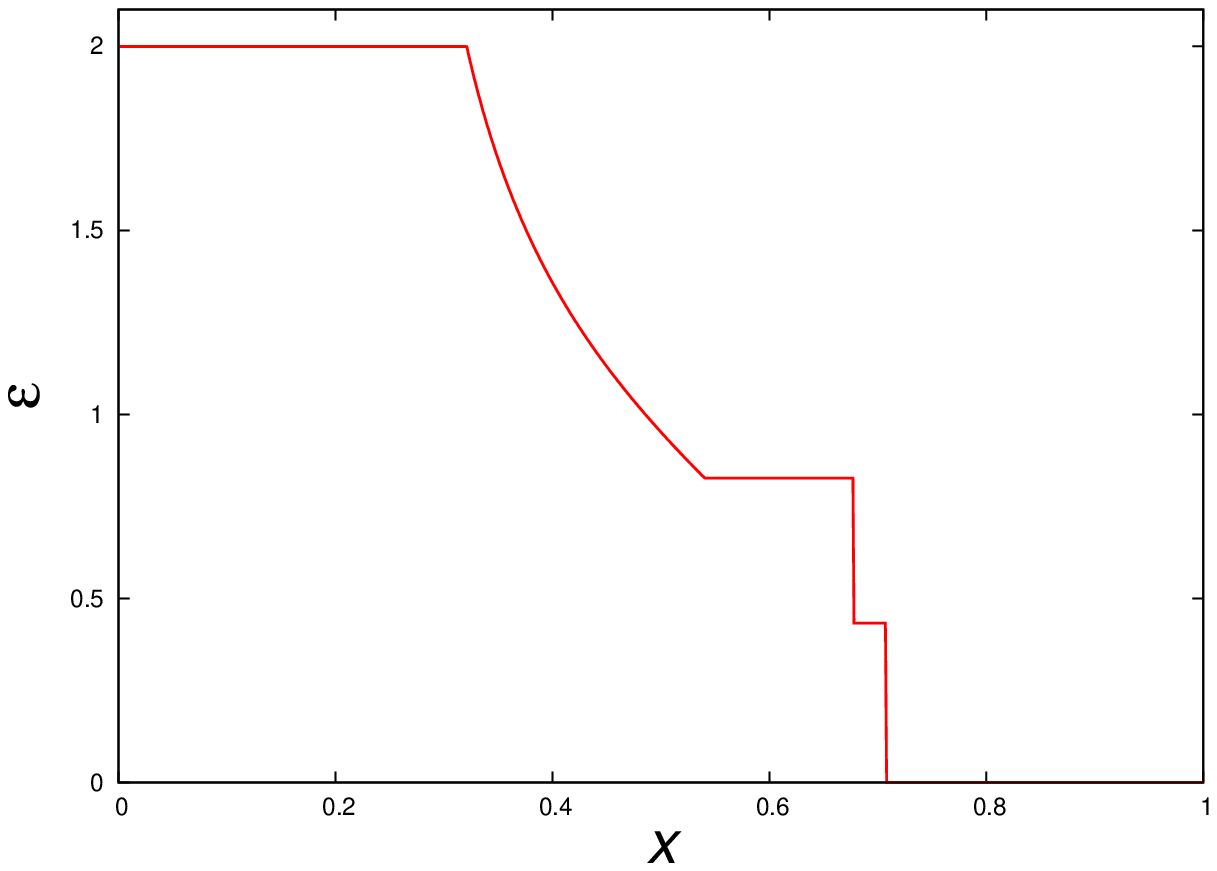}
\caption{\label{fig:Relativistic_RS} Exact solution for the Rarefaction-Shock case at time $t=0.35$ for the parameters in Table \ref{tab:relativistic}.}
\end{figure}

As an example we show in Fig. \ref{fig:Relativistic_RS} the primitive variables at $t=0.35$, for the initial parameters in Table \ref{tab:relativistic}.

\subsubsection{Case 2: Shock-Rarefaction}
\label{subsec:SR}

This is pretty much the previous case, except that one has to be careful at using the correct signs and conditions. We then start again with the contact wave conditions $v_3=v_4=v^*$ and $p_3=p_4=p^*$. The velocity of the gas in region 3 corresponds to the velocity on the state at the right from a shock moving to the left (\ref{eq:vRight})

\begin{equation}
v_3= \frac{h_1 W_1 v_1 + \frac{W_{s,2}}{j} (p_3-p_1)}{h_1W_1+(p_3-p_1)\left( \frac{W_{s,2} v_1}{j}+ \frac{1}{\rho_1 W_1}\right)},\label{eq:SR_vel3}
\end{equation}

\noindent where $W_{s,2}=1/\sqrt{1-V_{s,2}^2}$ is the Lorentz factor of the shock. In order to obtain $v_3$ in terms of $p_3$ and other known information we need to perform the following steps:

\begin{itemize}

\item The rest mass density is given in terms of $p_3$ using the expression (\ref{eq:densityR}) as
\begin{widetext}
\begin{eqnarray}
\frac{1}{\rho_3} &=& \frac{-[p_3 (2\sigma -1) + p_1] + \sqrt{[p_3 (2\sigma -1) + p_1]^2 + 4\zeta_{3}\sigma [p^2_3 (\sigma-1)+p_3 p_1]}}{2\sigma[p^2_3 (\sigma-1)+p_3 p_1]}, \label{eq:rho3_sr}\\
\zeta_{3}&=&\frac{1}{\rho_1}[p_1 (2\sigma -1) + p_3] + \frac{\sigma}{\rho^2_1} [p^2_1 (\sigma-1)+p_3 p_1],\qquad \text{where $\sigma=\frac{\Gamma}{\Gamma-1}$.}
\end{eqnarray}
\end{widetext}

\item Once $\rho_3$ is given in terms of $p_3$ it is possible to compute the enthalpy in region 3  as $h_3=1+\sigma \frac{p_3}{\rho_3}$. 

\item Then from equation (\ref{eq:Taubs1}) we obtain 

\begin{equation}\label{eq:j_rs}
j^2=-\frac{(p_3-p_1)}{\frac{h_3}{\rho_3}-\frac{h_1}{\rho_1}},
\end{equation}

\noindent where $h_1=1+\sigma \frac{p_1}{\rho_1}$. As the shock is moving to the left we consider the negative root of the above expression for $j$.

\item Once $j$ is obtained, the shock velocity can be found from expression (\ref{eq:ShockVelL}) in terms of $p_3$ as

\begin{equation}
\label{eq:vs_s_r}
V_{s,2}=\frac{\rho^2_1 W^2_1 v_1- |j| \sqrt{j^2+\rho^2_1}}{j^2+ \rho^2_1 W^2_1}.
\end{equation}

\item Finally one calculates $W_{s,2}=\frac{1}{\sqrt{1-V^2_{s,2}}}$ and in this way $v_3$ in terms of $p_3$ and the known state in region 1 using (\ref{eq:SR_vel3}).
\end{itemize}

The velocity in region 4 is given by equation (\ref{eq:rar_vel_L}) that provides the velocity on the state at the left from a rarefaction wave moving to the right:

\begin{equation}
v_4= \frac{(1+v_6)A^{-}_6-(1-v_6)A^{-}_4}{(1+v_6)A^{-}_6+(1-v_6)A^{-}_4},\label{eq:SR_vel4}
\end{equation}

\noindent where following (\ref{eq:fan1_r_A})

\begin{equation}
A^{-}_{(4,6)}=\left[ \frac{\sqrt{\Gamma-1}+c_{s,(4,6)}^{-}}{\sqrt{\Gamma-1}-c_{s,(4,6)}^{-}} \right]^{- 2 (\Gamma - 1)^{-1/2}}.
\end{equation}

\noindent Here $c^{-}_{s,6}:=c_s(p_6)=\sqrt{\Gamma p_6/(\rho_6 h_6)}$, $h_6=1+\frac{p_6 \Gamma}{\rho_6(\Gamma-1)}$ and $c^{-}_{s,4}:=c_s(p_4)$ is given by equation (\ref{eq:cs_square_p})

\begin{equation}
c^{-}_{s,4}(p_4)= \sqrt{\frac{\Gamma-1}{\frac{\Gamma-1}{K \Gamma} \left( \frac{p_4}{K}\right)^{\frac{1-\Gamma}{\Gamma}}+1}},\,\, ~~K=\frac{p_6}{\rho^\Gamma_6}.
\end{equation}

\noindent because $K$ is the same in regions 4 and 6.

We obtain a transcendental equation for $p^*$ using the contact discontinuity condition $v_3=v_4=v^*$, and equate (\ref{eq:SR_vel3}) and (\ref{eq:SR_vel4}):

\begin{eqnarray}\label{eq:newrap_rs}
\frac{(1+v_6)A^{-}_6-(1-v_6)A^{-}_4(p^*)}{(1+v_6)A^{-}_6+(1-v_6)A^{-}_4(p^*)} &-& \nonumber \\
\frac{h_1 W_1 v_1 + \frac{W_s}{j} (p^*-p_1)}{h_1W_1+(p^*-p_1)\left( \frac{W_s v_1}{j}+ \frac{1}{\rho_1 W_1}\right)}&=&0,
\end{eqnarray}

\noindent which has to be solved using a root finder.

Once this equation is solved, $p_3$ and $p_4$ are automatically known and $v_3$ and $v_4$ can be calculated using (\ref{eq:SR_vel3}) and (\ref{eq:SR_vel4}), respectively. It is possible to calculate $\rho_4$ using the fact that in the rarefaction zone the process is adiabatic and then $\rho_4=\rho_6(p_4/p_6)^{1/\Gamma}$. We can also  calculate $\rho_3$ using (\ref{eq:rho3_sr}). With this information it is already possible to construct the solution in the whole domain. 

Up to this point we have the known initial states $(p_1,v_1,\rho_1)$ and $(p_6,v_6,\rho_6)$, the solution in regions 3 and 4 given by $(p_3,v_3,\rho_3)$ and $(p_4,v_4,\rho_4)$, and $V_{s,2}$ which represents the velocity of propagation of the shock 2. The exact solution region by region is described next.

\begin{enumerate}

\item Region 1 is defined by the condition $x-x_0 < t V_{s,2}$, where $V_{s,2}$ is given by (\ref{eq:vs_s_r}) and the solution there is that of the initial state on the left chamber

\begin{eqnarray}
p_{exact} &=& p_1, \noindent\\
v_{exact} &=& v_1, \noindent\\
\rho_{exact} &=& \rho_1. \noindent
\end{eqnarray}

\item There is no region 2. Only the shock traveling with speed $V_{s,2}$.

\item Region 3 is defined by the condition $t V_{s,2} < x-x_0 < t V_{contact}$, where $V_{contact}=\lambda_o=v_3=v_4$. The solution is

\begin{eqnarray}
p_{exact} &=& p_3, \noindent\\
v_{exact} &=& v_3, \noindent\\
\rho_{exact} &=& \rho_3. \noindent
\end{eqnarray}

\item Region 4 is defined by the condition $tV_{contact} < x-x_0 <t \xi_{t}$, where according to (\ref{eq:tailxi}) $\xi_t$ is the characteristic value again, but this time evaluated at the tail of the rarefaction wave, that is $\xi_{t}=\frac{v_4+c_{s,4}}{1 + v_4 c_{s,4}}$. The solution in this region is

\begin{eqnarray}
p_{exact} &=& p_4, \noindent\\
v_{exact} &=& v_4, \noindent\\
\rho_{exact} &=& \rho_4. \noindent
\end{eqnarray}

\item Region 5 is defined by the condition $t \xi_t < x-x_0 < t \xi_h$, where according to (\ref{eq:headxi}) $\xi_{h}$ is the velocity of the head of the rarefaction wave traveling to the right $\xi_h=\frac{v_6 + c_{s,6}}{1 + v_6 c_{s,6}}$. In order to compute $v_5$ we use (\ref{eq:fan2_r})

\begin{equation}
\frac{1+v_6}{1-v_6} A_6^{-}- \frac{1+v_5}{1-v_5} A_5^{-}(v_5)=0,	\label{eq:SR_vel2}
\end{equation}

\noindent  whoch requires the information in (\ref{eq:cs_rel_classic}), (\ref{eq:fan1_r_A}) and (\ref{eq:fan_2}):

\begin{eqnarray}
A^{-}_{(5,6)} &=& \left[ \frac{\sqrt{\Gamma-1}+c_{s,(5,6)}^{-}}{\sqrt{\Gamma-1}-c_{s,(5,6)}^{-}} \right]^{- 2 (\Gamma - 1)^{-1/2}}, \\
c^{-}_{s,6} &=& \sqrt{\frac{\Gamma p_6}{ \rho_6 h_6}}, \, \, h_6=1+\frac{p_6}{\rho_6}\left( \frac{\Gamma}{\Gamma -1}\right)\\
c^{-}_{s,5} &=& \frac{v_{5}-\xi}{1-v_{5} \xi} ~~\Rightarrow~~ v_5 = \frac{\xi - c^{-}_{s,5}}{1- c^{-}_{s,5} \xi}. \label{eq:cs_square_5}
\end{eqnarray}

\noindent where $\xi=(x-x_0)/t$.  In this way, equation (\ref{eq:SR_vel2}) is transcendental and has to be solved equivalently for $v_5$ or for $c^{-}_{s,5}$ using a root finder for each point of region 5. We recommend solving for $c^{-}_{s,5}$ and then construct $v_5$ using (\ref{eq:cs_square_5}). Finally we calculate $\rho_5$ using equation (\ref{eq:rho_rel_rar}):

\begin{equation}
\rho_5=\frac{1}{\left[ K \Gamma  \left( \frac{1}{(c^{-}_{s,5})^2}- \frac{1}{\Gamma-1} \right) \right]^{\frac{1}{\Gamma-1}}},\, \, ~~K= \frac{p_6}{\rho_6^\Gamma},
\end{equation}

\noindent since $K$ is the same in regions 5 and 6, and by the same reason we obtain $p_5$ using

\begin{equation}
p_5=p_6\left( \frac{\rho_5}{\rho_6} \right)^\Gamma.
\end{equation}

\item Region 6 is defined by $t \xi_{h} < x-x_0$. In this region the solution is simply

\begin{eqnarray}
p_{exact} &=& p_6, \noindent\\
v_{exact} &=& v_6, \noindent\\
\rho_{exact} &=& \rho_6. \noindent
\end{eqnarray}

\end{enumerate}

\begin{figure}[htp]
\includegraphics[width=4cm]{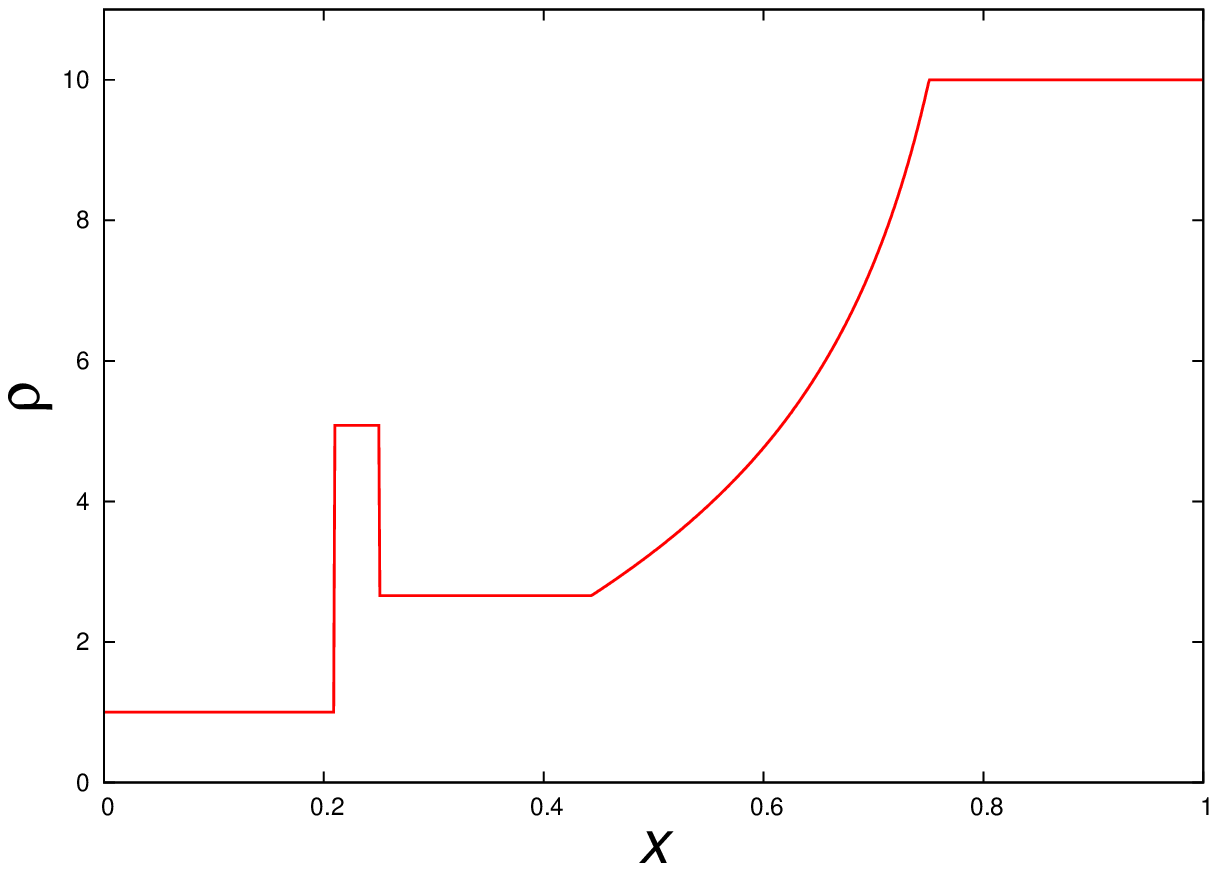}
\includegraphics[width=4cm]{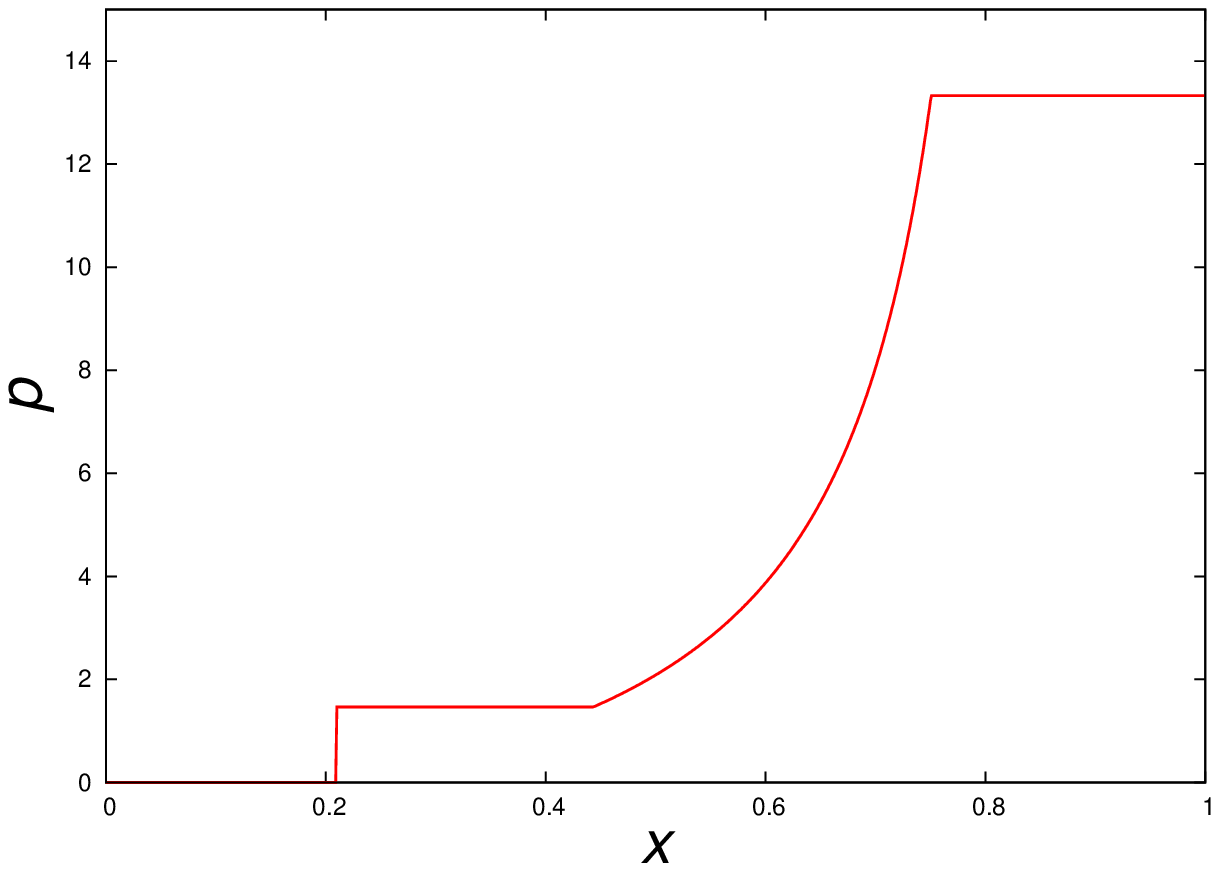}
\includegraphics[width=4cm]{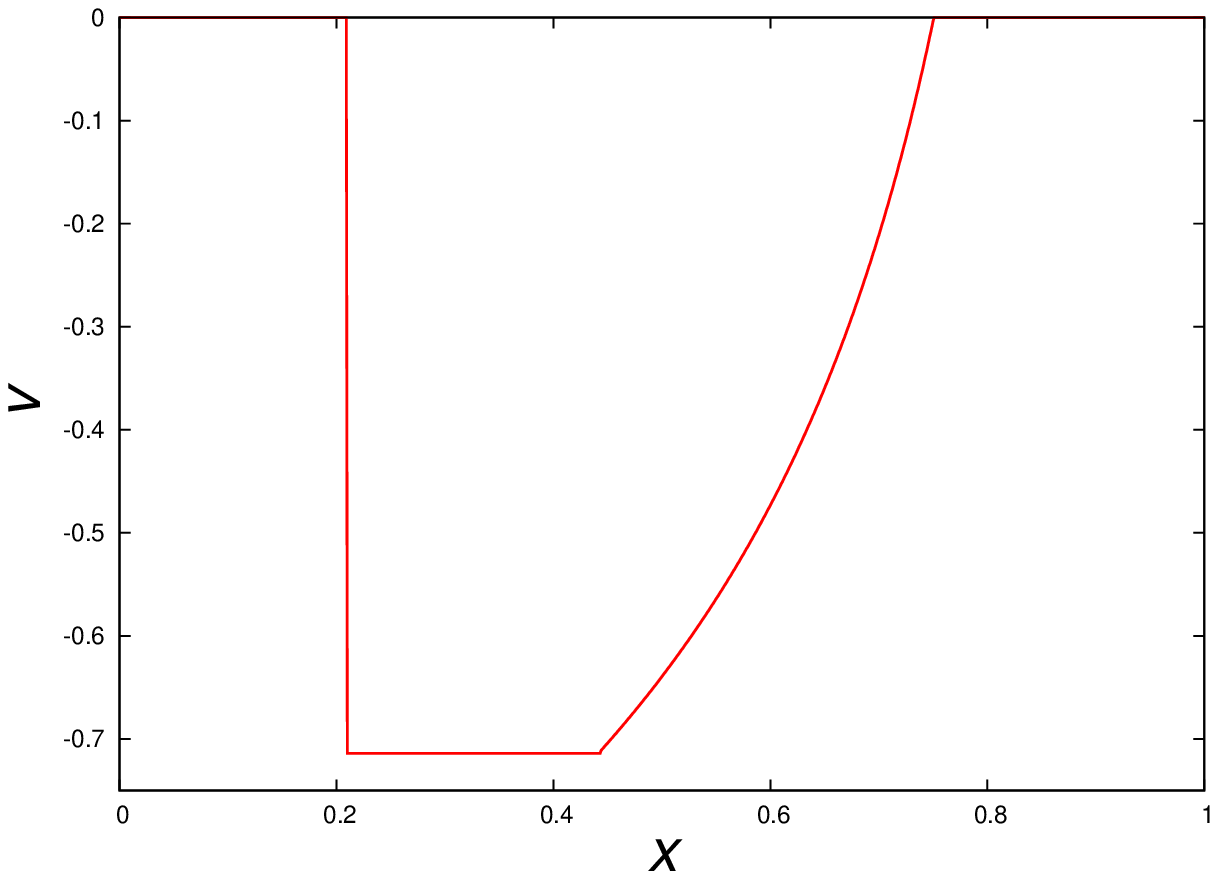}
\includegraphics[width=4cm]{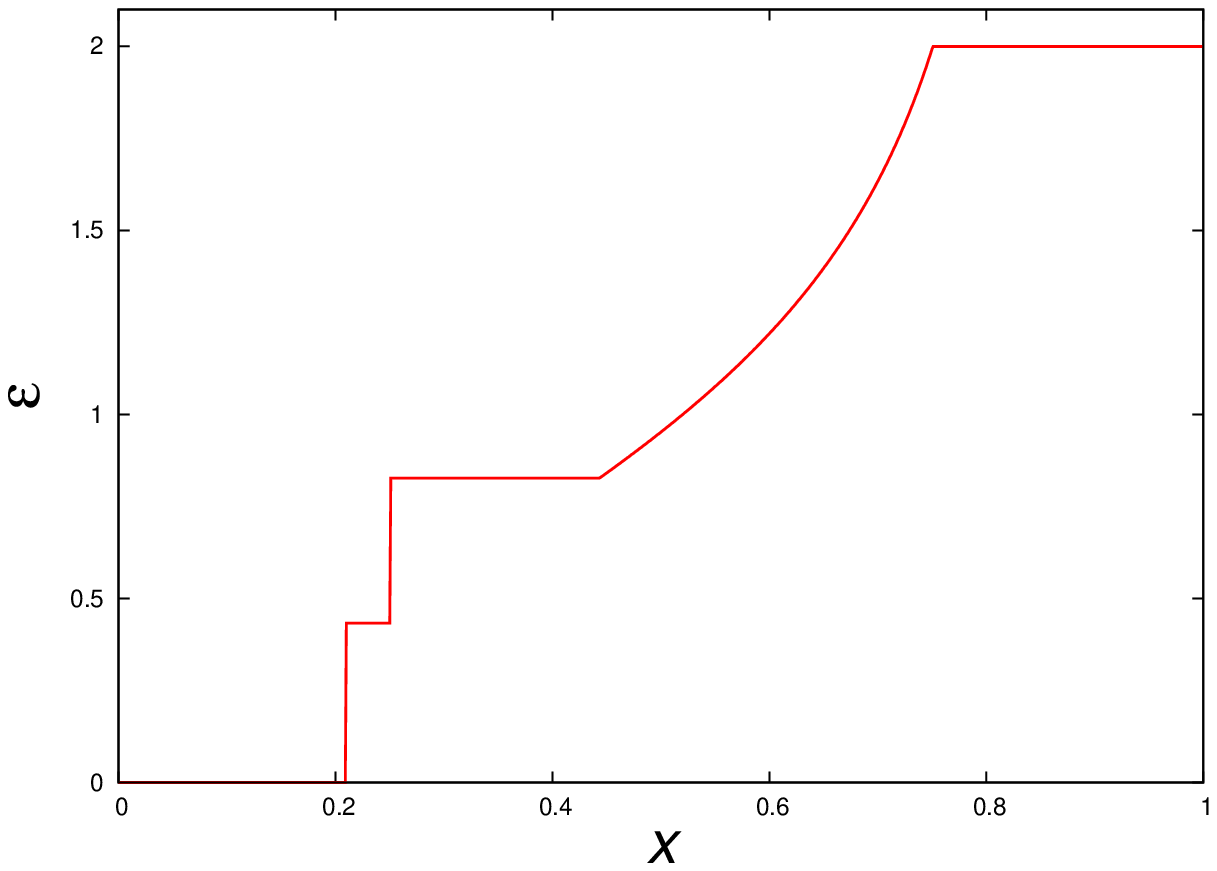}
\caption{\label{fig:Relativistic_SR} Exact solution for the Shock-Rarefaction case at time $t=0.35$ for the parameters in Table \ref{tab:relativistic}.}
\end{figure}

As an example we show in Fig. \ref{fig:Relativistic_SR} the primitive variables at $t=0.35$ for the initial data in Table \ref{tab:relativistic}.

\subsubsection{Case 3: Rarefaction-R¿arefaction}
\label{subsec:RR}

In this case the transcendental equation for the pressure at the contact discontinuity is given again by the condition $v_3=v_4$ where both velocities are constructed using the information of the unknown state aside rarefaction waves. The velocity in region 3 is given by equation (\ref{eq:velR_rarL}) for the velocity on the state at the right from a rarefaction wave moving to the left:

\begin{equation}
v_3= \frac{(1+v_1)A^{+}_1-(1-v_1)A^{+}_3}{(1+v_1)A^{+}_1+(1-v_1)A^{+}_3},\label{eq:RR_v3}
\end{equation}

\noindent where according to (\ref{eq:fan1_r_A})

\begin{equation}
A^{+}_{(1,3)}=\left[ \frac{\sqrt{\Gamma-1}+c_{s,(1,3)}^{+}}{\sqrt{\Gamma-1}-c_{s,(1,3)}^{+}} \right]^{+ 2 (\Gamma - 1)^{-1/2}}.
\end{equation}

\noindent Here $c^{+}_{s,1}:=c_s(p_1)=\sqrt{\Gamma p_1/(\rho_1 h_1)}$, $h_1=1+\frac{p_1 \Gamma}{\rho_1(\Gamma-1)}$
and $c^{+}_{s,3}:=c_s(p_3)$ is given by equation (\ref{eq:cs_square_p})

\begin{equation}
c^+_{s,3}(p_3)= \sqrt{\frac{\Gamma-1}{\frac{\Gamma-1}{K \Gamma} \left( \frac{p_3}{K}\right)^{\frac{1-\Gamma}{\Gamma}}+1}},\,\, ~~K=\frac{p_1}{\rho^\Gamma_1}.
\end{equation}

On the other hand the velocity of the gas in region 4 corresponds to the velocity on the state at the left of a rarefaction wave moving to the right (\ref{eq:rar_vel_L})

\begin{equation}
v_4= \frac{(1+v_6)A^{-}_{6} - (1-v_6)A^{-}_{4}}{(1+v_6)A^{-}_{6} + (1-v_6)A^{-}_{4}}, \label{eq:RR_v4}
\end{equation}

\noindent where according to (\ref{eq:fan1_r_A})

\begin{equation}
A^{-}_{(4,6)}=\left[ \frac{\sqrt{\Gamma-1}+c_{s,(4,6)}^{-}}{\sqrt{\Gamma-1}-c_{s,(4,6)}^{-}} \right]^{- 2 (\Gamma - 1)^{-1/2}},
\end{equation}

\noindent and the speed of sound in region 4 is given by

\begin{equation}
c^{-}_{s,4}(p_4)= \sqrt{\frac{\Gamma-1}{\frac{\Gamma-1}{K \Gamma} \left( \frac{p_4}{K}\right)^{\frac{1-\Gamma}{\Gamma}}+1}},\,\, ~~K=\frac{p_6}{\rho^\Gamma_6}.
\end{equation}

Then using the contact discontinuity condition $v_3=v_4=v^*$, we equate (\ref{eq:RR_v3}) and (\ref{eq:RR_v4}) and obtain a transcendental equation for $p^*$:

\begin{eqnarray}\label{eq:newrap_rs}
\frac{(1+v_1)A^{+}_1-(1-v_1)A^{+}_3(p^*)}{(1+v_1)A^{+}_1+(1-v_1)A^{+}_3(p^*)} &-& \nonumber \\
\frac{(1+v_6)A^{-}_{6} - (1-v_6)A^{-}_{4}(p^*)}{(1+v_6)A^{-}_{6} + (1-v_6)A^{-}_{4}(p^*)}
&=&0,
\end{eqnarray}

\noindent which has to be solved using a root finder.

Once this equation is solved, $p_3$ and $p_4$ are automatically known and $v_3$ and $v_4$ can be calculated using (\ref{eq:RR_v3}) and (\ref{eq:RR_v4}), respectively. As in the previous two cases, it is possible to calculate $\rho_3$ and $\rho_4$ using the fact that in the rarefaction zone the process is adiabatic and then $\rho_3=\rho_1(p_3/p_1)^{1/\Gamma}$ and $\rho_4=\rho_6(p_4/p_6)^{1/\Gamma}$.  Thus we have the known initial states $(p_1,v_1,\rho_1)$, $(p_6,v_6,\rho_6)$  and the solution in regions 3 and 4 given by $(p_3,v_3,\rho_3)$ and $(p_4,v_4,\rho_4)$. The solution in each of the fan regions aside the  rarefaction zones has to be constructed in terms of the position and time $\xi=(x-x_0)/t$ as described below for regions 2 and 5.

\begin{enumerate}

\item Region one is defined by the condition $x-x_0 < t \xi_{h2}$, where according to (\ref{eq:headxi}) $\xi_{h2}$ is the velocity of the head of the rarefaction wave traveling to the left $\xi_{h2}=\frac{v_1 - c_{s,1}}{1 - v_1 c_{s,1}}$. The values of the physical variables are known from the initial conditions:

\begin{eqnarray}
p_{exact} &=& p_1, \noindent\\
v_{exact} &=& v_1, \noindent\\
\rho_{exact} &=& \rho_1. \noindent
\end{eqnarray}

\item Region 2 is defined by the condition $t \xi_{h2} < x-x_0 < t \xi_{t2}$, where according to (\ref{eq:tailxi}) $\xi_{t2}$ is the characteristic value again, but this time evaluated at the tail of the rarefaction wave, that is $\xi_{t2}=\frac{v_3-c_{s,3}}{1 - v_3 c_{s,3}}$.  In order to compute $v_2$ we use (\ref{eq:fan1_r})

\begin{equation}\label{eq:forv2}
\frac{1+v_1}{1-v_1} A_1^{+}- \frac{1+v_2}{1-v_2} A_2^{+}(v_2)=0,	
\end{equation}

\noindent where using (\ref{eq:cs_rel_classic}), (\ref{eq:fan1_r_A}) and (\ref{eq:fan_2})

\begin{eqnarray}
A^{+}_{(1,2)} &=& \left[ \frac{\sqrt{\Gamma-1}+c_{s,(1,2)}^{+}}{\sqrt{\Gamma-1}-c_{s,(1,2)}^{+}} \right]^{+ 2 (\Gamma - 1)^{-1/2}}, \\
c^+_{s,1} &=& \sqrt{\frac{\Gamma p_1}{ \rho_1 h_1}}, \, \, h_1=1+\frac{p_1}{\rho_1}\left( \frac{\Gamma}{\Gamma -1}\right)\\
c^+_{s,2} &=& \frac{v_{2}-\xi}{1-v_{2} \xi} ~~\Rightarrow~~ v_2 = \frac{\xi + c^+_{s,2}}{1+ c^+_{s,2} \xi}. \label{eq:cs_square_2}
\end{eqnarray}

\noindent where $\xi=(x-x_0)/t$.  In this way, equation (\ref{eq:forv2}) is transcendental and has to be solved equivalently for $v_2$ or for $c^+_{s,2}$ using a root finder for each point of region 2. We solve  for $c^+_{s,2}$ and construct $v_2$ using (\ref{eq:cs_square_2}). Finally we calculate $\rho_2$ using equation (\ref{eq:rho_rel_rar}):

\begin{equation}
\rho_2=\frac{1}{\left[ K \Gamma  \left( \frac{1}{(c^+_{s,2})^2}- \frac{1}{\Gamma-1} \right) \right]^{\frac{1}{\Gamma-1}}},\, \, ~~K= \frac{p_1}{\rho_1^\Gamma}.
\end{equation}

\noindent Finally we obtain $p_2$ using
\begin{equation}
p_2=p_1\left( \frac{\rho_2}{\rho_1} \right)^\Gamma.
\end{equation}

\item Region 3 is defined by the condition $t \xi_{t2} < x-x_0 < t V_{contact}$, where $V_{contact}=\lambda_o=v_3=v_4$. The solution there reads

\begin{eqnarray}
p_{exact} &=& p_3, \noindent\\
v_{exact} &=& v_3, \noindent\\
\rho_{exact} &=& \rho_3. \noindent
\end{eqnarray}

\item Region 4 is defined by the condition $t V_{contact}< x-x_0 < t \xi_{t5}$, where $\xi_{t5}$ is the third characteristic value calculated at the tail of rarefaction moving to the right, and according to (\ref{eq:tailxi}) $\xi_{t5}=\frac{v_4+c_{s,4}}{1 + v_4 c_{s,4}}$. In this region thus

\begin{eqnarray}
p_{exact} &=& p_4, \noindent\\
v_{exact} &=& v_4, \noindent\\
\rho_{exact} &=& \rho_4. \noindent
\end{eqnarray}

\item Region 5 is defined by the condition $t \xi_{t5} < x - x_0 < t \xi_{h5}$, where $\xi_{h5}=\frac{v_6 + c_{s,6}}{1 + v_6 c_{s,6}}$ according to (\ref{eq:headxi}). In order to compute $v_5$ we use (\ref{eq:fan2_r})

\begin{equation}
\frac{1+v_5}{1-v_5} A_5^{-}(v_5)- \frac{1+v_6}{1-v_6} A_6^{-}	=0,\label{eq:RR_vel5}
\end{equation}

\noindent where according to (\ref{eq:cs_rel_classic}), (\ref{eq:fan1_r_A}) and (\ref{eq:fan_2}) 

\begin{eqnarray}
A^{-}_{(5,6)}&=&\left[ \frac{\sqrt{\Gamma-1}+c_{s,(5,6)}^{-}}{\sqrt{\Gamma-1}-c_{s,(5,6)}^{-}} \right]^{- 2 (\Gamma - 1)^{-1/2}},\\
c^{-}_{s,6} &=& \sqrt{\frac{\Gamma p_6}{\rho_6 h_6}},~~ h_6 = 1 + \frac{p_6}{\rho_6} \left( \frac{\Gamma}{\Gamma-1} \right),\\
c_{s,5}^{-}&=& - \frac{v_{5}-\xi}{1-v_{5} \xi} ~~ \Rightarrow ~~ v_5 = \frac{\xi - c^{-}_{s,5}}{1-c^{-}_{s,5}\xi},\label{eq:c5min_RR}
\end{eqnarray}

\noindent where $\xi=(x-x_0)/t$. Again (\ref{eq:RR_vel5}) is a transcendental equation either for $v_5$ or for $c_{s,5}^{-}$. Once $c^{-}_{s,5}$ has been calculated use (\ref{eq:c5min_RR}) to construct $v_5$ or directly solve (\ref{eq:RR_vel5}) for $v_5$. It is possible to calculate $\rho_5$ using (\ref{eq:rho_rel_rar}):

\begin{equation}
\rho_5=\frac{1}{\left[ K \Gamma  \left( \frac{1}{(c^{-}_{s,5})^2}- \frac{1}{\Gamma-1} \right) \right]^{\frac{1}{\Gamma-1}}},\, \, ~~K= \frac{p_6}{\rho_6^\Gamma},
\end{equation}

\noindent and finally the pressure

\begin{equation}
p_5=p_6\left( \frac{\rho_5}{\rho_6} \right)^\Gamma.
\end{equation}

\item Region 6 is defined by $t \xi_{h5} < x-x_0$. In this region the solution is simply

\begin{eqnarray}
p_{exact} &=& p_6, \noindent\\
v_{exact} &=& v_6, \noindent\\
\rho_{exact} &=& \rho_6. \noindent
\end{eqnarray}

\end{enumerate}

\begin{figure}[htp]
\includegraphics[width=4cm]{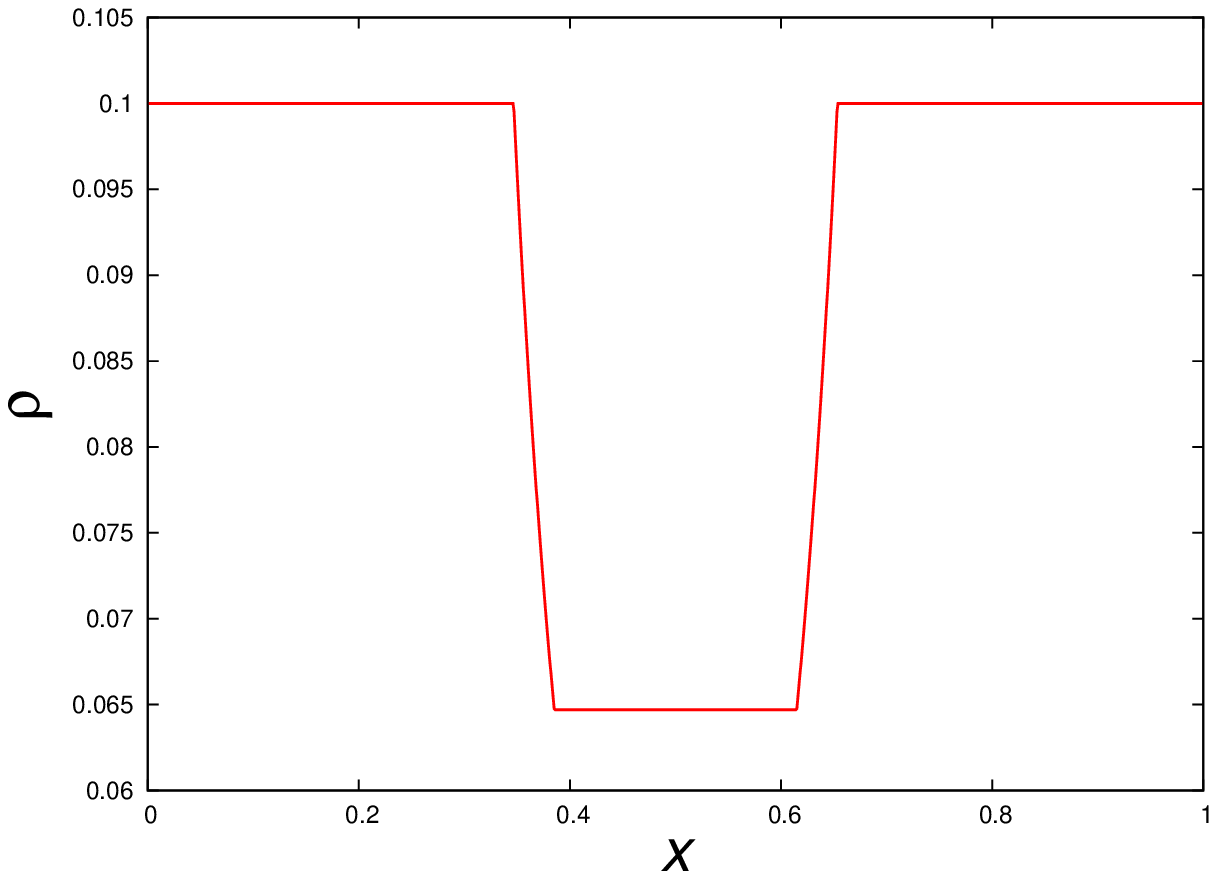}
\includegraphics[width=4cm]{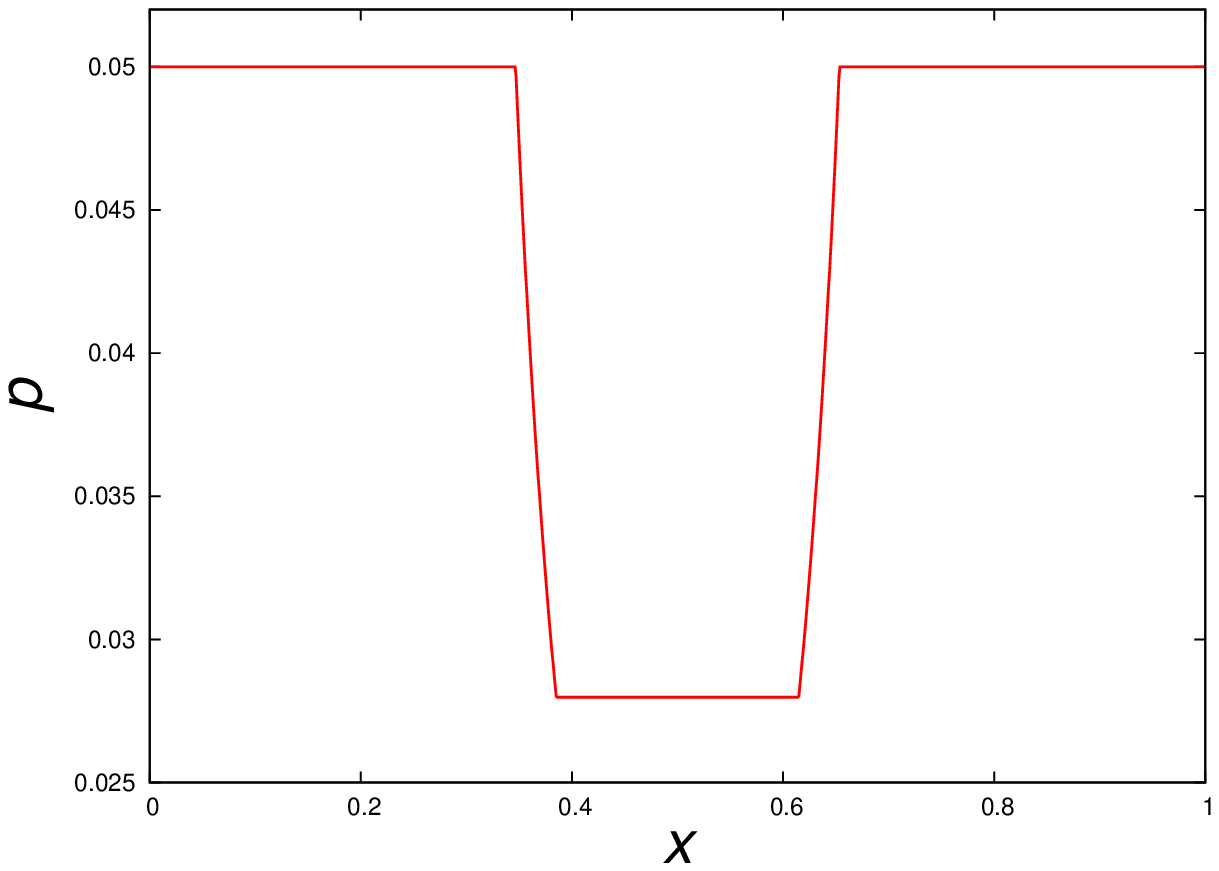}
\includegraphics[width=4cm]{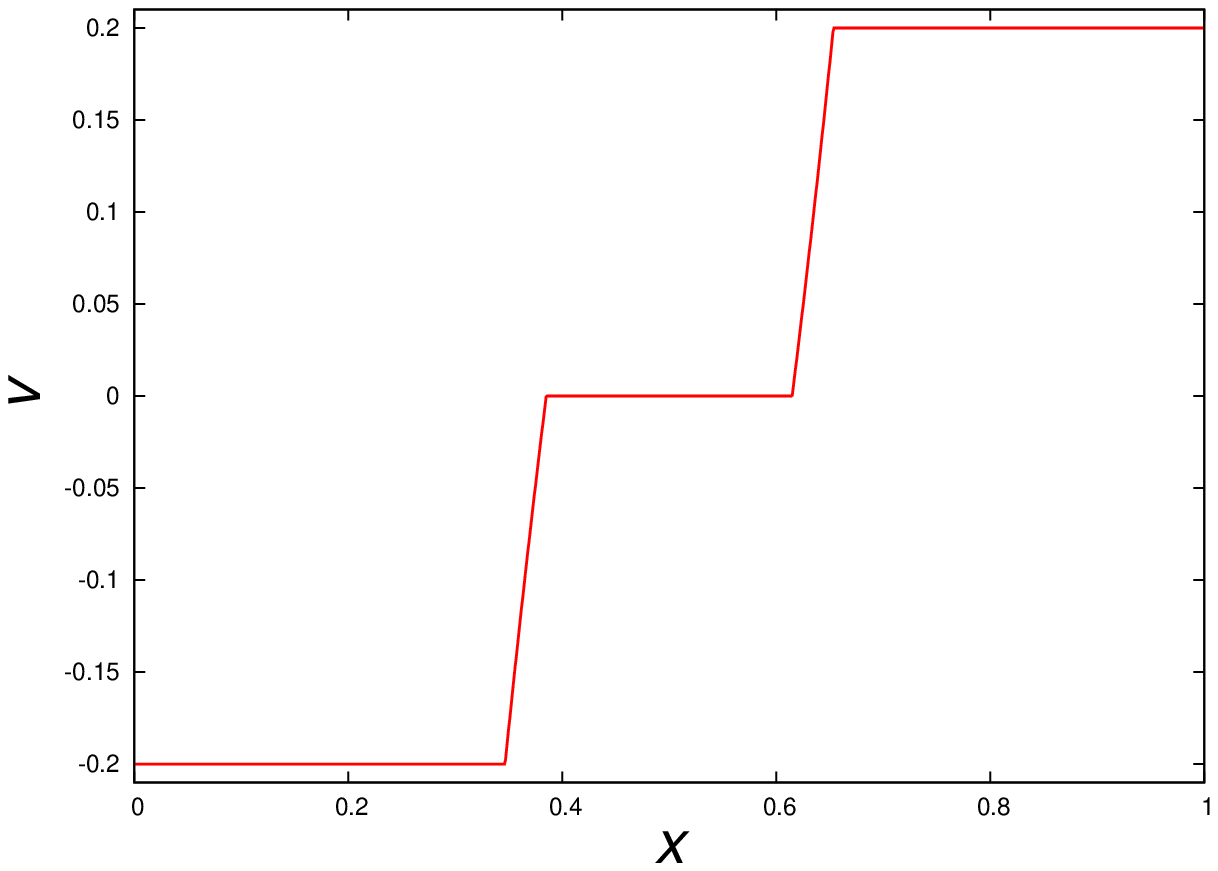}
\includegraphics[width=4cm]{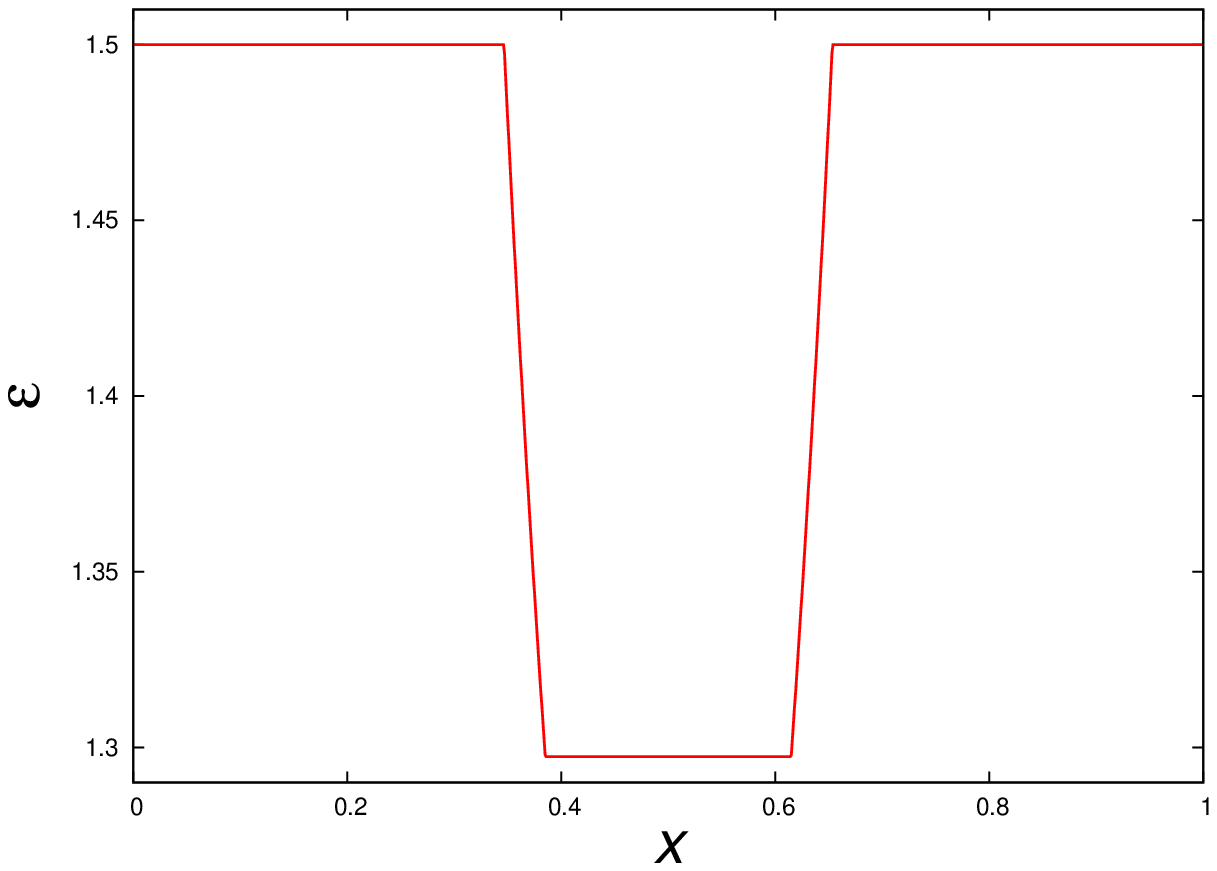}
\caption{\label{fig:Relativistic_RR} Exact solution for the Rarefaction-Rarefaction case at time $t=0.25$ for the parameters in Table \ref{tab:relativistic}.}
\end{figure}

As an example we show in Fig. \ref{fig:Relativistic_RR} the primitive variables at $t=0.25$, for the initial parameters in Table \ref{tab:relativistic}.

\subsubsection{Shock-Shock}

We proceed as always, by establishing a relationship between the velocity in regions 3 and 4. We start by expressing $v_3$ as the velocity of the gas on a region at the right from a shock moving to the left, that is, according to (\ref{eq:vRight})

\begin{equation}\label{eq:ss-v3}
v_3= \frac{h_1 W_1 v_1 + \frac{W_{s,2}}{j_2} (p_3-p_1)}{h_1W_1+(p_3-p_1)\left( \frac{W_{s,2} v_1}{j_2}+ \frac{1}{\rho_1W_1}\right)},
\end{equation}

\noindent where $W_{s,2}=1/\sqrt{1-V_{s,2}^2}$ is the Lorentz factor of the shock moving to the left. In this particular case we distinguish between the two values of $j$ depending using the subindices 2 and 5. In order to obtain $v_3$ in terms of $p_3$ we can proceed following these steps:

\begin{itemize}

\item The rest mass density is given in terms of $p_3$ using the expression (\ref{eq:densityR}) as

\begin{widetext}
\begin{eqnarray}
\frac{1}{\rho_3} &=& \frac{-[p_3 (2\sigma -1) + p_1] +\sqrt{[p_3 (2\sigma -1) + p_1]^2 + 4\zeta_{3}\sigma [p^2_3(\sigma-1)+p_3 p_1]}}{2\sigma[p^2_3 (\sigma-1)+p_3 p_1]}, 
\label{eq:rho3_ss}\\
\zeta_{3}&=&\frac{1}{\rho_1}[p_1 (2\sigma -1) + p_3] + \frac{\sigma}{\rho^2_1} [p^2_1 (\sigma-1)+p_3 p_1],\qquad \text{where $\sigma=\frac{\Gamma}{\Gamma-1}$.}
\end{eqnarray}
\end{widetext}

\item Once $\rho_3$ is given in terms of $p_3$ it is possible to compute enthalpy in region 3 as $h_3=1+\sigma \frac{p_3}{\rho_3}$. 

\item Then from equation (\ref{eq:Taubs1}) we obtain 

\begin{equation}\label{eq:j_rs}
j^{2}_{2}=-\frac{(p_3-p_1)}{\frac{h_3}{\rho_3}-\frac{h_1}{\rho_1}},
\end{equation}

\noindent where we choose $j_2$ to be the negative root since the shock is moving to the left; here $h_1=1+\sigma \frac{p_1}{\rho_1}$.

\item Once $j_2$ is obtained, the shock velocity can be found from expression (\ref{eq:ShockVelL}) in terms of $p_3$ as

\begin{equation}\label{eq:vs_l_s}
V_{s,2}=\frac{\rho^2_1 W^2_1 v_1- |j_2| \sqrt{j^{2}_{2}+\rho^2_1}}{j^{2}_{2}+ \rho^2_1 W^2_1}.
\end{equation}

\item Finally we calculate $W_{s,2}=\frac{1}{\sqrt{1-V^2_{s,2}}}$ and thus $v_3$ in terms of $p_3$ and the known state in region 1 using (\ref{eq:ss-v3}).

\end{itemize}


Using the information of the shock moving to the right we obtain the velocity at the left from the shock, that is $v_4$ using (\ref{eq:vleft})

\begin{equation}\label{eq:ss-v4}
v_4= \frac{h_6 W_6 v_6 + \frac{W_{s,5}}{j_5} (p_4-p_6)}{h_6W_6+(p_4-p_6)\left( \frac{W_{s,5} v_6}{j_5}+ \frac{1}{\rho_6 W_6}\right)},
\end{equation}

\noindent where $W_{s,5}=1/\sqrt{1-V_{s,5}^2}$ is the Lorentz factor of the shock. In order to obtain $v_4$ in terms of $p_4$ we need to perform the following steps:

\begin{itemize}

\item The rest mass density is given in terms of $p_4$ using the expression (\ref{eq:densityL}) as

\begin{widetext}
\begin{eqnarray}
\frac{1}{\rho_4} &=& \frac{-[p_4 (2\sigma -1) + p_6] + \sqrt{[p_4 (2\sigma -1) + p_6]^2 + 4\zeta_{4}\sigma [p^2_4 (\sigma-1)+p_4 p_6]}}{2\sigma[p^2_4 (\sigma-1)+p_4 p_6]}, \label{eq:rho4_ss}\\
\zeta_{4}&=&\frac{1}{\rho_6}[p_6 (2\sigma -1) + p_4] + \frac{\sigma}{\rho^2_6} [p^2_6 (\sigma-1)+p_4 p_6],\qquad \text{where $\sigma=\frac{\Gamma}{\Gamma-1}$.}
\end{eqnarray}
\end{widetext}

\item Once $\rho_4$ is given in terms of $p_4$, we are able to compute enthalpy in region 4 as $h_4=1+\sigma \frac{p_4}{\rho_4}$. 

\item Then equation (\ref{eq:Taubs1}) reads
 
\begin{equation}\label{eq:j_rs}
j^{2}_{5}=-\frac{(p_4-p_6)}{\frac{h_4}{\rho_4}-\frac{h_6}{\rho_6}},
\end{equation}

\noindent here $h_6=1+\sigma \frac{p_6}{\rho_6}$. In this case, since the shock is moving to the right we choose the $j_5$ to be the positive root.

\item Once $j_5$ is obtained, the shock velocity can be found from expression (\ref{eq:ShockVelR}) 

\begin{equation}\label{eq:vs_r_s}
V_{s,5}=\frac{\rho^2_6 W^2_6 v_6+ |j_5| \sqrt{j^{2}_{5}+\rho^2_6}}{j^{2}_{5}+ \rho^2_6 W^2_6}.
\end{equation}

\item Finally we calculate $W_{s,5}=\frac{1}{\sqrt{1-V^2_{s,5}}}$ and in this way we can obtain $v_4$ in terms of $p_4$ with (\ref{eq:ss-v4}) and the known state in region 6.

\end{itemize}

According to the contact discontinuity condition $v_3=v_4=v^*$, we equate (\ref{eq:ss-v3}) and (\ref{eq:ss-v4}) and obtain a transcendental equation for $p^*$:

\begin{eqnarray}\label{eq:newrap_ss}
 \frac{h_1 W_1 v_1 + \frac{W_{s,2}}{j_2} (p^*-p_1)}{h_1W_1+(p^*-p_1)\left( \frac{W_{s,2} v_1}{j_2}+ \frac{1}{\rho_1W_1}\right)}&-& \nonumber \\
\frac{h_6 W_6 v_6 + \frac{W_s}{j_5} (p^*-p_6)}{h_6W_6+(p^*-p_6)\left( \frac{W_s v_6}{j_5}+ \frac{1}{\rho_6 W_6}\right)}&=&0,
\end{eqnarray}

\noindent which has to be solved using a root finder.

Once this equation is solved, $p_3$ and $p_4$ are automatically known,  and $v_3$ and $v_4$ can be calculated using (\ref{eq:ss-v3}) and (\ref{eq:ss-v4}), respectively. It is possible to calculate $\rho_3$ and $\rho_4$ using (\ref{eq:rho3_ss}) and (\ref{eq:rho4_ss}), respectively. With this information it is already possible to construct the solution in the whole domain. 

Up to this point we have the known initial states $(p_1,v_1,\rho_1)$ and $(p_6,v_6,\rho_6)$, the solution in regions 3 and 4 given by $(p_3,v_3,\rho_3)$ and $(p_4,v_4,\rho_4)$, together with $V_{s,2}$ and $V_{s,5}$  which represent the velocities of propagation of the shocks. 

\begin{enumerate}

\item Region 1 is defined by the condition $x-x_0 < t V_{s,2}$, where the velocity of the shock is (\ref{eq:vs_l_s}). The solution there is that of the initial values of the variables on the left chamber:

\begin{eqnarray}
p_{exact} &=& p_1, \nonumber\\
v_{exact} &=& v_1, \nonumber\\
\rho_{exact} &=& \rho_1. \nonumber
\end{eqnarray}

\item There is no region 2, only the shock wave traveling at speed $V_{s,2}$.

\item Region 3 is defined by the condition $t V_{s,2} < x - x_0 < t V_{contact}$, where the velocity of the contact discontinuity is the characteristic value $\lambda^0=v$ evaluated in this region $V_{contact} = v_3 = v_4 = v^{*}$.

\begin{eqnarray}
p_{exact} &=& p_3,\nonumber\\
v_{exact} &=& v_3\nonumber\\
\rho_{exact} &=& \rho_3.\nonumber
\end{eqnarray}

\item Region 4 is defined by $t V_{contact} < x-x_0 < tV_{s,5}$ and the solution is

\begin{eqnarray}
p_{exact} &=& p_4, \nonumber\\
v_{exact} &=& v_4, \nonumber\\
\rho_{exact} &=& \rho_4. \nonumber
\end{eqnarray}
\item There is no region 5, only the shock wave traveling with speed $V_{s,5}$.

\item Finally region 6 is defined by the condition $V_{s,5} < x-x_0$. The exact solution is given by the initial values at the chamber at the right:

\begin{eqnarray}
p_{exact} &=& p_6, \nonumber\\
v_{exact} &=& v_6, \nonumber\\
\rho_{exact} &=& \rho_6. \nonumber
\end{eqnarray}

\end{enumerate}

\begin{figure}[htp]
\includegraphics[width=4cm]{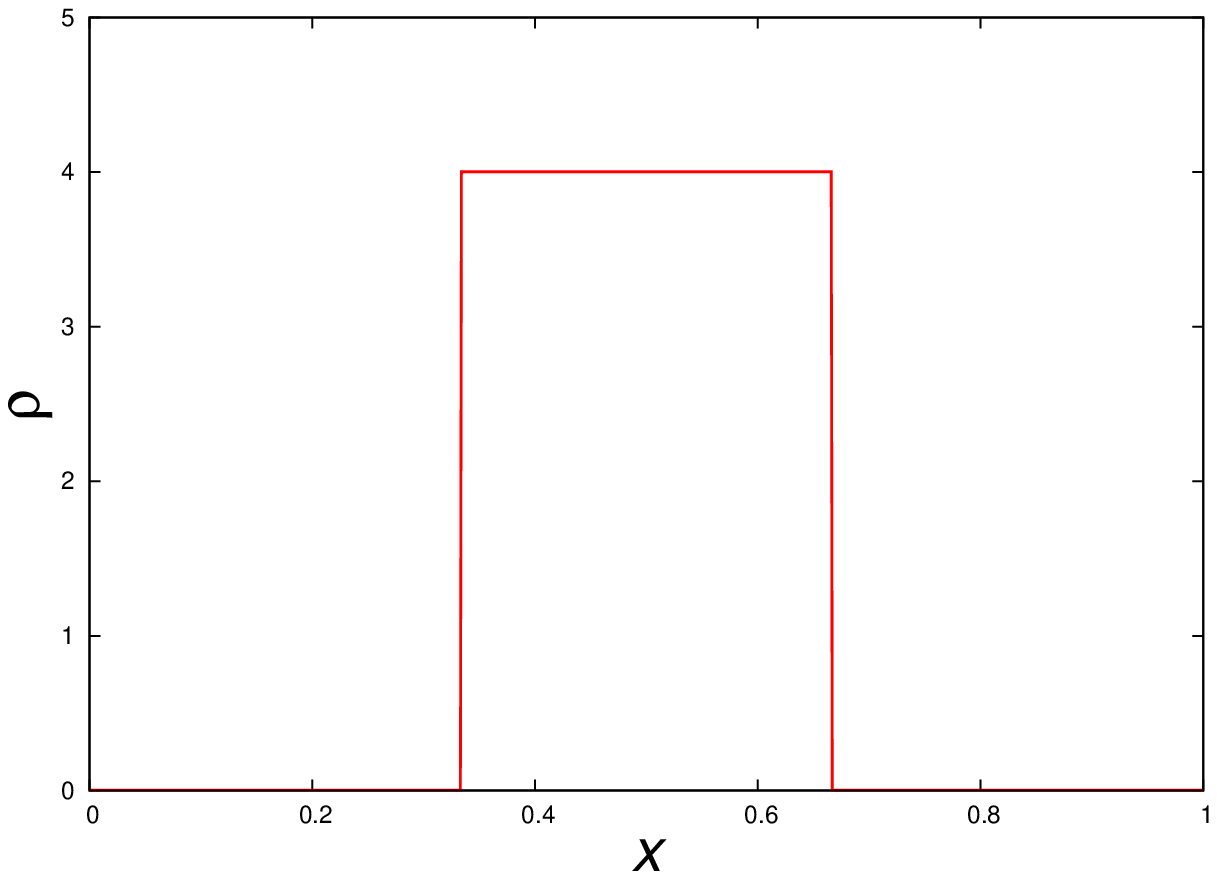}
\includegraphics[width=4cm]{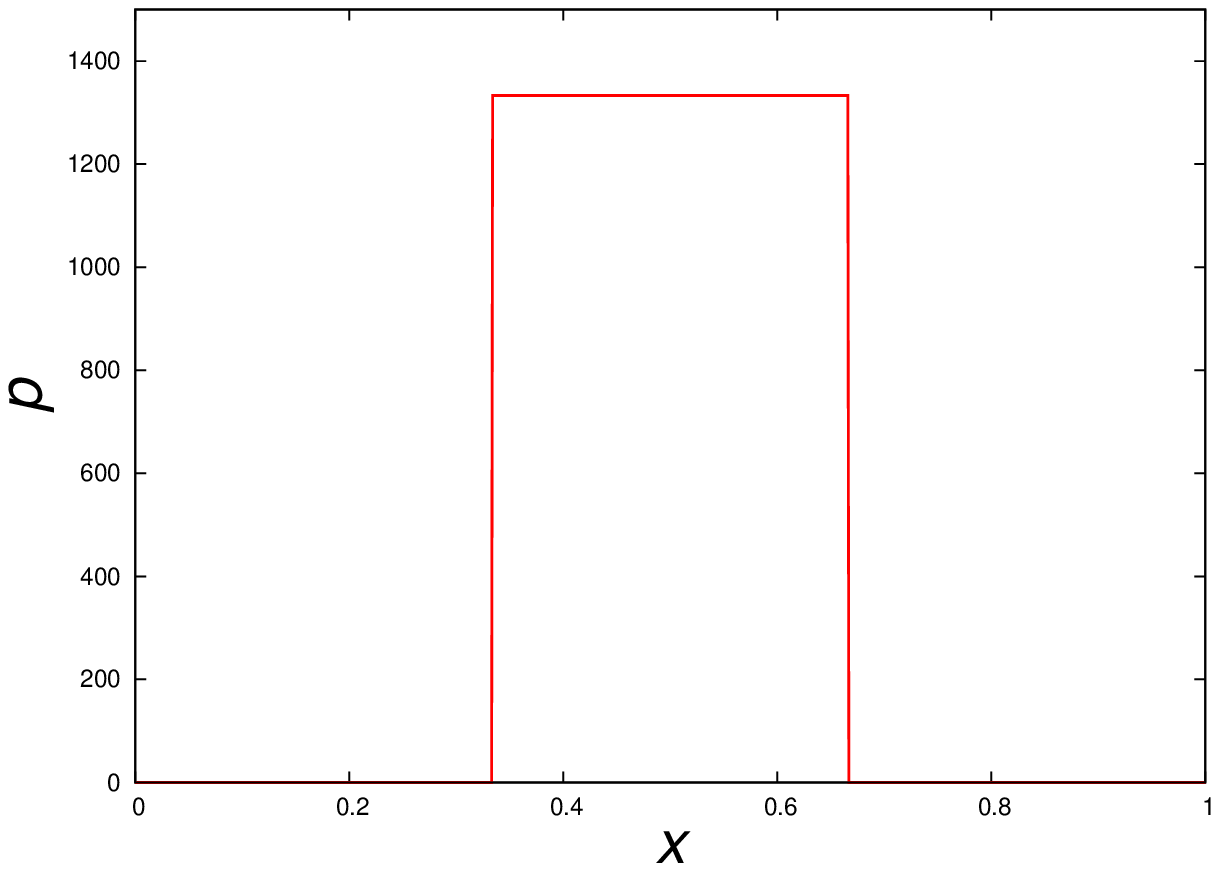}
\includegraphics[width=4cm]{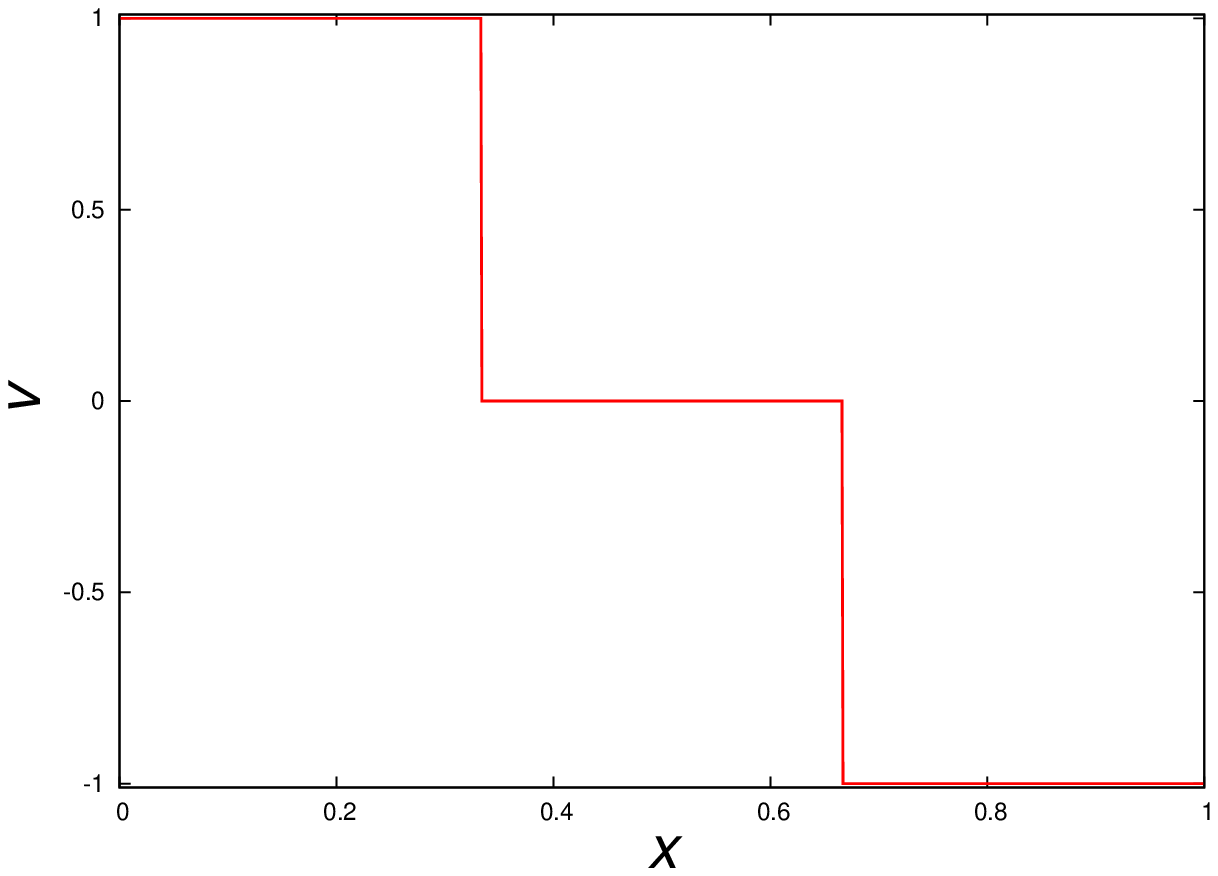}
\includegraphics[width=4cm]{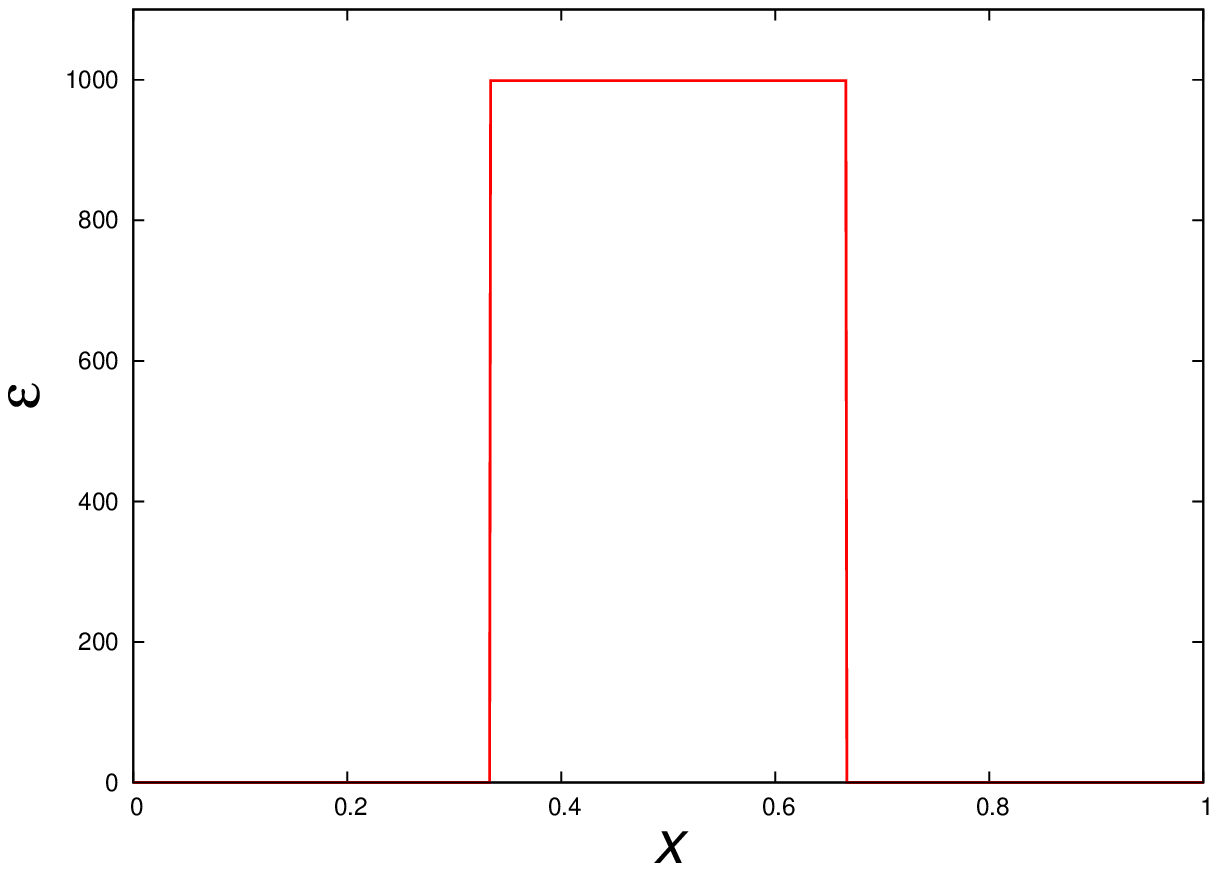}
\caption{\label{fig:Relativistic_SS} Exact solution for a Shock-Shock case at time $t=0.5$ for the parameters in Table \ref{tab:relativistic}.}
\end{figure}

As an example we show in Fig. \ref{fig:Relativistic_SS} the primitive variables at $t=0.55$, for the initial parameters in Table \ref{tab:relativistic}.

\section{FInal comments}
\label{sec:final}

In this academic article we have described in detail the implementation of the exact solution of the 1D Riemann in the newtonian and relativistic regimes, which according to our experience is not presented in a straightforward enough recipe in literature.

The contents in this article can be used in various manners, specially to:
i) test numerical solutions of the Newtonian Riemann problem in basic courses of hydrodynamics,
ii) test numerical implementations of codes solving hydrodynamical relativistic equations,
iii) understand the different properties of the propagation of the different type of waves developing in a gas and the different conditions on the hydrodynamical variables in each case.

It is also helpful because with our approach it is possible to straightforwardly implement the exact solution, and this will save some time to a student starting a career in astrophysics involving hydrodynamical processes.


\section*{Acknowledgments}

This research is partly supported by grants: 
CIC-UMSNH-4.9,4.23 and 
CONACyT 106466.
(J.P.C-P and F.D.L-C) acknowledge support from the CONACyT scholarship program.


\end{document}